\documentclass[12pt]{article}
\usepackage{amsfonts,amssymb}
\usepackage{mathtools}
\usepackage{graphicx}
\usepackage{wrapfig}
\usepackage{caption}
\usepackage{feynmp-auto}
\usepackage{comment}
\usepackage[colorlinks,linkcolor=black]{hyperref}

\makeatletter
\newcommand*\wt[1]{\mathpalette\wthelper{#1}}
\newcommand*\wthelper[2]{%
        \hbox{\dimen@\accentfontxheight#1%
                \accentfontxheight#11.15\dimen@
                $\m@th#1\widetilde{#2}$%
                \accentfontxheight#1\dimen@
        }%
}

\newcommand*\accentfontxheight[1]{%
        \fontdimen5\ifx#1\displaystyle
                \textfont
        \else\ifx#1\textstyle
                \textfont
        \else\ifx#1\scriptstyle
                \scriptfont
        \else
                \scriptscriptfont
        \fi\fi\fi3
}
\makeatother

\newcommand*{\sgn}{\ensuremath{\mathrm{sgn}}}
\renewcommand*{\mod}{\ensuremath{\mathrel{\mathrm{mod}}}}
\renewcommand*{\Re}{\ensuremath{\mathrm{Re}}}
\renewcommand*{\Im}{\ensuremath{\mathrm{Im}}}
\newcommand*{\Z}{\ensuremath{{\mathbb{Z}}}}
\newcommand*{\R}{\ensuremath{{\mathbb{R}}}}
\newcommand*{\cp}{\ensuremath{\cup}}
\newcommand*{\capp}{\ensuremath{\cap}}

\newcommand*{\vt}{\ensuremath{{\mathrm{v}}}}
\newcommand*{\lk}{\ensuremath{{\mathrm{l}}}}
\newcommand*{\pl}{\ensuremath{{\mathrm{p}}}}
\newcommand*{\cb}{\ensuremath{{\mathrm{c}}}}
\newcommand*{\td}{\ensuremath{{\mathrm{t}}}}

\renewcommand*{\a}{\ensuremath{\mathrm{a}}}
\renewcommand*{\b}{\ensuremath{\mathrm{b}}}
\renewcommand*{\c}{\ensuremath{\mathrm{c}}}
\renewcommand*{\d}{\ensuremath{\mathrm{d}}}
\newcommand*{\e}{\ensuremath{\mathrm{e}}}

\newcommand*{\M}{\ensuremath{{\mathcal{M}}}}
\newcommand*{\N}{\ensuremath{{\mathcal{N}}}}
\newcommand*{\lat}{\ensuremath{{\mathrm{lat.}}}}

\begin{document}

\title{\Large \bf Abelian Topological Order on Lattice Enriched with Electromagnetic Background}
\author{ \large Jing-Yuan Chen \\ \\
{\small\em Stanford Institute for Theoretical Physics, Stanford, CA 94305, USA}}

\date{}

\maketitle

\begin{abstract}
In topological phases of matter, the interplay between intrinsic topological order and global symmetry is an interesting task. In the study of topological orders with discrete global symmetry, an important systematic approach is the construction of exactly soluble lattice models. However, for continuous global symmetry, in particular the electromagnetic $U(1)$, the lattice approach has been less systematically developed. In this paper, we introduce a systematic construction of effective theories for a large class of abelian topological orders on three-dimensional spacetime lattice with electromagnetic background. We discuss the associated topological properties, including the Hall conductivity and the spin-c nature of the electromagnetic background. Some of these effective spacetime lattice theories can be readily mapped to microscopic Hamiltonians on spatial lattice; others may also shed light on their possible microscopic Hamiltonian realizations. Our approach is based on the gauging of $1$-form $\mathbb{Z}$ symmetries. Our construction is naturally related to the continuum path integral of (doubled) $U(1)$ Chern-Simons theory, through the latter's formal description in terms of Deligne-Beilinson cohomology; when the global symmetry is dropped, our construction can be reduced to the Dijkgraaf-Witten model of associated abelian topological orders, as expected.
\end{abstract}

\pagebreak

\parskip .0ex
\tableofcontents

\pagebreak

\parskip .1ex

\section{Introduction}

Topological phases of matter is an important subject of modern condensed matter physics. It connects between fascinating phenomena in experiments and profound structures in mathematics. One of the earliest and most prominent example which opened the field was the fractional quantum Hall effect, whose exotic properties observed in experiments found their natural interpretations in Chern-Simons theory \cite{Wen:1995qn}. Chern-Simons theory \cite{Witten:1988hf}, on its own, is among the richest subjects in theoretical physics and mathematical physics over the past three decades, and has to led to much cross fertilization at the frontiers of these fields.

Since the quantum Hall effect, the general concept of topological phases of matter, along with the interplay with global symmetry, has been greatly developed. During the course of the development, one important approach has been the construction of exactly soluble lattice models, for examples the spacetime lattice models in the topological field theory perspective \cite{Dijkgraaf:1989pz, Turaev:1992hq, Barrett:1993ab}, the spatial lattice models in the condensed matter perspective \cite{Levin:2004mi, Kitaev:2006lla} and in the quantum computation perspective \cite{Kitaev:1997wr}. The lattice models provide an explicit means to compute and understand the topological properties, and (for spatial lattice models), at least in principle, show the realizability of the phases in strongly interacting solid state systems.

While the construction of exactly soluble lattice models is such an important approach, it is believed that not all topological phases admit exactly soluble lattice models. The topological phases that are known to admit exactly soluble lattice models mainly belong two kinds: those that admit gappable boundary conditions \cite{Turaev:1992hq, Barrett:1993ab, Levin:2004mi, Kirillov:2010nh, Kirillov:2011mk, kitaev2012models, Bhardwaj:2016clt, Cong:2017ffh}, and those that are close reminiscences of free fermions \cite{Kitaev:2006lla, Levin:2011hq}. It is then natural to ask, for topological phases beyond these kinds, what can we say about the lattice implementation and the (loss of) solubility? Under this general theme, in this paper we explore a particular class of topological orders with global symmetry -- abelian topological order in three spacetime dimensions, coupled to a global electromagnetic $U(1)$ symmetry background. This class of topological orders is relatively simple and intuitive to understand, and provides some interesting prospect into our general question above.

To introduce our problem in better details, let us first recall that, in general, a topological phase of matter is characterized by two universal aspects, its intrinsic topological order, and its enrichment with global symmetry \cite{Wen:2016ddy}. The intrinsic topological order is probed by inserting extended observables, such as Wilson loops, which amounts to creating topological excitations in the system, and then examining their braiding and fusion. On the other hand, the interplay with global symmetry is probed by turning on a background gauge field associated with the symmetry, and then examining the appropriate responses of the system; the fractionalized electric charge and the Hall conductivity in quantum Hall systems, for instance, are probes of such kind, where the background electromagnetic field is associated with the $U(1)$ charge conservation of the electronic system.

The topological phases we have in mind have intrinsic abelian topological orders, and they are further enriched by $U(1)$ global symmetries. They are interesting in the sense that, if we ignore their global symmetry enrichment and consider their intrinsic topological orders only, then they admit gappable boundary conditions, and the associated exactly soluble lattice models have already been constructed and extensively studied \cite{Kapustin:2010hk, Wang:2012am, Levin:2013gaa, Barkeshli:2013yta, Kapustin:2013nva, Lin:2014aca}. However, once we take into account the global $U(1)$ symmetry, the situation could have changed so substantially that an exactly soluble lattice model is no longer possible \cite{kapustin2018local}. Therefore, it would be particularly interesting to understand in details what has happened upon the introduction of the global symmetry, and how we can model the system on the lattice after the introduction of the global symmetry.

It turns out that a key to understand why this can happen is associated with the breakdown of an important principle commonly applied in exactly soluble lattice models. The principle is the ``fixed-point'' principle \cite{Levin:2004mi}, which can be phrased as that: a topological lattice theory can be \emph{equally well} viewed either as a (toy) microscopic lattice description, or as an effective description in the coarse-grained topological limit of a gapped system; indeed, the name ``fixed-point'' refers to the identification of an effective description as a microscopic description. Upon the inclusion of the $U(1)$ global symmetry, however, an effective description on a coarse-grained ``lattice'' can no longer be identified with a microscopic description on a lattice; there will be a non-trivial difference between this two views. It turns out that the effective description on a coarse-grained ``lattice'' is still exactly soluble, but non-trivial treatment must be involved to turn this into any legitamate microscopic description, and this process makes the latter not exactly soluble in certain cases. 

The main body of this paper will be devoted towards a systematic construction of the \emph{effective description} on three-dimensional spacetime lattice. This is done systematically by gauging $1$-form $\Z$ symmetries \cite{Gaiotto:2014kfa}. During this process, we also gain some additional insights: first, we are able to manifest the spin-c \cite{Seiberg:2016rsg} nature of the background electromagnetic field when the topological order is fermionic \cite{Gu:2012ib, Gu:2013gma, Tarantino:2016qfy, Gaiotto:2015zta, Bhardwaj:2016clt}; second, our effective description on coarse-grained spacetime lattice can be explicitly mapped to the continuum presentation of the associated topological orders in terms of (twisted) doubled Chern-Simons theory, via the language of Deligne-Beilinson double cohomology \cite{Carey:2004xt, Bauer:2004nh, Guadagnini:2008bh, Guadagnini:2014mja, Mathieu:2016txt}.

After the effective description, we will discuss what process it takes to turn an effective description on a coarse-grained spacetime lattice to a legitimate (toy) microscopic description on lattice. In some cases -- when the Hall conductivity vanishes -- such map can be readily done in an exactly soluble manner \cite{Levin:2011hq}, while in other cases -- when the Hall conductivity is non-trivial \cite{kapustin2018local} -- our process suggests microscopic lattice models which, although not exactly soluble, may be controllably soluble, as we will elaborate on in later works.

This paper is organized as the following. In Section \ref{sect_bosonic} we consider the bosonic phases. We show how to start with the doubled $\R$ Chern-Simons on spacetime lattice and obtain the doubled $U(1)$ Chern-Simons which describe the bosonic abelian topological orders; we discuss the main properties such as anyon braiding and Hall conductivity. In Section \ref{sect_fermionic} we extend to fermionic topological orders and demonstrate the spin-c nature of the electromagnetic field. In Section \ref{sect_DB} we review the Deligne-Beilinson description of Chern-Simons theories in the continuum, and show how the continuum description can be exactly mapped to our lattice effective description. Doubled abelian Chern-Simons theories admit gapped boundary conditions, which are often studied, while in Section \ref{sect_boundary} we study a gapless boundary condition on the lattice, which may be protected by the electromagnetic $U(1)$ on the boundary. Although our spacetime lattice construction is an effective description, in Section \ref{sect_Hamiltonian} we map it to microscopic lattice Hamiltonian for the cases without Hall conductivity, and make proposals for such a realization for the cases with Hall conductivity.

\section{Bosonic Phases}
\label{sect_bosonic}

In this section we introduce our systematic construction for a large class of bosonic abelian topological orders on $3D$ spacetime lattice with electromagnetic background.

In the continuum, these phases are known to be described by (twisted) doubled $U(1)$ Chern-Simons (CS) theories \cite{Kapustin:2010hk, Wang:2012am, Levin:2013gaa}. Such a $U(1)$ CS theory has $2m$ many $U(1)$ gauge fields $a_i, \, b_i, \: (i=1, \cdots, m)$. On a $3D$ oriented spacetime manifold $\M$ with electromagnetic background $A$, the action is
\begin{align}
S = \int_\M \left( \sum_{i,j = 1}^m \sum_{i=1}^m \frac{n_i}{2\pi}  a^i \d b^i + \frac{k_{ij}}{4\pi} b^i \d b^j  - \sum_i \left( \frac{q_i}{2\pi} a^i + \frac{p_i}{2\pi} b^i \right) \d A \right)
\label{doubledU1CS_cont}
\end{align}
which corresponds to a $K$-matrix and charge vector
\begin{align}
K_{IJ}= \left[ \begin{array}{cc} K_{a^i a^j} = 0 & K_{a^i b^j} = n_i \delta_{ij} \\ K_{b^i a^j} = n_i \delta_{ij} & K_{b^i b^j} = k_{ij} \end{array} \right], \ \ \ \ \ \ Q_I = \left[ \begin{array}{c} Q_{a^i} = q_i \\ Q_{b^i} = p_i \end{array} \right].
\end{align}
The theory is well-defined only if the matrix elements and the electric charges are integers; moreover, $k_{ii}$ must be even for the theory to be bosonic. These requirements are because the $U(1)$ gauge configuration might be topologically non-trivial so that the gauge fields have discontinuities, making the action above ambiguous. A usual way to define the action unambiguously is to express it in a $4D$ bulk whose boundary is $\M$, and demand the theory to be independent of the choice of the $4D$ bulk \cite{Dijkgraaf:1989pz, Witten:2003ya, Seiberg:2016rsg}; an equivalent alternative way is to include correction terms to the action that involve transition functions \cite{Guadagnini:2014mja}. The details of both methods are reviewed in Section \ref{ssect_DBrev}, and this leads to the said quantization condition of the $K$-matrix elements and the electric charges.

In this section our task is to realize this action on the lattice, including the electromagnetic background. We start with an introduction to doubled $\R$ CS on lattice in Section \ref{ssect_RRCS}. The lattice can be either a cubic lattice or an arbitrary simplicial complex (a special case, the $\R$-valued BF theory, has been previously studied \cite{Kantor:1991ty, Adams:1996yf}
\footnote{The construction in \cite{Adams:1996yf} has incorporated some $U(1)$ aspects by hand (that cannot be implemented locally on the lattice), such as the normalization of the large gauge transformations, but other $U(1)$ features are missing, for instance the level quantization is not required by consistency in this construction, and there is no summation over flat connection in spacetime with torsion. Therefore we refer to this construction still as $\R$-valued.}
). In Section \ref{ssect_U1U1CS} we gauge the $1$-form global $\Z$ symmetries of the $\R$ gauge fields to obtain the (twisted) doubled $U(1)$ CS on the lattice. The quantization of the $K$-matrix elements is required by the consistency of the procedure, i.e. by the cancellation of the anomalies of the $1$-form symmetries \cite{Gaiotto:2014kfa}. We then introduce the background electromagnetic field and compute the Hall conductivity. We show our construction reduces to the Dijkgraaf-Witten construction \cite{Dijkgraaf:1989pz} with discrete gauge group $G=\Z_{n_1} \oplus \cdots \oplus \Z_{n_m}$ (by a flip of $a_i$ we can make all $n_i$ positive, and this is understood in the below) when the electromagnetic background is trivial. In the discussion we will mainly take $m=1$, as the generalization to other values of $m$ is straightforward.

In Section \ref{ssect_U1RCS} we use a similar procedure to obtain a $U(1)\times \R$ CS, which is interesting that its observables coupled to the $U(1)$ sector only are completely the same as those in a single $U(1)$ CS theory. Moreover, the complex phase of its partition function is the same as a that of single $U(1)$ CS theory (in canonical framing), though the overall norm is different.

\subsection{Doubled $\R$ Chern-Simons on Spacetime Lattice}
\label{ssect_RRCS}

\subsubsection{On Cubic Lattice}

\begin{figure}
\centering
\includegraphics[width=.5\textwidth]{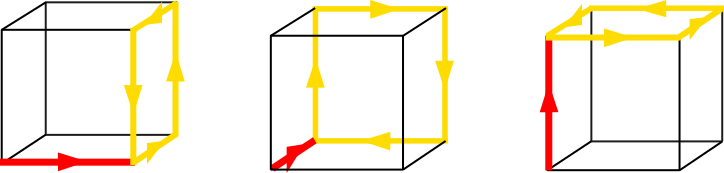}
\caption{$b \cp db$ on each cube is a sum of three terms. For each term, $b$ on the red link is multiplied to $db$ around the yellow plaquette, respecting the right-hand-rule. That is, $b$ on each link is multiplied to $db$ around the plaquette centered at $\hat{x}/2+\hat{y}/2+\hat{z}/2$ away from the center of the link.}
\label{f2-1_bdb}
\end{figure}

We first provide the view on cubic lattice, since it is easier to picture. The cubic lattice might be infinite or finite without boundary. There are two dynamical gauge fields in the doubled CS theory (assuming $m=1$), $a$ and $b$, which are 1-forms in the continuum. They appear in the lattice gauge theory as the following:
\begin{itemize}
\item
$b$ takes $\R$ value on each (directed) link. $db$ takes $\R$ value on each (oriented) plaquette, given by the (signed) sum of the $b$'s on the four links around the plaquette.
\item
$a$ takes $\R$ value on each (oriented) plaquette, i.e. on each link of the dual lattice. $d^\star a$ takes $\R$ value on each (directed) link, i.e. on each plaquette of the dual lattice, given by the (signed) sum of the $a$'s on the four plaquettes around the link.
\end{itemize}
The doubled $\R$ CS theory on lattice is given by
\begin{align}
Z & = \int [Da]^\R \: [Db]^\R \ e^{iS}, \nonumber \\[.2cm]
S & =  \frac{n}{2\pi} \int a \cdot db + \frac{k}{4\pi} \int b \cp db \equiv \frac{n}{2\pi} \sum_{\mathrm{plaq.} \: \pl} a_\pl (db)_\pl + \frac{k}{4\pi} \sum_{\mathrm{cubes} \: \cb} (b \cp db)_\cb
\end{align}
where $[Da]^\R$ means the Faddeev-Popov measure for $\R$-valued gauge field; we will comment more on the normalization later in Section \ref{sssect_U1R_partition}, while for now we only compute expectation values. It does not matter whether we think of the Lorentzian or Euclidean signature, since the topological theory stays the same. The term $a\cdot db$ is associated to each lattice plaquette. The term $b \cp db$, associated to each cube, is explained by Figure \ref{f2-1_bdb}. Essentially, we multiple $b$ on each link $\lk$ to $db$ around the plaquette centered $\hat{x}/2+\hat{y}/2+\hat{z}/2$ away from the center of the link $\lk$. More generally, for $X \cp Y$, the $X$ and $Y$ fields live on objects whose dimensions add up to $3$ (e.g. $X$ may be a field on cubes and $Y$ on vertices), and the $X$ field is multiplied to the $Y$ field $\hat{x}/2+\hat{y}/2+\hat{z}/2$ away. (We can choose other combinations of signs, as long as we use the same choice for all the cubes.) Clearly the $\cp$ will become the cup product when we consider a simplicial complex.

We should check the gauge invariance of the theory. The gauge transformation $\varphi$ associated with $a$ takes real value on each cube, i.e. on each vertex of the dual lattice, such that
\begin{align}
a \ \rightarrow \ a + d^\star \varphi,
\end{align}
where for each plaquette, $d^\star \varphi$ is the difference of the gauge transformations $\varphi$ on the cubes on the two sides of the plaquette. Under this gauge transformation, one can explicitly check the action changes by $n/2\pi$ times
\begin{align}
-\int \varphi \cdot d(db) \equiv -\sum_{\mathrm{cubes} \: \cb} \varphi_c \  \left( d (db) \right)_\cb = 0.
\end{align}
The gauge transformation $\kappa$ associated with $b$ takes real value on each vertex, such that
\begin{align}
b \ \rightarrow \ b + d\kappa,
\end{align}
where for each link, $d\kappa$ is the difference of the gauge transformations $\kappa$ on the vertices at the two ends of the link. Under this gauge transformation, $db$ is invariant, and one can explicitly check the action changes by $k/2\pi$ times
\begin{align}
-\int \kappa \cp d(db) \equiv -\sum_{\mathrm{cubes} \: \cb} \kappa_{\cb-\hat{x}/2-\hat{y}/2-\hat{z}/2} \  \left( d(db) \right)_\cb = 0.
\end{align}
This justifies our identification of $a$ and $b$ as $\R$-valued gauge fields.

We emphasize that, in an $\R$-valued CS theory, the parameters $n$ and $k$ have \emph{no} reason to be quantized to integers. Indeed, in the $\R$ CS theory we may change the values of $n$ and $k$ just by rescaling our definitions of $a$ and $b$. Only when we reduce the theory to $U(1)$ will their quantization become necessary.

Let's now consider some Wilson loop observables. The action becomes
\begin{align}
S[W, V] =  \frac{n}{2\pi} \int a \cdot db + \frac{k}{4\pi} \int b \cp db - \int a \cdot W - \int b \cp V 
\label{theory_RR_cubic}
\end{align}
where the Wilson loops are
\begin{itemize}
\item
$W$ takes integer value on each (oriented) plaquette, such that $dW=0$, i.e. the sum of $W$'s coming out of any cube is zero. Equivalently, $W$ takes integer value on each link of the dual lattice, such that the divergence vanishes at each vertex of the dual lattice.
\item
The same for $V$. 
\end{itemize}
\footnote{In the previous literature on BF theory \cite{Kantor:1991ty, Adams:1996yf}, it is customary to let $V$ live on the links rather than plaquettes, and then our $\int b\cp V$ term becomes $\int b\cdot V$. As we proceed we shall see there is no substantial difference.}
In the simplest cases, $W, V$ will take the shape of one or more disjoint closed loops going through plaquettes and cubes. However, we can also consider configurations that involve the intersecting, merging and splitting of Wilson loops, which have no continuum counterpart but are nevertheless well-defined on the lattice. We will come back to this issue later.

We emphasize that, in an $\R$-valued CS theory, the $W$ and $V$ observables have \emph{no} reason to take integer values. When we reduce the theory to $U(1)$, however, they must take integer values. At this point we are just assuming integer values to make convenient connection to the $U(1)$ theory later.

\begin{figure}
\centering
\includegraphics[width=.2\textwidth]{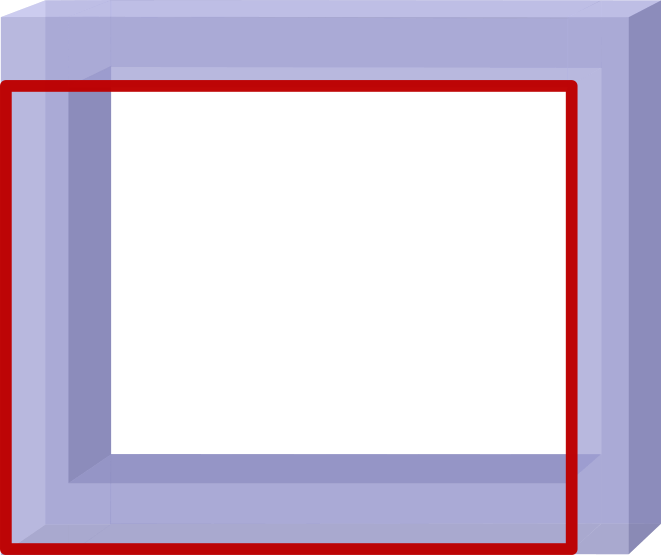}
\caption{A Wilson loop $W$ (blue) going through plaquettes and cubes, and its associated frame $W_f$ (red) going through links and vertices. $W_f$ is located $-\hat{x}/2-\hat{y}/2-\hat{z}/2$ away from the central line of $W$.}
\label{f2-1_framing}
\end{figure}

\begin{figure}
\centering
\includegraphics[width=.5\textwidth]{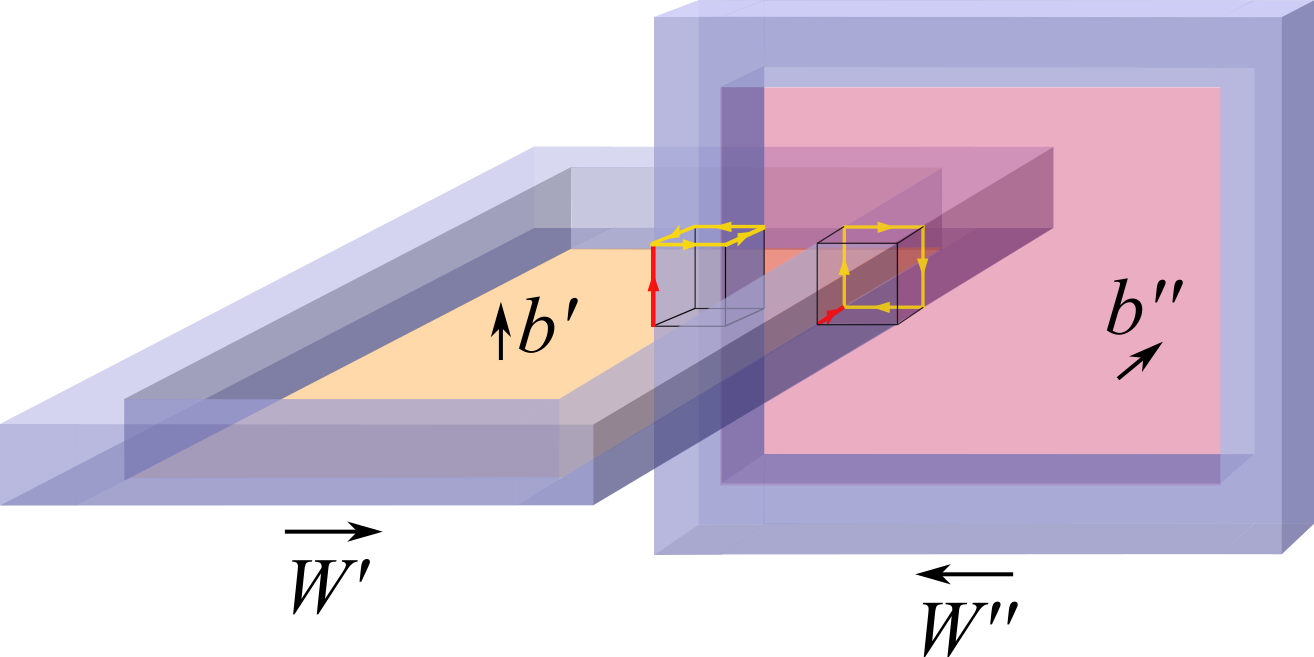}
\caption{A Hopf link consists of two disjoint Wilson loops $W=W'+W''$. Let's say they take values $w'$ and $w''$ (their \emph{charges}) respectively on the plaquettes they go through. We can make $b=b'+b''$ such that $db'=(2\pi/n)W'$, $db''=(2\pi/n)W''$. In the Hopf link example, we can choose $b'=(2\pi/n)w'$ on those $z$-direction links which start on the yellow surface bounded by $W'$, and choose $b''=(2\pi/n)w''$ on the those $y$-direction links which start on the pink surface bounded by $W''$. There are two cubes being indicated. The left one is where $b'\cp db'' = (2\pi/n)^2 w' w''$, capturing $\mathrm{L}(W', W''_f)=w' w''$ (how the framing arises can be seen from the red link on the indicated cube). The right one is where $b''\cp db' = (2\pi/n)^2 w' w''$, capturing $\mathrm{L}(W'', W'_f)=w' w''$. One can also see $b'\cp db' = 0 =b''\cp db''$ everywhere in this example. Thus, in total, $\int b\cp d\b=(2\pi/n)^2 \mathrm{L}(W, W_f)=(2\pi/n)^2 2w' w''$.}
\label{f2-1_hopf}
\end{figure}

\begin{figure}
\centering
\includegraphics[width=.4\textwidth]{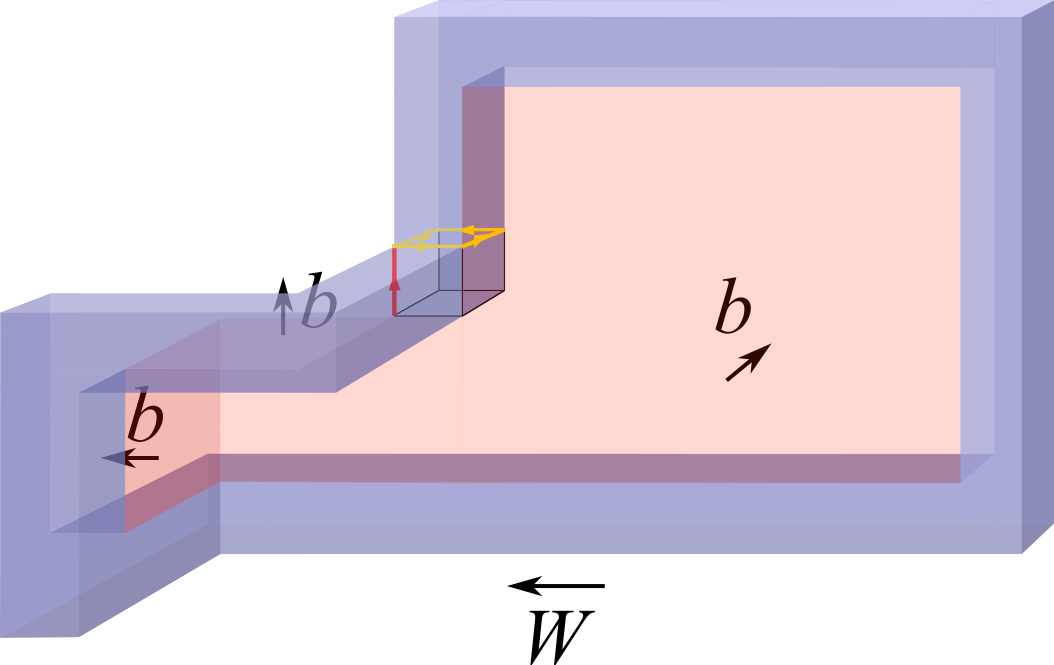}
\caption{A Wilson loop $W$ with a coil. Let's say it takes value $w$ on the plaquettes it goes through. We can choose $b=(2\pi/n)w$ on links which start on the orange surface bounded by $W$. The indicated cube is where $b\cp db = (2\pi/n)^2 w^2$, capturing the self-linking number $\mathrm{L}(W, W_f)=w^2$ (how the framing arises can be seen from the red link on the indicated cube).}
\label{f2-1_coil}
\end{figure}

Now we compute the Wilson loop observables. Let's set $V=0$ first, and consider non-trivial $W$. A crucial feature of doubled CS is that the $a$ field is a Lagrange multiplier. Integrating out the $a$ field leads to the constraint
\begin{align}
db = \frac{2\pi}{n} W
\end{align}
on each plaquette.
We focus on the local aspects of the theory for now, so let's assume $W$ is contractible, and therefore the solution for $b$ exists; in fact, the local aspects are completely the same for $\R$ and $U(1)$ CS. (On the other hand, if the cubic lattice is a three-torus spacetime and $W$ wraps around a periodic direction, there would be no solution for $b$. This is related to the fact that our theory is $\R$-valued. We will leave the discussion of the topological aspects until we have the $U(1)$ instead of $\R$ theory.) Then, the $\int b\cp db$ term just measures the self-linking number of $W$. Let's make this more precise. Let $W_f$, the \emph{frame} of $W$, be the loop(s) going through the links and vertices that are $-\hat{x}/2-\hat{y}/2-\hat{z}/2$ away from the plaquettes and cubes that $W$ goes through; see Figure \ref{f2-1_framing} for instance.
\footnote{Such notion of framing on cubic lattice appeared in the ``loop model'' \cite{Fradkin:1996xb}, which was a non-local model proposed to have some $U(1)$ CS anyon observables. Historically, the ``loop model'' was incorrectly regarded as not to carry topological spin, which is not true as we will see soon in the example of Figure \ref{f2-1_coil}.}
Then $\int b\cp db$ is equal to $(2\pi/n)^2$ times the linking number $\mathrm{L}(W, W_f)$. We demonstrate this with the examples of a Hopf link in Figure \ref{f2-1_hopf} and a coiled loop in Figure \ref{f2-1_coil}. Therefore, we find the Wilson loop observable is
\begin{align}
\left\langle e^{-i\int a\cdot W} \right\rangle \equiv \frac{Z[W]}{Z[0]} = \exp \left(i\frac{\pi k}{n^2} \mathrm{L}(W, W_f)\right).
\end{align}
Indeed, abelian CS theories in the continuum compute linking numbers that require a frame prescription \cite{Witten:1988hf} (again, the local aspects do not care about whether the theory is $U(1)$ or $\R$).
\footnote{Here we are referring to Witten's topological, or artificial point-splitting, frame prescription. Another choice of prescription, Polyakov's geometrical frame prescription \cite{Polyakov:1988md}, requires a metric and is therefore non-topological, and will not be considered in this paper. (Here we mention that the geometrical frame prescription also admits a lattice version. On cubic lattice, using the \emph{average} over $\pm\hat{x}/2\pm\hat{y}/2\pm\hat{z}/2$ in place of $\hat{x}/2+\hat{y}/2+\hat{z}/2$ in the definition of $\cp$ implements the geometrical prescription with flat metric. On simplicial complex, the same idea leads to a super-commuting, i.e. wedge product like, variant of cup product.) \label{geo_framing}}
In our lattice construction, a ``standardized choice'' of Wilson loop frame is provided by our notion of $\cp$. (If we use other choices, for instance $\hat{x}/2+\hat{y}/2-\hat{z}/2$, in our notion of $\cp$, the linking number may change.) For the example of Figure \ref{f2-1_hopf}, the phase is interpreted as the anyon braiding phase, while for the example of Figure \ref{f2-1_coil}, the phase is interpreted as the anyon statistical phase or the anyon topological spin.

It is easy to include the $V$ loop, still assuming it contractible for now:
\begin{align}
\left\langle e^{-i\int a\cdot W - i\int b\cdot V} \right\rangle \equiv \frac{Z[W, V]}{Z[0]} = \exp \left(i\frac{\pi k}{n^2} \mathrm{L}(W, W_f) - i\frac{2\pi}{n} \mathrm{L}(W, V_f) \right).
\end{align}
The linking number coefficients are given by $K^{-1}$, the inverse of the $K$-matrix, as expected from the continuum. The case of $k=0$ is the $\R$-valued lattice BF theory that has been previously studied \cite{Kantor:1991ty, Adams:1996yf}, except our $\int b\cp V$ is expressed in the equivalent form $\int b\cdot V_f$ there (and dropping the $f$ subscript).

At this point we would like to comment on a subtlety that occurs on the lattice. In the continuum, we assume the Wilson loop observables never self-intersect. On the lattice, we can consider configurations where the Wilson loops intersect, split or merge, as long as the charge is conserved, $dW=0, dV=0$. The linking numbers $\mathrm{L}(W, W_f)$ and $\mathrm{L}(W, V_f)$ are nevertheless still well-defined, because $W, V$ are on the plaquettes while $W_f, V_f$ are on the links. This lattice scenario does not have a continuum counterpart, because in the continuum, we always assume the framed loop $W_f$ to be arbitrarily close to the loop $W$, while this might not be true in the lattice scenario with self-intersecting $W$.
\footnote{To better demonstrate this point, consider the Figure \ref{f2-1_hopf} example, in which $\mathrm{L}(W'', W'_f)=\mathrm{L}(W', W''_f)$. Indeed, this equality always holds in the continuum, since a frame is ``arbitrarily close'' to the original loop. Is there a case on the lattice that they are unequal? One can check that in the absence of boundary, $\int b' \cp db'' = \int db' \cp b''$ holds (as a lattice Leibniz rule), but unlike the continuum wedge product, it is not necessarily true that $\int db' \cp b''$ equals $\int b'' \cp db'$. In fact, for $dY=0$, $\int Y \cp X \neq \int X \cp Y$ can happen only if $dX$ geometrically intersects $Y$ (this is also true when we later consider cup product on simplicial complex; this fact is related to the notion of Steenrod product \cite{Gaiotto:2015zta}). Consider the Figure \ref{f2-1_hopf} example. Let's slowly ``pull apart'' the two loops $W'$ and $W''$, so that the linking numbers change from $\mathrm{L}(W'', W'_f)=\mathrm{L}(W', W''_f) = w'w''$ to $\mathrm{L}(W'', W'_f)=\mathrm{L}(W', W''_f) = 0$. But there is an intermediate stage, when the two loops intersect, at which we have unequal $\mathrm{L}(W'', W'_f)=0$, $\mathrm{L}(W', W''_f)=w' w''$. This intermediate stage may also be related to the anyon exchange phase if $w'=w''$, that half of a braiding is an exchange.}
This subtlety does not cause a substantial problem in the remaining of the paper.

Before we end the discussion on the cubic lattice, we briefly explain why our construction works for \emph{doubled} CS but not for \emph{single} CS. In fact, it is widely speculated that a single CS, which has chiral central charge, may not be realizable in a lattice theory that has strictly local correlations \cite{Kitaev:2006lla}. In our construction, the problem arises in the following way (the problem is unrelated to whether the theory is $U(1)$ or $\R$). Suppose we implement the single CS $S= (k/4\pi) \int_\M b \d b$ in the continuum by the lattice action $S=(k/4\pi) \int b\cp db$ in our notation. The lattice equation of motion reads $(db)_\pl + (db)_{\pl + \hat{x}+\hat{y}+\hat{z}} = 0$ on any plaquette $\pl$. However, this is not the $db=0$ expected from the continuum theory. This lattice equation of motion does not have a unique solution; the issue stays as we include Wilson loops. In the momentum space, this problem is equivalent to saying there are extra undesired zero modes for any $(p_x, p_y, p_z)$ satisfying $p_x+p_y+p_z=\pi$.
\footnote{If one let the cubic lattice be a three-torus with an odd number of vertices in each direction, then $p_x+p_y+p_z=\pi$ cannot be satisfied, and the solution to the equation of motion is unique. However, this is not considered a resolution because it is not generic, and moreover there are still modes with undesired small eigenvalues for $p_x+p_y+p_z\approx \pi$.}
In fact, this issue is not due to our specific definition of $\cp$. As long as the lattice action is local, quadratic in $b$, and strictly odd under reflection, this issue is generic \cite{Berruto:2000dp} due to the famous mode doubling on lattice \cite{Nielsen:1980rz}. This explains why the method does not construct a single CS; if a lattice Maxwell term is added, the doubling mode would be removed \cite{Berruto:2000dp}, but the lattice theory would not be exactly topological and exactly soluble anymore. On the other hand, for doubled CS, the equation of motion from the Lagrange multiplier $a$ field successfully constraints $db$ in a topological manner, as we have seen.

\subsubsection{On Simplicial Complex}

While the cubic lattice is convenient to picture, and also convenient for later computations (Section \ref{sect_boundary}) on the non-topological boundary, the topologies of the spacetime manifold that can be discretized as cubic lattice are rather limited. Therefore, to explore the theory to the full extent, we should consider the construction on an arbitrary simplicial triangulation of the spacetime.

The construction on simplicial complex is obvious given what we have introduced on cubic lattice. Basically, the theory is \eqref{theory_RR_cubic} with $\cp$ understood as the \emph{cup product} on a simplicial complex, which depends on our choice of an ordering for all the vertices.

Let's be more detailed so that we can introduce some notations. We denote the simplicial complex that triangulates the spacetime manifold $\M$ as $M$. Moreover, we use $\mathrm{M}$ for the $3$-chain of $\M$ in the simplicial complex. The dynamical variables of the path integral are:
\begin{itemize}
\item
$b\in C^1(M; \R)$, with gauge invariance $b \rightarrow b + d\kappa$ for $\kappa \in C^0(M; \R)$.
\item
$a\in C_2(M; \R)$, with gauge invariance $a \rightarrow a + \partial \varphi$ for $\varphi \in C_3(M; \R)$. Fixed the simplicial complex, there is an automatic identification of $C_n$ with $C^n$, so we may also view $a$ as a field $a\in C^2(M; \R)$, and $\varphi\in C^3(M; \R)$, and the gauge invariance is then $a \rightarrow a + d^\star \varphi \equiv a + \star d \star \varphi$ using the Poincar\'{e} dual complex.
\end{itemize}
The Wilson loops are:
\begin{itemize}
\item
$W, V \in Z^2(M; \Z)$.
\end{itemize}
(Recall that for $\R$ CS that we consider now, there is no particular reason for them to value in $\Z$, though they have to be when we consider $U(1)$ CS later.) The theory is
\begin{align}
Z[W, V] & = \int [Da]^\R \: [Db]^\R \ e^{iS[W,V]}, \nonumber \\[.2cm]
S[W, V] & = \int a \cdot \left( \frac{n}{2\pi} db - W \right) + \int \left( \frac{k}{4\pi} b \cp db - b \cp V \right).
\label{theory_RR}
\end{align}
Here, $\cp$ is the usual cup product on the simplicial complex,
\footnote{It is important to recall that $X \cp Y - (-1)^{\delta_X \delta_Y} Y\cp X = 0$ (where $\delta$ denotes the degree of the cochains) only holds in the cohomology class context. That is, when both cochains $X, Y$ are closed, the right-hand-side is always exact but does not necessarily vanish as a cochain; when $X, Y$ are not closed, the right-hand-side is in general not closed.}
$\int \, \cdot$ is the defining map from $C_n(M; \R) \times C^n(M; \R)$ to $\R$, and $\int$ of a $C^3(M; \R)$ element means $\int \, \cdot$ of this $C^3(M; \R)$ element with the $3$-chain $\mathrm{M}$ of $\mathcal{M}$. The gauge invariance may be easily checked. Again, we emphasize that for $\R$ CS that we consider now, $n, k$ have no reason to be quantized yet.

We shall now discuss the Wilson loop frame. Previously we mentioned that in Figure \ref{f2-1_bdb} if we replace $\hat{x}/2+\hat{y}/2+\hat{z}/2$ by, say $\hat{x}/2+\hat{y}/2-\hat{z}/2$, we have changed the default Wilson loop frame on the cubic lattice. On a simplicial complex, we have a much larger freedom of doing so while maintaining gauge invariance: we just change the vertex ordering on the simplicial complex, which changes the detailed notion of cup product. More precisely, considering contractible Wilson loops only, the expectation value is the familiar
\begin{align}
\left\langle e^{-i\int a\cdot W - i\int b\cp V} \right\rangle \equiv \frac{Z[W, V]}{Z[0]} = \exp \left(i\frac{\pi k}{n^2} \mathrm{L}(W, W_f) - i\frac{2\pi }{n} \mathrm{L}(W, V_f)\right).
\end{align}
In the linking number, the frame is given by the \emph{cap product}:
\footnote{Recall that $\mathrm{M}\capp W$ and $W\capp\mathrm{M}$ differs by a boundary in general.}
\begin{align}
W_f = \mathrm{M} \capp W \in Z_1(M; \Z), \ \ \ \ \ V_f = \mathrm{M} \capp V \in Z_1(M; \Z)
\label{frame_capp}
\end{align}
where $\mathrm{M}$ is the $3$-chain of $\mathcal{M}$. 
The choice of the vertex ordering corresponds to the choice of the default Wilson loop frame prescription. In particular, if a change of vertex ordering involves the vertices next to some $2$-simplexes that $W$ goes through, then the frame changes by some $\delta W_f = \partial \omega\in B_1(M; \Z)$, where $\omega \in C_2(M; \Z)$ may be non-zero only on those $2$-simplexes that $W$ goes through.

\begin{figure}
\centering
\includegraphics[width=.85\textwidth]{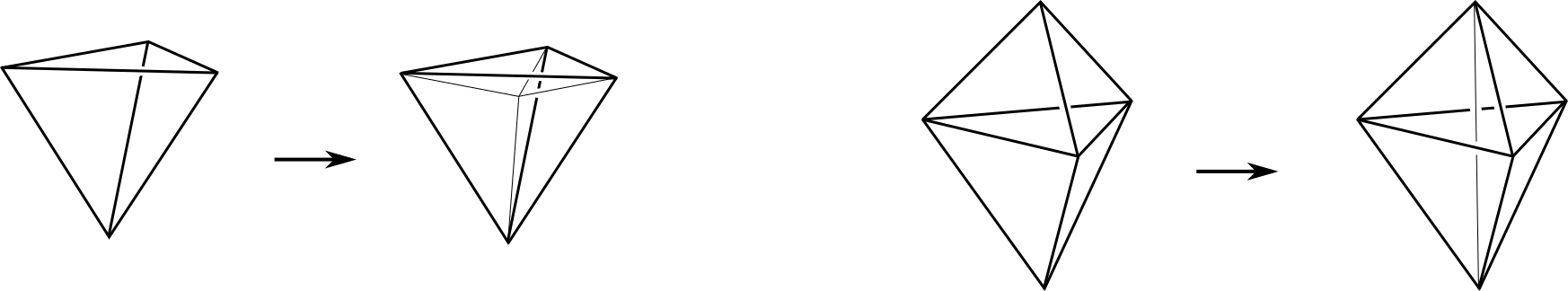}
\caption{The $1-4$ Pachner move and the $2-3$ Pachner move.}
\label{f2-1_Pachner}
\end{figure}

Let's now consider a change of the triangulation of the spacetime. Since the simplicial complex has changed, the allowed possibilities for $W, V$ insertions also change. We would like to have a means to keep track of the Wilson loop observables as we change the triangulation. Starting with two different triangulations of the same $\M$, we can always refine them to arrive at a common finer triangulation, through finite steps of subdivision using the $1-4$ and $2-3$ Pachner moves illustrated in Figures \ref{f2-1_Pachner}. We shall just discuss how we keep track of the Wilson loops under the Pachner moves. The Lagrangian multiplier $a$ always imposes $db = (2\pi/n) W$. This constraint is understood in the below. Either the $1-4$ or the $2-3$ Pachner move involves five tetrahedra and five vertices. Let's say the vertices are labeled by $1, \cdots, 5$. We may consider an abstract $4$-simplex spanned by these five vertices involved in the Pachner move, and the five tetrahedra involved are just the five facets of this abstract $4$-simplex. Then, the change of $b \cp db$ under a Pachner move can be understood as
\begin{align}
\delta \int b \cp db = \int_{4D} db \cp_{4D} db = \pm (db)_{123} \  (db)_{345} = \pm \left(\frac{2\pi}{n}\right)^2 W_{123} \ W_{345}
\end{align}
where $_{4D}$ means the operation is defined on the abstract $4$-simplex, and the sign depends on the orientations of the $123$ and $345$ triangles. Consider the $1-4$ Pachner move first. Since a new vertex has been introduced, we can always let this new vertex be vertex $5$, so the triangle $345$ is a new triangle that did not exist before the Pachner move. It is easy to see that, no matter how $W$ is originally configured before the subdivision, after the subdivision we can always choose the $W$ configuration on the new, internal triangles so that $W_{345}=0$, meanwhile preserving the $dW=0$ condition on each of the four new tetrahedra. In this way, a $1-4$ Pachner move does not change any expectation value. Next let's consider the $2-3$ Pachner move. There is no new vertex being introduced. If triangle $123$ (or $345$) is a new, internal triangle that arises after the subdivision, we can always choose the internal $W$ configuration after the subdivision so that $W_{123}=0$ (or $W_{345}=0$), and no expectation value is changed. However, if both triangle $123$ and $345$ are triangles from the two original tetrahedra, then we cannot set $W_{123}$ or $W_{345}$ to zero. In this case, the self-linking number changes by an integer under the $2-3$ Pachner move. The scenario is therefore a reminiscent of the dependence of self-linking number on the vertex ordering. In this sense, the theory depends on the triangulation \emph{no more than} it does on the vertex ordering, which is the lattice counterpart of the dependence of the continuum CS theory on the Wilson loop frame prescription. The normalization of the partition function of the doubled $\R$ CS is also invariant under the Pachner moves by including appropriate local counter-terms. We will leave this to Section \ref{sssect_U1R_partition}.
\footnote{Later, when we consider doubled $U(1)$ CS and $U(1)\times \R$ CS, there will be \emph{fractional} self-linking numbers when the spacetime topology has torsion, and the fractional ($\mod \Z$) part of the fractional self-linking numbers (related to the complex phase of the partition function) is invariant under the Pachner moves.}

We should also emphasize that the mode doubling problem that prohibited us from defining a single CS on the cubic lattice using $\int b\cp db$ persists on a generic simplicial complex. The equation of motion $\mathrm{M} \capp db + db \capp \mathrm{M} = 0$ in general has non-unique solutions, because the two terms are unequal in general (differ by an exact term). Therefore this construction is still only good for doubled CS, as expected.

Finally, we mention that usually the physical spacetime is taken to be $\M=\Sigma\times S^1$ where $\Sigma$ is an oriented $2D$ space and $S^1$ is the periodic time. In this case it is convenient to discretize the spacetime into prisms extending in the time direction, with the trianglar faces triangulating $\Sigma$. The notion of $\cp$ can also be defined on the prism discretization, mixing the ways we defined it on cubic lattice and simplicial complex. Our general discussions apply to prism discretization as well. In Section \ref{sect_Hamiltonian} we will consider prism lattice when mapping our spacetime lattice Lagrangian to spatial lattice Hamiltonian.

\subsection{Doubled $U(1)$ Chern-Simons on Spacetime Lattice}
\label{ssect_U1U1CS}

In the above we have introduced the doubled $\R$ CS theory on lattice. As we will see, the local aspects of the $\R$-valued theory and the $U(1)$-valued theory are the same, but the $U(1)$-valued theory has non-trivial topological aspects. The relevant basics of the homology and cohomology groups of a $3D$ oriented manifold are summarized in Appendix \ref{app_alg_topo}.

The lattice construction in the below can be understood either on a simplicial complex, or a cubic lattice (if it can discretize the spacetime manifold $\M$).

\subsubsection{Dirac Strings from Gauging $1$-Form Global $\Z$ Symmetries}

To obtain the doubled $U(1)$ CS, we gauge the $1$-form global $\Z$ symmetries following the general procedure \cite{Gaiotto:2014kfa}; the basic idea of the procedure is similar to the Villain model in condensed matter physics.
Consider the variations $b \rightarrow b + \delta b, \ a \rightarrow a+ \delta a$. It is easy to see the doubled $\R$ CS theory \eqref{theory_RR} is invariant if and only if
\begin{align}
d(\delta b) =0, \ \ \ \ \ \ \partial(\delta a) = 0 \ \ (\mbox{i.e.} \ \ d^\star (\delta a) = 0).
\label{global_R_1-form_symm}
\end{align}
This does \emph{not} necessarily mean they are gauge transformations $\delta b = d\kappa$, $\delta a = \partial \varphi$, because $H^1(M; \R) \cong H_2(M; \R) =\R^{B_1}$ might be non-trivial in general (see Appendix \ref{app_alg_topo} for a summary of the algebraic topology we use). Therefore, we should view them as extra symmetries. They are known as $1$-form global $\R$ symmetries: ``$1$-form'' because the transformation $\delta b \in C^1(M; \R)$ corresponds to a $1$-form field in the continuum, and the transformation $\delta a \in C_2(M; \R)$, if viewed from the dual lattice, also corresponds to a $1$-form field; ``global'' because the transformations are symmetries only if they vanish under $d$ (or $d^\star$, through the dual lattice). This notion is analogous to the usual ($0$-form) global symmetry, where the transformation is a function ($0$-form field) in the continuum, and the transformation is a symmetry only if it vanishes under derivative.

Now we gauge the $1$-form global symmetries. But we only gauge the $2\pi\Z$ parts of the symmetries, i.e. the parts where $\delta b, \delta a$ take $2\pi\Z$ values, rather than the full $\R$ symmetries (otherwise the resulting CS theory will be anomalous as we will see later). Recall that for a usual ($0$-form) global symmetry, to gauge it means that: while the original theory is not invariant when the $0$-form transformation is local, in the new theory we introduce a new dynamical $1$-form gauge field to absorb the local transformation, so that the new theory is invariant, and the local transformation becomes gauge equivalence; the new dynamical gauge field might have its own, non-universal dynamics, such as a Maxwell term in electrodynamics or a vortex fugacity in the Villain model.
\footnote{It may be helpful to review the gauging of $0$-form global $\Z$ symmetry in the Villain model. We start with an $\R$-valued scalar field $\theta$ on the lattice vertices and assume the (Euclidean) action only depends on $d\theta$, say $(-1/2T)\sum_{\mathrm{links}\, \lk}(d\theta)_\lk^2$. Gauging the symmetry of shifting $\theta$ by $2\pi\Z$ leads to $(-1/2T)\sum_{\mathrm{links}\, \lk}(d\theta - 2\pi m)_\lk^2$ where $\theta$ is now $U(1)$-valued and $m$ is an integer gauge field on the links. The integer valued $dm$ on the plaquettes is interpreted as vortex cores, and we may introduce a vortex fugacity term $(-T'/2) \sum_{\mathrm{plaq.} \, \pl}(dm)_\pl^2$ \cite{Peskin:1977kp}. When $T'$ is small, the theory behaves as the XY model, i.e. $U(1)$ non-linear sigma model $(1/T)\sum_{\mathrm{links}\, \lk}\cos(d\theta)_\lk$. When $T'$ is large, the theory locally behaves as the original $\R$-valued $\theta$ theory, though there are global differences: the holonomy of $d\theta-2\pi m$ around a non-contractible circle maybe non-zero but quantized to $2\pi\Z$, revealing the theory is still a $U(1)$ theory. \label{footnote_Villain_model}}
The procedure is analogous for $1$-form global symmetry. We introduce new dynamical $2$-form gauge fields:
\begin{itemize}
\item
$s\in C^2(M; \Z)$, with $1$-form $\Z$ gauge invariance
\begin{align}
b \ \rightarrow \ b + 2\pi \beta, \ \ \ \ \ s \ \rightarrow \ s + 2\pi d\beta, \ \ \ \ \ \ \beta\in C^1(M; \Z),
\end{align}
\item
$l\in C_1(M; \Z)$ ($2$-form in the dual lattice sense), with $1$-form (in the dual lattice sense) $\Z$ gauge invariance
\begin{align}
a \ \rightarrow \ a + 2\pi \alpha, \ \ \ \ \ l \ \rightarrow \ l + 2\pi \partial\alpha, \ \ \ \ \ \ \alpha\in C_2(M; \Z).
\end{align}
\end{itemize}
With these invariances, we can safely let $a, b$ take values in $[0, 2\pi)$ rather than $\R$, so they indeed become $U(1)$ gauge fields. Local quantities that are invariant under the $1$-form gauge transformations include
\begin{align}
db - 2\pi s, \ \ \ \ \ \ \partial a - 2\pi l, \ \ \ \ \ \ ds, \ \ \ \ \ \ \partial l. 
\label{manifestly_1-form_inv}
\end{align}
The former two are understood as \emph{total} fluxes for the $b$ and $a$ fields respectively, and the integer valued $s$ and $l$ are understood as the Dirac strings with Dirac quantization condition.
\footnote{To correspond to $U(1)$ gauge theory in the continuum, we should more precisely think of $-2\pi s$ as a narrow, \emph{visible} $2\pi\Z$ flux of the $U(1)$ gauge field, which can be possibly non-exact due to the presence of an \emph{invisible} Dirac string of strength $2\pi s$ running right next to this narrow flux. This picture will be made concrete in Figure \ref{f4-2_free_rep} in Section \ref{ssect_retrieveU1U1} when we show the connection between our lattice construction and the continuum path integral. In lattice gauge theory terms, we just refer to $s$ as the ``Dirac string'' variable, though what we really mean is its associated visible flux $-2\pi s$. Likewise for $l$.}
In turn, their $3$-form integer ``fluxes'' $ds\in B^3(M; \Z)$ and $\partial l \in B_0(M; \Z)$ are understood as Dirac monopoles. The non-universal dynamics of $s$ and $l$ are associated with the $3$-form fluxes $ds$ and $\partial l$ (just like the Maxwell term in electrodynamics or the vortex fugacity in the Villain model). We \emph{choose} to impose the ``no Dirac monopole'' constraints on them:
\begin{align}
ds=0, \ \ \ \ \ \partial l =0
\label{ds_dl_closed}
\end{align}
These constraints can be imposed by $U(1)$ Lagrange multipliers on the tetrahedra and the vertices respectively.
\footnote{The constraints may equivalently well be imposed by infinitely strong quadratic suppression terms, but it will turn out the Lagrange multipliers are more convenient.}
Under these constraints, we may view $s$ and $l$ as two sets of \emph{conserved} integer charges, so they are in turn associated with their own $U(1)$ global symmetries that will allow us to couple the theory to the electromagnetic background in Section \ref{sssect_EM_Hall}.

Given the constraint $ds=0$, the total flux is conserved, $d(db - 2\pi s)=0$ (the $a$ field is analogous and we will not repeat). While the $db$ part is exact, the integer $s$ part might be closed but non-exact, and this leads to the Dirac-quantized non-trivial flux configurations in a $U(1)$ gauge theory, as desired. Therefore, the flux $db-2\pi s$ for a $U(1)$ theory is locally the same as the $db$ flux for an $\R$ theory, but can have non-trivial topological aspects. Historically, the flux $db - 2\pi s$ satisfying $ds=0$ is called a ``non-compact'' $U(1)$ flux; a ``non-compact'' Maxwell term refers to $(-1/4e^2) \sum_{\mathrm{plaq.} \, \pl} (db - 2\pi s)_{\pl}^2$ with $ds=0$ in Euclidean signature. This historical term ``non-compact'' is because of its local resemblance to an $\R$ gauge theory, despite the important topological difference. In contrast, the historical term ``compact'' $U(1)$ gauge field refers to when $ds$ is subjected to no constraint, or a sufficiently weak constraint, in which case the monopole configurations may fluctuate and lead to confinement \cite{Polyakov:1975rr, Peskin:1977kp}.
\footnote{One can recognize that a ``compact'' Maxwell term $(-1/4e^2) \sum_{\mathrm{plaq.} \, \pl} (db - 2\pi s)_{\pl}^2$ with arbitrary $s$ as the Villain version of $(1/2e^2) \sum_{\mathrm{plaq.} \, \pl} \cos(db)_\pl$, in a manner similar to footnote \ref{footnote_Villain_model}.}
Since the physical electromagnetism does not confine and monopole fluctuation is not observed, the ``non-compact'' Maxwell is the lattice implementation used for physical electromagnetism.

To summarize, the Dirac string variables with the closedness constraints are
\begin{itemize}
\item
$s\in Z^2(M; \Z)$, with $1$-form $\Z$ gauge invariance
\begin{align}
b \ \rightarrow \ b + 2\pi \beta, \ \ \ \ \ s \ \rightarrow \ s + d\beta, \ \ \ \ \ \ \beta\in C^1(M; \Z),
\label{s_gauge_transf}
\end{align}
\item
$l\in Z_1(M; \Z)$, with $1$-form $\Z$ gauge invariance
\begin{align}
a \ \rightarrow \ a + 2\pi \alpha, \ \ \ \ \ l \ \rightarrow \ l + \partial\alpha, \ \ \ \ \ \ \alpha\in C_2(M; \Z).
\label{l_gauge_transf}
\end{align}
\end{itemize}
Using the $1$-form gauge invariances, we can then reduce $a, b$ to $U(1)$ gauge fields:
\begin{itemize}
\item
$b\in C^1(M; U(1))$, with $U(1)$ gauge invariance
\begin{align}
b \ \rightarrow \ b + d\kappa \ \mod 2\pi, \ \ \ \ \ \ \kappa\in C^0(M; U(1)),
\label{b_gauge_transf}
\end{align}
\item
$a\in C_2(M; U(1))$, with $U(1)$ gauge invariance
\begin{align}
a \ \rightarrow \ a + \partial \varphi \ \mod 2\pi, \ \ \ \ \ \ \varphi\in C_3(M; U(1)).
\label{a_gauge_transf}
\end{align}
\end{itemize}
For definiteness, the range $[0, 2\pi)$ is understood when mapping the $U(1)$ into $\R$.

If the original $\R$ theory only involved $db$ and $\partial a$ in the action, we can simply replace them by the first two invariant combinations in \eqref{manifestly_1-form_inv} and obtain a $U(1)$ theory; this is the case for e.g. a ``non-compact'' $U(1)$ Maxwell theory mentioned above. But this is \emph{not} the case for CS theory of our interest. It is non-trivial to ensure the $1$-form $\Z$ gauge invariances \eqref{s_gauge_transf} and \eqref{l_gauge_transf}, which lead to the quantization conditions. Let's see this in detail. The doubled $U(1)$ CS theory on lattice is given by
\begin{align}
Z[W, V] & = \int [Da]^{U(1)} \: [Db]^{U(1)} \: [Dl]^\Z_{\partial l =0} \: [Ds]^\Z_{ds=0} \ e^{iS[W,V]}, \nonumber \\[.2cm]
S[W, V] & = \int a \cdot \left( \frac{n}{2\pi} db - ns - W\right) - n \int l \cdot b \nonumber \\[.2cm] 
& \phantom{=} + \frac{k}{4\pi} \int \left( b \cp db - b\cp 2\pi s - 2\pi s \cp b \right) - \int b \cp V
\label{theory_U1U1_no_A}
\end{align}
(in comparison to \eqref{theory_RR}). We will motivate this expression soon through \eqref{theory_RR_4D}. Here $[Dl]^\Z_{\partial l=0}$ means to sum over the integer $l$ field on each link, subjected to the local constraint $\partial l=0$, and likewise for $[Ds]^\Z_{ds=0}$; for definiteness we impose the constraints by $U(1)$ Lagrange multipliers $\theta\in C^0(M; U(1))$ and $\varpi \in C_3(M; U(1))$ respectively:
\begin{align}
\int \theta \cdot \partial l = \int l \cdot d\theta, \ \ \ \mbox{and} \ \ \ \int \varpi \cdot ds = \int s \cdot \partial\varpi.
\label{closedness_Lagrange_multiplier}
\end{align}
A more detailed description of the path integral measure is given by \eqref{measure_U1U1}.

It is easy to see that under the $1$-form $\Z$ gauge transformations \eqref{s_gauge_transf} and \eqref{l_gauge_transf}, the action changes by
\begin{align}
\delta S = -2\pi \int \alpha \cdot \left( ns + W\right) - 2\pi n \int l \cdot \beta + k\pi \int \left( \beta \cp d\beta - \beta \cp s - s \cp \beta \right) - 2\pi \int \beta \cp V.
\label{deltaS_U1U1}
\end{align}
Since the action must be gauge invariant up to $2\pi$, the theory \eqref{theory_U1U1_no_A} is well-defined if and only if $n\in \Z$, $k\in 2\Z$,
\footnote{Note that $\int \beta \cp s$ is in general not equal to $\int s\cp \beta$ unless $\beta$ is closed, and the point the gauge to the $1$-form global $\Z$ symmetry is to allow non-closed $\beta$.}
and $W, V$ are integer valued. This is how the quantization conditions for the $K$-matrix elements and the observables arise. In the continuum theory \eqref{doubledU1CS_cont} (with $m=1$), one may apply an $SL(2, \Z)$ redefinition of the $(a, b)$ fields and the theory stays equivalent though the $K$-matrix is transformed; on the lattice scale this equivalence is not exact simply because $a$ and $b$ live on plaquettes and links respectively. This lattice theory of doubled $U(1)$ CS allows consistent coupling to the electromagnetic background, which will be the task of Section \ref{sssect_EM_Hall}. The previous lattice Hamiltonian for fractional topological insulator \cite{Levin:2011hq} has close relation with our present spacetime lattice construction, as we will explain in Section \ref{sect_Hamiltonian}.

How to motivate the action \eqref{theory_U1U1_no_A}? One natural way is to let our $3D$ simplicial complex $M$ be the boundary of an oriented $4D$ simplicial complex $N$ -- clearly this lattice $4D$ perspective is related to the continuum version mentioned below \eqref{doubledU1CS_cont} which we will elaborate on in Section \ref{ssect_DBrev}. Thanks to the lattice Stoke's theorem, we can express the doubled $\R$ theory \eqref{theory_RR} in $N$ as (here we ignore $W, V$ for convenience, but one may also include them)
\begin{align}
S= \frac{n}{2\pi} \int_N (\partial a)\cdot (db) + \frac{k}{4\pi} \int_N (db)\cp(db)
\label{theory_RR_4D}
\end{align}
where the fields extended into $N$ are $b\in C^1(N; \R)$ and $a\in C_3(N; \R)$. Since the action consists of $db$ and $\partial a$ only, when gauging the $1$-form global $\Z$ symmetries, we can simply replace them by $db-2\pi s$ and $\partial a - 2\pi l$. Using the Stoke's theorem again, the doubled $U(1)$ theory is
\begin{align}
S=\left( \mbox{Eq.}\eqref{theory_U1U1_no_A} \mbox{ on } M \right) + 2\pi n \int_N l \cdot s + \pi k \int_N s \cp s.
\label{theory_U1U1_4D}
\end{align}
The last two terms cannot be expressed on the $3D$ simplicial complex $M$, but the conditions $n\in \Z$, $k\in 2\Z$ allow us to drop them.
\footnote{For arbitrary (non-quantized) $n$ and $k$, the transformation of these two terms in $N$ indeed cancel \eqref{deltaS_U1U1}.}
This $4D$ perspective provides a natural motivation for the expression of our doubled $U(1)$ CS \eqref{theory_U1U1_no_A}, and is helpful when we study the fermionic phases with odd $k$ in Section \ref{sect_fermionic}.

The theory \eqref{theory_U1U1_no_A} has a residual $1$-form global $\Z_n$ symmetry for $a$:
\begin{align}
\partial(\delta a) = 0, \ \ \ \delta a \in C_2(M; (2\pi/n)\Z_n).
\label{a_res_1-form_symm}
\end{align}
We may further gauge some $\Z_{n'}$ subgroup of this $\Z_n$ symmetry where $n'$ divides $n$. But this is obviously equivalent to rescaling the original $\R$-valued $a$ by $1/n'$ before we reduce it to $U(1)$-valued, and therefore the resulting theory is just to replace $n$ by another integer $n/n'$. The theory \eqref{theory_U1U1_no_A} also has a residual $1$-form global $\Z_g$ symmetry for $b$, where $g=\mathrm{gcd}(n, k)$:
\begin{align}
d(\delta b) = 0, \ \ \ \delta b \in C^1(M; (2\pi/g) \Z_g).
\label{b_res_1-form_symm}
\end{align}
\footnote{$g=\mathrm{gcd}(n, k)$ rather than the weaker $g=\mathrm{gcd}(n, k/2)$ because $\int X\cp Y = \int Y\cp X$ if both $X, Y$ are closed, for $X=\delta b, Y=s$ here.}
Suppose we want to further gauge some $\Z_{g'}$ subgroup of this $Z_g$ symmetry where $g'$ divides $g$. Again this is equivalent to rescaling the original $\R$-valued $b$ by $1/g'$ before we reduce it to $U(1)$-valued. The resulting theory is to replace $n$ by another integer $n/g'$ which is fine since $g'$ divides $n$, and to replace $k$ by $k/g'^2$ which may be problematic since $k/g'^2$ might not be an even integer. When this problem occurs, we say the $1$-form global $\Z_{g'}$ symmetry is anomalous and cannot be gauged \cite{Gaiotto:2014kfa}. Obviously, it would also be anomalous to gauge the full $1$-form global $\R$ symmetries \eqref{global_R_1-form_symm}.

A particularly interesting case of anomalous $1$-form global $\Z_{g'}$ residual symmetry, however, is when $n=2n_f$, $k=4k_f$ for some \emph{odd} integer $k_f$ and some integer $n_f$. In this case, if we gauge a residual $\Z_{g'=2}$ symmetry of $b$, we will obtain a CS described by $(n_f, k_f)$, which is anomalous since $k_f$ is odd and leads to $\pi$ ambiguity. The ambiguity, however, can be fixed by the introduction of fermionic variables -- the task of Section \ref{sect_fermionic}. In this case, the original bosonic theory described by $(n, k)$ is the ``bosonic shadow'' \cite{Bhardwaj:2016clt} of the fermionic theory described by $(n_f, k_f)$.

What we have constructed so far is the lattice realization of the doubled $U(1)$ CS theory \eqref{doubledU1CS_cont} with $m=1$ pair of the $a$ and $b$ fields. The generalization to multiple pairs is obvious, leading to integer matrix elements and in particular even $k_{ii}$. In the below it still suffices for us to demonstrate our main ideas with $m=1$. We however note that the form \eqref{doubledU1CS_cont} does \emph{not} encompass all boundary gappable abelian topological orders, for instance there are CS-like theories with three pairs of abelian gauge fields which detects configurations of Borromean rings (Milnor triple linking) and hosts anyons with non-abelian braiding \cite{ferrari2015topological, He:2016xpi, Putrov:2016qdo}. They also have their DW descriptions \cite{deWildPropitius:1995cf}. These theories will be mentioned in the Conclusion.

\subsubsection{Wilson Loop Observables}
\label{sssect_WL_by_central_Z}

Now that we have constructed the doubled $U(1)$ CS theory \eqref{theory_U1U1_no_A} on lattice, to understand the theory better we compute the Wilson loop observables. We focus on the topological aspects, since the local aspects are the same as in the previous doubled $\R$ theory (and we will verify this). The relevant algebraic topology and notations are summarized in Appendix \ref{app_alg_topo}.

The first thing to note is the $l$ field in \eqref{theory_U1U1_no_A} just imposes the constraint
\begin{align}
b \in C^1(M; (2\pi/n) \Z_n)
\end{align}
up to $U(1)$ gauge equivalence.
\footnote{Recall that the $\partial l=0$ condition is imposed by \eqref{closedness_Lagrange_multiplier} for $\theta\in C^0(M; U(1))$. Summing over $l$ on each link, we have $nb-d\theta$ is valued in $2\pi \Z_n$. By a gauge choice we can remove the $d\theta$.}
The field $(n/2\pi) b$ has therefore become a $\Z_n$ gauge field. From here, the computation can proceed in two slightly different perspectives: a linking number perspective, and a $\Z_n$ Dijkgraaf-Witten (DW) perspective. For now we will introduce the linking number perspective, which is more intuitive and has more of the ``Chern-Simons flavor''. In Section \ref{sssect_HS_PF_DW} we will rigorously reduce the theory (including the normalization of path integral measure) to $\Z_n$ DW model \cite{Dijkgraaf:1989pz}. Formally speaking, the $(k/4\pi)$ term in \eqref{theory_U1U1_no_A} is the DW term, and the linking number method we perform now is to compute the DW term via central extension \cite{Bhardwaj:2016clt} by $\Z$.
\footnote{Yet another linking number perspective, without $l$ being summed over from the beginning, will be mentioned above \eqref{PF_U1U1_alt}.}

In the linking number perspective, we exploit \eqref{s_gauge_transf} and define
\begin{align}
& s = s_\gamma - dy, \ \ [s_\gamma]=[s]=\gamma\in H^2(M; \Z), \ \ \ y\in C^1(M; \Z) \nonumber \\[.1cm]
& \wt{b} \equiv \frac{n}{2\pi} \left(b + 2\pi y\right) \in C^1(M; \Z)
\label{central_extension}
\end{align}
where $s_\gamma$ is a \emph{fixed} representative for the topological classes $[s]=\gamma$ (see Appendix \ref{app_alg_topo}). Upon this redefinition of variables, we may use the integer gauge field $\wt{b}$ and and topological class $\gamma$ as the variables in the lattice partition function:
\begin{align}
Z[W, V] & \propto \int [Da]^{U(1)} \: [D\wt{b}]^{\Z} \: \sum_{\gamma\in H^2(M;\Z)} \ e^{iS[W,V]}, \nonumber \\[.2cm]
S[W, V] & = \int a \cdot \left(d\wt{b} - n s_\gamma - W\right) + \frac{\pi k}{n^2} \int \left( \wt{b} \cp d\wt{b} -\wt{b}\cp ns_\gamma - ns_\gamma \cp \wt{b} \right) - \frac{2\pi}{n}\int \wt{b} \cp V.
\label{theory_U1U1_no_A_b_s}
\end{align}
(Note that we have used $\propto$ for the partition function, because there are topology dependent singular factors arising from the measure.
\footnote{For instance, consider a closed, non-exact $y$ field in the redefinition \eqref{central_extension}. In the new field $\wt{b}$, such $y$ contributes to the $n\Z$ part of the $\Z$-valued flat holonomies (which belongs to $H^1(M; \Z)$ after removing gauge redundancy) around the non-contractible free directions. On the other hand, in the original $s$, such $y$ is equivalent to $0$ and should not be considered part of the original theory; related to this, the original $\Z_n$ gauge field $(n/2\pi) b$ only has $\Z_n$-valued flat holonomies, while the $n\Z$ part is considered large gauge transformation ($2\pi \Z$ rescaled by $n/2\pi$). This difference leads to singular factors in the partition function. Another source of singular factor is the identification of global symmetry, which was originally $\Z_n$ but now $\Z$. We will elaborate on this point later in Section \ref{sssect_HS_PF_DW} when we discuss the partition function.}
Let's ignore the proportionality factors for now since we are computing the expectation values of observables.) The constraint from $a$ reads
\begin{align}
d\wt{b} - n s_\gamma = W.
\label{a_EoM}
\end{align}
Since $\wt{b}$, $s_\gamma$ and $W$ are all integer valued, we can map this constraint from $Z^2(M; \Z)$ into $H^2(M; \Z)$ and further into $\mathcal{F}\equiv \mathrm{Hom}(H_2(M; \Z); \Z)$, and find 
\begin{align}
-n\gamma = [W] \in H^2(M; \Z), \ \ \ \ \ -n[\gamma] = [[W]] \in \mathcal{F}\equiv \mathrm{Hom}(H_2(M; \Z); \Z).
\label{a_EoM_class}
\end{align}
In particular, $[[W]] \in \mathcal{F}$ must be a multiple of $n$, in order for the observable not to vanish. The physical interpretation is familiar: a $W$ ``magnetic'' anyon carries a $U(1)$ flux $db-2\pi s=2\pi/n$, and since the total flux on a closed $2D$ manifold must be Dirac quantized, the total number of $W$ anyons must be a multiple of $n$. This relation uniquely determines $[\gamma]=-[[W]]/n \in\mathcal{F}$, since $\mathcal{F}=\Z^{B_1}$ is free. However, this may not uniquely determine $\gamma$ when torsion is present. To proceed, in the below we first consider spacetime topologies that are free, $H^2(M; \Z) = \mathcal{F}$ (e.g. three-trous where $\mathcal{F}=\Z^3$), then topologies that have torsion only, $H^2(M; \Z)=\mathcal{T}$ (e.g. $\R\mathrm{P}^3$ where $\mathcal{T}=\Z_2$), and finally general topologies $H^2(M; \Z)=\mathcal{F}\oplus\mathcal{T}$.

When $H^2(M; \Z) = \mathcal{F}$, not only $[\gamma]\in\mathcal{F}$ is determined by $W$, but also $\gamma\in H^2(\M; \Z)$ itself: $\gamma=\gamma_W\equiv -[W]/n$. We may write the solution of $\wt{b}$ to \eqref{a_EoM} as
\begin{align}
\wt{b} = \wt{b}^{sol.} + \wt{b}^{flat} 
\end{align}
up to $\Z$ gauge equivalence, where $\wt{b}^{sol.}$ is a \emph{fixed} solution to the constraint \eqref{a_EoM}, while $\wt{b}^{flat}$ is closed but might be non-exact. Thus $[\wt{b}^{flat}] \in H^1(M; \Z) \cong \mathcal{F}$ is not fixed by $W$ and remains to be summed over. Then \eqref{theory_U1U1_no_A_b_s} becomes
\begin{align}
Z[W, V] & \propto \sum_{[\wt{b}^{flat}] \in H^1(M; \Z)} \ e^{iS[W,V]}, \nonumber \\[.2cm]
S[W, V] & = \frac{\pi k}{n^2} \int \left( \wt{b}^{sol.} \cp d\wt{b}^{sol.} - \wt{b}^{sol.} \cp ns_{\gamma_W} - ns_{\gamma_W} \cp \wt{b}^{sol.} \right) - \frac{2\pi}{n}\int \wt{b}^{sol.} \cp V\nonumber \\[.2cm]
& \phantom{=} - \frac{2\pi}{n} \int \wt{b}^{flat} \cp (ks_{\gamma_W} + V).
\end{align}
\footnote{In the last line we used the fact that $\int X\cp Y = \int Y\cp X$ if both $X, Y$ are closed.}
The summation over $\wt{b}^{flat}$ is non-vanishing if and only if the class $k[\gamma_W]+[[V]] \in \mathcal{F} \cong \mathrm{Hom}(H^1(M;\Z), \Z)$ is a multiple of $n$. Again the physical interpretation is familiar: by the equation of motion of $b$, a $V$ ``electric'' anyon simultaneously carries a $U(1)$ flux $da-2\pi l =2\pi/n$ and a $U(1)$ flux $db-2\pi s = 2\pi/k$, and Dirac quantization of these fluxes lead to the constraint on $[[V]]$. On the other hand, denote the contractible loop $d\wt{b}^{sol.}=W+ns_{\gamma_W} \equiv \mathcal{W}$, the first line of $S$ include linking numbers involving the contractible loop $\mathcal{W}$.
\footnote{The integer linking number is well-defined as long as one of the two loops involved is contractible.}
We find
\begin{align}
& \phantom{= \ } \left\langle e^{-i\int a\cdot W - i \int b\cdot V} \right\rangle = \frac{Z[W, V]}{Z[0, 0]} \nonumber \\[.2cm]
& = \ \exp\left(  i\frac{\pi k}{n^2} \mathrm{L}(\mathcal{W}, \mathcal{W}_f) - i\frac{\pi k}{n} \left(  \mathrm{L}(\mathcal{W}, {s_{\gamma_W}}_f) + \mathrm{L}(s_{\gamma_W}, \mathcal{W}_f) \right) - i\frac{2\pi}{n} \mathrm{L}(\mathcal{W}, V_f) \right)
\end{align}
with the constraints that $[[W]]$ is a multiple of $n$ and so is $-k[[W]]/n + [[V]]$. When $W, V$ are contractible, we have $\mathcal{W}=W$, and the expectation value reduces to that in the doubled $\R$ theory, as we claimed.

Now we turn to $H^2(M; \Z) = \mathcal{T}$. With the presence of torsion, even when $W=0$, the constraint $d\wt{b} = ns_\tau$ is not so trivial. For each $\tau\in\mathcal{T}$, there is some finite (minimal) period $\mathrm{p}_\tau$ such that $\mathrm{p}_\tau \tau = 0 \in \mathcal{T}$, i.e. $\mathrm{p}_\tau s_\tau = d\xi_\tau \in B^2(M; \Z)$ for some $\xi_\tau \in C^1(M; \Z)$. Thus, when $\mathrm{p}_\tau$ divides $n$, denoted as $\mathrm{p}_\tau | n$, there is a solution
\begin{align}
\wt{b}_\tau = \frac{n}{\mathrm{p}_\tau} \xi_\tau
\end{align}
up to $\Z$ gauge equivalence. We have to sum over all such $\tau$. Substituting the solution into \eqref{theory_U1U1_no_A_b_s} leads to a complex factor in $Z[0, 0]$:
\footnote{In fact, this determines the complex phase of $Z[0, 0]$. The value of the partition function $Z[0, 0]$ will be discussed in Section \ref{sssect_HS_PF_DW}.}
\begin{align}
\sum_{\tau\in \mathcal{T}, \ \mathrm{p}_\tau | n} \ \exp\left( -i\frac{\pi k}{\mathrm{p}_\tau} \int s_\tau \cp \xi_\tau \right) = \sum_{\tau\in \mathcal{T}, \ \mathrm{p}_\tau | n} \ \exp\left( -i\frac{\pi k}{\mathrm{p}_\tau} \mathrm{L}(s_\tau, (\mathrm{p}_\tau s_\tau)_f) \right) 
\end{align}
Note that both $\xi_\tau$ and $s_\tau=d\xi_\tau/\mathrm{p}_\tau$ are integer valued, so the linking number $\mathrm{L}(s_\tau, (\mathrm{p}_\tau s_\tau)_f) =  \int s_\tau \cp \xi_\tau$ is integer valued; in other words, it is well-defined because $\mathrm{p}_\tau s_\tau$ is contractible. We may define $\mathrm{L}(s_\tau, (s_\tau)_f) \equiv (1/\mathrm{p}_\tau) \mathrm{L}(s_\tau, (\mathrm{p}_\tau s_\tau)_f)$, known as the \emph{fractional} linking number, valued in $\Z/\mathrm{p}_\tau$. An important point is that the fractional part of this fractional linking number does \emph{not} depend on the choice of framing. Suppose we change the framing prescription (change the vertex ordering), which changes $(s_\tau)_f$ by a contractible loop $\partial \omega$, according to the discussion below \eqref{frame_capp}. The fractional linking number $\mathrm{L}(s_\tau, (s_\tau)_f)$ changes only by an integer $\mathrm{L}(s_\tau, \partial \omega)$. Thus, the frame prescription is only needed when the coefficient of $\mathrm{L}$ is not a multiple of $2\pi$. We may therefore drop the $_f$ subscript and use $\mathrm{L}(s_\tau, s_\tau)$ to mean the fractional part. The factor above becomes (recall $k$ is even)
\begin{align}
\sum_{\tau\in \mathcal{T}, \ \mathrm{p}_\tau | n} \ \exp\left( -i\pi k\mathrm{L}(s_\tau, s_\tau) \right).
\label{U1U1_torsion_factor}
\end{align}
This property ensures the partition function $Z[0, 0]$ to be \emph{independent} of the frame prescription, i.e. the vertex ordering. In some cases $Z[0, 0]$ may vanish, for instance when $k=n=2$ and the spacetime is $\R\mathrm{P}^3$ (pictured in Figure \ref{fA_RP3} in Appendix \ref{app_alg_topo}) for which $\mathcal{T}=\Z_2$, $\mathrm{L}(s_1, s_1)=1/2 \mod \Z$.

To include Wilson loops $W$, we shall note that \eqref{a_EoM_class} has a solution if and only if there exist some $\gamma$ such that $-n\gamma = [W]$. Let $\gamma_W$ be one such choice, though it may not be unique since $\gamma_W + \tau$ where $\mathrm{p}_\tau | n$ is an equally good choice. Fixed a choice of $\gamma_W$, the solution to \eqref{a_EoM_class} for $\gamma=\gamma_W + \tau$ with $\mathrm{p}_\tau | n$ is given by
\begin{align}
& \wt{b} = \wt{b}^{sol.} + \wt{b}_\tau + n \wt{b}^{aux.}, \nonumber \\[.2cm]
& d\wt{b}^{sol.} = W + ns_{\gamma_W} \equiv \mathcal{W}, \ \ \ \ \ d\wt{b}_\tau = ns_\tau, \ \ \ \ \ d\wt{b}^{aux.}_\tau = s_{\gamma_W + \tau} - (s_{\gamma_W} + s_\tau),
\end{align}
where the auxiliary part $\wt{b}^{aux.}$ will drop out later since \eqref{s_gauge_transf} ensures the independence of the choice of representative; there is no $b^{flat}$ since we assumed trivial $\mathcal{F}$. Substituting the solution into \eqref{theory_U1U1_no_A_b_s} leads to the expectation value 
\begin{align}
& \phantom{= \ } \left\langle e^{-i\int a\cdot W - i \int b\cdot V} \right\rangle = \frac{Z[W, V]}{Z[0, 0]} \nonumber \\[.2cm]
& = \ \exp\left(  i\frac{\pi k}{n^2} \mathrm{L}(\mathcal{W}, \mathcal{W}_f) - i\frac{\pi k}{n} \left(  \mathrm{L}(\mathcal{W}, {s_{\gamma_W}}_f) + \mathrm{L}(s_{\gamma_W}, \mathcal{W}_f) \right) - i\frac{2\pi}{n} \mathrm{L}(\mathcal{W}, V_f) \right)  \nonumber \\[.2cm]
& \phantom{= \ } \cdot \frac{\sum_{\tau\in \mathcal{T}, \ \mathrm{p}_\tau | n} \ \exp\left( -i\pi k \mathrm{L}(s_\tau, s_\tau) - i 2\pi \mathrm{L}(s_\tau, (ks_{\gamma_W}+V)) \right)}{\sum_{{\tau'}\in \mathcal{T}, \ p_{\tau'} | n} \ \exp\left( -i\pi k \mathrm{L}(s_{\tau'}, s_{\tau'}) \right) }
\label{WL_expectation_general}
\end{align}
where we used the fact that $\mathrm{L}((\mathrm{p}_\tau s_\tau), (s_{\gamma_W})_f) - \mathrm{L}((s_{\gamma_W}), (\mathrm{p}_\tau s_\tau)) \in \mathrm{p}_\tau \Z$. Notice the last line does not depend on the frame prescription.
\footnote{When $k=0$ (BF theory), the summation over $\tau$ requires the class $[V]$ to be some $-n \gamma_V$ in order to be non-vanishing; this can be easily understood, since when $k=0$ the roles of the $a$ field and the $b$ field are symmetric. When $k\neq 0$, the summation over $\tau$ does \emph{not} require the class $k\gamma_W-[V]$ to be some $-n \gamma_V$ in general. For example, consider the spacetime being a Lens space $L_{\mathrm{p}/\mathrm{r}}$ for which $\mathcal{T}=\Z_\mathrm{p}$, $\mathrm{L}(s_\tau, s_\tau)=\tau^2 \mathrm{r}/\mathrm{p} \ \mod \Z$. Take $n=\mathrm{p}$ so that any $\mathrm{p}_\tau$ divides $n=\mathrm{p}$. Moreover, with the choice $n=\mathrm{p}$, the class $[W]=-n\gamma_W$ must be trivial, so for convenience we set $W=0$; in turn, it is impossible to have $[V]=-n\gamma_V$ unless $[V]$ is trivial. The numerator reads $\sum_{\tau=0}^{\mathrm{p}-1} \exp\left( - i\pi k \mathrm{r} \tau^2 / \mathrm{p} + i 2\pi \mathrm{r} \tau [V]/\mathrm{p}  \right)$, and the denominator is without the $[V]$ term. The result is in general non-vanishing for non-trivial $[V]$, for instance when $k=2$, $n=\mathrm{p}=4$, $\mathrm{r}=1$, $[V]=2$, the expectation value is $(2+2i)/(2-2i)=i$.}
When $W, V$ are contractible, $\mathcal{W}=W$ and $\mathrm{L}(s_\tau, (ks_{\gamma_W}+V)_f)\in\Z$, the expectation value agrees with that in a doubled $\R$ theory.

Having considered $H^2(M; \Z) = \mathcal{F}$ and $H^2(M; \Z) = \mathcal{T}$ separately, we are now ready for general topologies $H^2(M; \Z) = \mathcal{F} \oplus \mathcal{T}$. Again \eqref{a_EoM_class} requires there exist some $\gamma_W$ such that $-n\gamma_W = [W]$. Then any $\gamma=\gamma_W + \tau$ where $\mathrm{p}_\tau | n$ also admit solutions; note they share the same $[\gamma]=-[[W]]/n \in\mathcal{F}$. The solution of $\wt{b}$ is $\wt{b} = \wt{b}^{sol.} + \wt{b}_\tau + n \wt{b}^{aux.}+\wt{b}^{flat}$.
\footnote{Both $\wt{b}_\tau$ and $\wt{b}^{flat}$ are considered flat contributions, because both of their contributions to the total flux $db - 2\pi s$ vanish.}
The expectation value of Wilson loops takes the same form as \eqref{WL_expectation_general}, and in addition the summation over $[\wt{b}^{flat}] \in H^1(M; \Z) \cong \mathcal{F}$ demands $-k[[W]]/n + [[V]] \in \mathcal{F}$ to be a multiple of $n$.
\footnote{Again, when $k\neq 0$, there is no requirement for the class $-k[W]/n + [V]$ to be a multiple of $n$.}
This completes the computation of the Wilson loop observables.

\subsubsection{Electromagnetic Background and Hall Conductivity}
\label{sssect_EM_Hall}

As we claimed in the Introduction, a main motivation behind our method of construction is to include coupling to a $U(1)$ electromagnetic background, and demonstrate Hall conductivity. The main element has been mentioned below \eqref{ds_dl_closed}, that the conserved Dirac string variables are the conserved $U(1)$ charges. The $s$ and $l$ Dirac strings are two conserved charges, so in principle we can have two $U(1)$ background gauge fields $A$ and $B$. We expect them to couple to the conserved charges as $q\int l \cdot A + p\int s \cp B$ for integers $p, q$.

Such coupling, however, does not respect the dynamical $1$-form $\Z$ gauge invariances \eqref{s_gauge_transf}\eqref{l_gauge_transf}. This is in fact a reminiscent of an important issue mentioned in the Introduction. We may either interpret a topological lattice theory as a microscopic lattice model or as an effective description in the deep IR topological limit. Usually a same theory may suit both interpretations \cite{Levin:2004mi}. As we include the $U(1)$ background gauge field, however, the two interpretations pose \emph{different} conditions on the theory:
\begin{itemize}
\item[--]
if taken as a microscopic model, a background $U(1)$ gauge flux of magnitude $2\pi$ at the microscopic lattice scale is \emph{invisible} by definition;
\item[--]
if taken as the deep IR limit, a narrow $2\pi$ flux tube at the coarse-grained lattice scale is \emph{visible}, but \emph{indistinguishable} from certain Wilson loop insertions.
\end{itemize}
In the below we develop our construction according to the deep IR effective theory interpretation, in which \eqref{s_gauge_transf}\eqref{l_gauge_transf} are manifestly respected. The effective theory construction will be connected to the microscopic interpretation in Section \ref{sect_Hamiltonian}, through relaxing the $1$-form $\Z$ gauge invariances \eqref{s_gauge_transf}\eqref{l_gauge_transf} from being \emph{fundamental} to \emph{emergent}, as inspired by \cite{Levin:2011hq}.

A systematic procedure to obtain the effective background $U(1)$ couplings is the following. We turn back to the doubled $\R$ CS \eqref{theory_RR}, and include the couplings
\begin{align}
- \frac{q}{2\pi} \int a \cdot dA - \frac{p}{2\pi} \int b \cp dB
\end{align}
to two $\R$-valued background gauge fields, where
\begin{itemize}
\item
$A\in C^1(M; \R)$, with background gauge invariance $A \rightarrow A+ d\phi_A$ for $\phi_A\in C^0(M; \R)$,
\item
Likewise for $B$.
\end{itemize}
Again in the $\R$-valued theory the charges $p$ and $q$ have no reason to be quantized, since they can be rescaled as we rescale $A$ and $B$. Next we gauge the $1$-form global $\Z$ symmetries, including the ones for $A, B$, as if they are dynamical for now. This reduces $A, B$ to $U(1)$-valued, and at the same time introduces new background Dirac string variables $L_A, L_B \in C^2(M; \Z)$, which we demand to be closed. This procedure couples \eqref{theory_U1U1_no_A} to these background variables:
\begin{align}
S & = \int a \cdot \left( \frac{n}{2\pi} db - ns - W - \frac{q}{2\pi}(dA-2\pi L_A)\right) - n \int l \cdot b + q\int l \cdot A \nonumber \\[.2cm] 
& \phantom{=} + \frac{k}{4\pi} \int \left( b \cp db - b\cp 2\pi s - 2\pi s \cp b \right) - \int b \cp \left(V + \frac{p}{2\pi}(dB-2\pi L_B) \right) + p\int s\cp B
\label{theory_U1U1_AB}
\end{align}
The $1$-form $\Z$ invariances between $A$ and $L_A$ and between $B$ and $L_B$ are ensured if and only if $q, p\in\Z$, as witnessed by the $l\cdot A$ and $s\cp B$ terms. Now there comes a crucial point: we may completely absorb the background variables $L_A, L_B$ into $W, V$. In physical terms, this is just the familiar property that a narrow $dA$ flux of $2\pi$ is \emph{indistinguishable} from an anyon of charge $q$ under $a$, and likewise for $dB$ flux.

In a physical system there is usually only one background $U(1)$ gauge field, the electromagnetic field $A$. Under this assumption, without loss of generality we may set $B=A$, $L_B=L_A\equiv L$ in the above. Also, we absorb
\begin{align}
W - qL \ \rightarrow \ W, \ \ \ \ \ V - pL \ \rightarrow \ V
\label{electromagnetic_string_absorbed}
\end{align}
which physically means a narrow $2\pi$ electromagnetic flux is indistinguishable from an anyon with charges $w=q$, $v=p$ under $a$ and $b$. This leads to our doubled $U(1)$ CS lattice theory with electromagnetic background:
\begin{align}
Z[W, V, A] & = \int [Da]^{U(1)} \: [Db]^{U(1)} \: [Dl]^\Z_{\partial l =0} \: [Ds]^\Z_{ds=0} \ e^{iS[W,V, A]}, \nonumber \\[.2cm]
S[W, V, A] & = \int a \cdot \left( \frac{n}{2\pi} db - ns - W - \frac{q}{2\pi} dA\right) - \int l \cdot \left( nb - qA \right) \nonumber \\[.2cm] 
& \phantom{=} + \frac{k}{4\pi} \int \left( b \cp db - b\cp 2\pi s - 2\pi s \cp b \right) - \int b \cp V - \frac{p}{2\pi} \int \left(db-2\pi s\right) \cp A.
\label{theory_U1U1}
\end{align}
The $U(1)$ compactness of the electromagnetic field is ensured by the background $1$-form $\Z$ invariance
\begin{align}
A \ \rightarrow \ A + 2\pi \Delta, \ \ \ \ W \ \rightarrow \ W - q\, d\Delta, \ \ \ \ V \ \rightarrow \ V - p\, d\Delta, \ \ \ \ \ \ \ \Delta\in C^1(M; \Z)
\label{EMU1_compactness}
\end{align}
thanks to the charges $q, p$ being integers. Actually the invariance works for fractional $\Delta$ as well, as long as $q\, d\Delta$ and $p\, d\Delta$ are integer valued.

The $U(1)$ compactness of the electromagnetic field being preserved by an accompanied transformation of $2$-form operators (which in $3D$ are line operators on dual lattice) is familiar in lattice gauge theory. First, if we think of the electromagnetism as dynamical and include a ``non-compact'' Maxwell term \cite{Polyakov:1975rr} (introduced below \eqref{ds_dl_closed}) $(-1/4e^2) \sum_{\mathrm{plaq.} \, \pl} (dA - 2\pi L)_{\pl}^2, \ \ dL=0$, it is indeed $dA-2\pi L$ to be recognized as the unambiguous total electromagnetic flux, which might have closed, non-exact configurations that are Dirac quantized. Second, if we think of the electromagnetic field as a non-dynamical background, as we do in this paper, such accompanied transformation of the $2$-form operators is typical in Villain type of models in which global $\Z$ symmetries are gauged to obtain effective $U(1)$ variables. For example, in the $3D$ Villain model with vortex fugacity \cite{Peskin:1977kp}, a $2\pi\Z$ shift of the electromagnetic field must be accompanied by a shift of the vortex line observable.
\footnote{The $3D$ Villain model with vortex fugacity (constructed in footnote \ref{footnote_Villain_model}) in Euclidean signature is
\begin{align*}
-\frac{1}{2T}\sum_{\mathrm{links}\, \lk}(d\theta - 2\pi m - qA)_\lk^2 + i \sum_{\mathrm{links}\, \lk} (d\theta - qA)_\lk V'_\lk - \frac{T'}{2} \sum_{\mathrm{plaq.} \, \pl}(dm - V)_\pl^2
\end{align*}
where $\theta\in C^0(M; U(1))$ is an angular variable, and $m\in C^1(M; \Z)$ is an integer gauge field whose $dm\in B^2(M; \Z)$ is interpreted as the vortex core. $V' \in C_1(M; \Z)$ is the Wilson line observable which may or may not be closed so that $\partial V'$ is the particle insertion, $V\in C^2(M; \Z)$ is the vortex line observable which may or may not be closed, and $A\in C^1(M; U(1))$ is the background electromagnetic field with integer coupling $q$. The $U(1)$ compactness is preserved by
\begin{align*}
A \ \rightarrow \ A + 2\pi \Delta, \ \ \ \ V \ \rightarrow \ V - q\, d\Delta, \ \ \ \ \ \ \ \Delta\in C^1(M; \Z)
\end{align*}
which can be manifest via a shift of the dynamical variable $m \rightarrow m-q\Delta$. Under the exact lattice version of particle-vortex duality \cite{Peskin:1977kp}, the $3D$ Villain model is equivalent to the abelian Higgs model
\begin{align*}
-\frac{1}{2T'}\sum_{\mathrm{plaq.}\, \pl}(\partial\theta' - 2\pi m' - a)_\pl^2 + i \sum_{\mathrm{plaq.}\, \pl} (\partial\theta' - a)_\pl V_\pl - \frac{T}{2} \sum_{\mathrm{links} \, \lk}\left(\frac{\partial a}{2\pi} - l - V' \right)_\lk^2 - i\frac{q}{2\pi} \sum_{\mathrm{links} \, \lk} A_\lk \left( \partial a - 2\pi l \right)_\lk
\end{align*}
where $a\in C_2(M; U(1))$ is a dynamical $U(1)$ gauge field (on the dual lattice) with the associated closed Dirac string $l\in Z_1(M; \Z)$, and $\theta' \in C_3(M; U(1))$ is a Goldstone boson with the associated $m' \in C_2(M; \Z)$ whose $\partial m'$ is the vortex core of the Goldstone field. Note that in this dual theory, the original particle insertion $\partial V'$ is interpreted as the Dirac monopole insertion, while the original vortex line $V$ is interpreted as the Wilson line, hence the name ``particle-vortex'' duality.}

Usually one would demand an extra condition, the spin-charge relation which requires $q=0 \mod 2$, and $p=k \mod 2$, which is $0 \mod 2$ for even $k$. At this point the motivation for this extra condition is not so clear. We leave this point to Section \ref{sect_fermionic} when $k$ can be either even (bosonic theory) or odd (fermionic theory).

Now we compute the electromagnetic observables in the theory \eqref{theory_U1U1}, which will lead to the Hall conductivity and the fractional electric charges of the anyons. It is convenient to shift the $b$ field by $qA/n \mod 2\pi$ in \eqref{theory_U1U1}, which leads to
\begin{align}
&(\mbox{the action \eqref{theory_U1U1_no_A} without $A$}) \nonumber \\[.1cm]
& + \frac{k q}{2n^2} \int \left( A \cp W + W \cp A \right) - \frac{p}{n} \int W \cp A - \frac{q}{n} \int A \cp V \nonumber \\[.2cm]
& + \left(\frac{kq^2}{n^2} - \frac{2pq}{n}\right) \frac{1}{4\pi} \int A \cp dA
\label{theory_U1U1_shifted_A}
\end{align}
where in the second line we have used the constraint from $a$ (or equivalently, we have made an appropriate shift to $a$). The second line manifests the fractional electric charges of the anyons $W$ and $V$; as expected from the continuum theory, they are given by the $K^{-1} Q$. One may note there is some framing details based on the definition of $\cp$, but for an electromagnetic background that is varying slowly enough, $W\cp A$ and $A\cp W$ are almost equal.
\footnote{$W\cp A$ and $A\cp W$ only differ by terms of order $\sim W\cdot dA$, which is negligible in the continuum limit when $dA$ is smoothly distributed while the Wilson line is infinitely narrow. If one still wants to make the appearance of $W\cp A$ and $A\cp W$ more symmetric, one may replace the last term of \eqref{theory_U1U1} by $(p/4\pi) \int \left( (db-2\pi s)\cp A + A \cp (db-2\pi s) \right)$, and the invariance under \eqref{EMU1_compactness} is ensured if we impose the spin-charge relation which requires $p=k\mod 2$ to be even. On the other hand, for the $V$ line, we may simply understand any $X\cp V$ as $X\cdot V_f$ and use $V_f$ instead of $V$ in all discussions.}
The third line is the desired fractional Hall conductivity, given by $Q^T K^{-1} Q$, as expected in the continuum. While we explained in Section \ref{ssect_RRCS} that the \emph{dynamical} $b\cp db$ cannot be recognized as a single CS due to mode doubling, for \emph{background} $AdA$ there is no such problem and we may indeed recognize it as the Hall conductivity.

We emphasize that the separate interpretation for the second line and the third line is only locally applicable to slowly varying electromagnetic background. When $dA$ approaches order $2\pi$ at the lattice scale, the $1$-form $\Z$ invariance \eqref{EMU1_compactness} becomes more pertinent, so the three lines cannot be separately discussed.
\footnote{One noteworthy scenario is the following. Consider a three-torus spacetime discretized as cubic lattice. The $x, y$-directions are regarded as the two-torus space and the $t$-direction the periodic time; the vertices' coordinates $(x, y, t)$ take integer values. At each time slice, a flat electromagnetic holonomy $(\Theta_x(t), \Theta_y(t))$ threads across the spatial $x, y$-directions, and they may change slowly in time, hence there is a weak electric field. For concreteness, we let the $A$ field on the links centered at coordinates $(1/2, y, t)$ take value $\Theta_x(t)$, and those on the links centered at $(x, 1/2, t)$ take value $\Theta_y(t)$, hence the weak electric field in the $x$- and $y$-directions exist on the plaquettes centered at $(1/2, y, t+1/2)$ and $(x, 1/2, t+1/2)$ respectively. Suppose over a finite period of time, $\Theta_y(t)$ stays unchanged while $\Theta_x(t)$ slowly increases through $2\pi$. Then there must be a time $t^\ast$ at which $\Theta_x$ jumps from slightly below $2\pi$ at time $t^\ast$ to slightly above $0$ at time $t^\ast+1$. The flux $dA$ on these $(1/2, y, t^\ast+1/2)$ plaquettes is therefore not small, but close to $-2\pi$. To remedy this large electric flux, we should insert an $L$ loop (recall \eqref{electromagnetic_string_absorbed}) of charge $-1$ through these plaquettes, so that the total flux $dA-2\pi L$ remains small, representing a weak electric field. This scenario is related to the setting of \cite{kapustin2018local} which we will briefly discuss below.}
An equivalent, but perhaps more illuminating way of saying this is: given the theory \eqref{theory_U1U1_no_A} without $A$, and imposed the $1$-form $\Z$ invariance \eqref{EMU1_compactness} between the observables, the appearance of $A$ in the theory is completely determined to be \eqref{theory_U1U1_shifted_A}, as long as we ignore non-topological terms such as a Maxwell term. Such consistency requirement between anyon braiding, anyon electric charge and Hall conductivity underlies the classification of abelian CS theories based on observables \cite{Belov:2005ze} (though our lattice realization is only limited to the doubled ones).

Finally we comment on the following important issue. It has been recently shown \cite{kapustin2018local} that local commuting projector Hamiltonians cannot host non-trivial Hall conductivity. On the other hand, our construction \eqref{theory_U1U1_no_A}, in the absence of electromagnetic field, can be reduced to the $\Z_n$ DW model \cite{Dijkgraaf:1989pz} (see Section \ref{sssect_HS_PF_DW}), which has natural Hamiltonian realization as local commuting projector Hamiltonians \cite{Kirillov:2011mk, hu2013twisted, mesaros2013classification, Lin:2014aca}. This might seem to suggest our construction \eqref{theory_U1U1}, now with electromagnetic field incorporated, may potentially be contradictory to the conclusion of \cite{kapustin2018local}. Of course, this comparison cannot be readily made because our construction is an exactly topological Lagrangian on the spacetime, while \cite{kapustin2018local} is about microscopic Hamiltonian models which are topological in their low energy sector; the former does not have a unique realization in the latter's kind, and hence a contradiction is not implied. Here we emphasize the reason why there is no contradiction.
The $U(1)$ compactness of the electromagnetic field is preserved in our construction by an accompanied $1$-form $\Z$ invariance, \eqref{electromagnetic_string_absorbed} and \eqref{EMU1_compactness}, because we took the lattice theory as an effective description in the deep IR topological limit; while in \cite{kapustin2018local} the $U(1)$ compactness stands on its own, because the lattice theory is taken as microscopic.
\footnote{The setting in the previous footnote is essentially that considered in \cite{kapustin2018local}, where it is shown that the energy eigenstates cannot have a non-trivial Chern number over the $(\Theta_x, \Theta_y)$ two-torus parameter space, hence no Hall conductivity \cite{niu1985quantized, avron1985quantization}. In our construction, the periodicity of this parameter space is only preserved via the insertion of the extra loop operator $L$. We may view the extra operator $L$ inserted when $\Theta_{x,y}$ jump from $2\pi$ to $0$ as essentially replacing the role of the non-trivial transition function acting on the energy eigenstate, which is responsible for the Chern number.}
Importantly, the former approach to the $U(1)$ compactness of the electromagnetic background can be converted to the latter approach, at the cost of the gauge invariance of the dynamical $U(1)$ gauge fields becoming \emph{emergent} rather than \emph{fundamental}.  This is what happens in the ``charging Hamiltonian'' in \cite{Levin:2011hq} (though this Hamiltonian, a local commuting projector one, does not have Hall conductivity; it corresponds to $k=0$, $q=0$, $p=2$ in our notations). A reminiscent of the $1$-form $\Z$ invariance, however, still persists in this Hamiltonian, in its reduced flux periodicity. We will discuss the concrete mapping between our work and \cite{Levin:2011hq} in better details in Section \ref{sect_Hamiltonian}, which may shed light on possible microscopic Hamiltonian realization of our deep IR Lagrangian construction for more general cases, especially when $q\neq 0$ which hosts Hall conductivity (one of our Hamiltonian proposal is closely related to \cite{geraedts2013exact}).

\subsubsection{Hilbert Space, Partition Function and Dijkgraaf-Witten Theory}
\label{sssect_HS_PF_DW}

So far we have seen the physical observables of the doubled $U(1)$ CS -- the anyon statistics, the anyon electric charge and the Hall conductivity --  realized on the lattice. Three formal questions remain to be addressed: First, a quantum field theory is not only to compute expectation values, but also to assign quantum states to a time slice. Second, related to the first point, the spacetime path integral should produce properly normalized partition function. Third, as we claimed before, we should see how our construction reduces to the DW model in the absence of electromagnetic background; in particular, the normalization of the partition function must agree. 

Previously we have considered closed oriented spacetime manifolds without boundary. To discuss the quantum states, we consider an oriented spacetime manifold $\M$ with boundary $\partial \M$. Upon discretization of $\M$ into $M$, the $2D$ boundary $\partial \M$ is also discretized into $\partial M$ which consists of vertices, links and plaquettes. We may think of $\partial \M$ as a ``time slice'', even though $\M$ may not be decomposable into a product of space and time. The fields on $\partial M$ must be appropriately specified as the boundary condition; each possible specification is assigned with a quantum state, which together form an orthonormal basis. For each boundary condition, i.e. each basis state, the path integral over $M$ evaluates an amplitude, thereby forming a wavefunction that describes the quantum state on $\partial M$ ``built'' by the spacetime $M$ (the wavefunction may or may not be normalized for a general $\M$). If two discretized manifolds share the same discretized boundary, they may ``glue'' along the boundary (with appropriate choice of orientation) to form a closed oriented manifold, which corresponds to taking the inner product in the Hilbert space. These ideas can be naturally described in the language of category theory, see \cite{Atiyah:1989vu, Dijkgraaf:1989pz, Turaev:1992hq} for more details.

For our construction, on the discretized boundary $\partial M$, there are the $a$ and the $s$ fields on the boundary plaquettes, the $b$ and the $l$ fields on the boundary links, and the $\theta$ field (see \eqref{closedness_Lagrange_multiplier}) on the boundary vertices. We may set the $a$ field on the boundary plaquettes (``perpendicular'' field) to be $0$ by a $U(1)$ gauge transformation on the tetrahedra right beneath the boundary. On the other hand, the $b$ field on the boundary links (``parallel'' field) is \emph{not} subjected to the gauge transformations \eqref{b_gauge_transf} and \eqref{s_gauge_transf}. This is because a local transformation that takes place on the boundary (in particular the boundary vertices for \eqref{b_gauge_transf} and boundary links for \eqref{s_gauge_transf}) is \emph{not} counted as gauge redundancy \cite{Elitzur:1989nr}. Since \eqref{s_gauge_transf} is not counted as a gauge redundancy on the boundary links, it is convenient not to gauge fix $b$ to $[0, 2\pi)$, but let it be $\R$-valued.
\footnote{The local Hilbert space for $b$ over $\R$ is normalizable in the same sense as a quantum mechanical single particle in free propagation or under periodic potential on an infinite line.}

Although the local transformations \eqref{b_gauge_transf} and \eqref{s_gauge_transf} on the boundary are not counted as gauge redundancies, we have the Gauss's constraints:
\begin{itemize}
\item[--]
Consider two basis states described by $\theta, b, s$ and $\theta', b' ,s'$ on the boundary respectively, such that $s' = s + d\beta$, $b'=b+2\pi\beta+d\kappa$, $\theta'=\theta+n\kappa \mod 2\pi$. Any state constructed from the path integral will have equal amplitudes in front of these two basis states. (It is understood that the variables and $d$ are restricted within the boundary $\partial M$.)
\end{itemize}
The Gauss's constraints are not imposed on the theory; rather, they are automatic reminiscent of the gauge invariances \eqref{b_gauge_transf} and \eqref{s_gauge_transf}.
\footnote{If the local $\kappa$ transformation occurs on the boundary vertices, the cup product $\cp$ does not remain invariant. Such local transformation is not counted as gauge redundancy so there is no violation of gauge invariance. Related to this, when $db\neq 0$ (due to $W$ or $dA$ insertion), in general $\int b\cp db \neq \int db\cp b$ due to differences on the boundary. We may view our choice, $\int b\cp db$, as a fixed prescription for the path integral.}
Recall that in this presentation we have let $b$ take $\R$ value in the local Hilbert space; letting $b$ take value in $[0, 2\pi)$ would be to look at a gauge fixed Hilbert space.

As for the observable insertions, the $A$ field may present on any link including the boundary links. The $W$ and $V$ Wilson loops are closed in $M$ but might have end points on the boundary $\partial M$, creating anyon excitations on boundary plaquettes on the ``time slice''.

In the lattice path integral \eqref{theory_U1U1}, we separate the terms into those that take places inside $M$, and those that take places on the boundary $\partial M$. Note that in addition to the terms explicitly written in the action \eqref{theory_U1U1}, the terms \eqref{closedness_Lagrange_multiplier} are also understood. Obviously any term with $\cp$ must occur on the tetrahedra and hence not on the boundary; same for the $\int \varpi\cdot ds$ term. The $\int a\cdot (\cdots)$ terms occur on the plaquettes, but as we mentioned above, $a$ can be set to zero on the boundary plaquettes, hence we may view this term as not occurring on the boundary either. We are left with the terms
\begin{align}
\int l\cdot (d\theta-nb+qA)
\end{align}
on the links. These are the only terms that may occur on the boundary $\partial M$. Terms that take places inside $M$ are the terms that ``build'' the quantum state on $\partial M$, whilst the terms on $\partial M$ involving the boundary fields only are viewed as the ``potential energy'' terms on the ``time slice''. Normally, a potential energy term on a single time slice is proportional to the infinitesimal time step parameter $\delta t$ (such as $-(k/2)x(t)^2 \, \delta t$ for a harmonic oscillator), and hence negligible when considering this single time slice. This is not the case here -- the finite rather than infinitesimal term on the ``time slice'' suggests an infinitely strong potential energy. This is not unexpected, because in an exactly topological theory (or in the deep IR topological limit of any gapped theory), any energy is either zero or infinite. The interpretation here is that the $l$ field is a Lagrange multiplier field on the ``time slice'' which is equivalent to an infinitely strong Higgs potential
\begin{align}
\mathcal{U}' \left(1-\cos(d\theta-nb+qA)_\lk\right), \ \ \ \mathcal{U}' \rightarrow \infty \ \ \ \ \Longrightarrow \ \ \ \ (d\theta-nb+qA)_\lk \in 2\pi\Z
\label{Higgsing_potential}
\end{align}
on each link $\lk$ on $\partial M$. More precisely, since $\theta$ and $b$ are continuous here, in order for the Higgs potential to yield an energy gap, a kinetic energy term smearing $b$ is needed; the kinetic term can be made arbitrarily small by taking $\mathcal{U}'\rightarrow \infty$.
\footnote{Suppose the kinetic term has coefficient $K$. Taking the harmonic oscillator approximation, the energy gap $\Delta E$ is of order $\sqrt{K\mathcal{U}'}$, and the spread $\Delta b^2$ is of order $\sqrt{K/\mathcal{U}'}$. We want to make $K$ small and $\mathcal{U}'$ large so that $\Delta E \rightarrow \infty$ and $\Delta b^2 \rightarrow 0$. The Higgs vacua of $(d\theta-nb+qA)_\lk \in 2\pi\Z$ are almost degenerate up to a tunneling energy split that is exponentially suppressed by $1/\Delta b^2$.}
In this sense, the lattice theory is not exactly topological, but we are solving it in the infinite potential energy and infinitesimal kinetic energy topological limit, which is equivalent to having the Lagrange multiplier $l$. This is closely related to our discussion about microscopic lattice Hamiltonian in Section \ref{sect_Hamiltonian}.

Piecing up the above, the quantum state on $\partial M$ built by the spacetime $M$ lives in the local product Hilbert space described by the local degrees of freedom
\begin{itemize}
\item
$\theta\in C^0(\partial M; U(1))$, the conjugate momentum variable of the $l$ field;
\item
$b\in C^1(\partial M; \R)$;
\item
$s\in C^2(\partial M; \Z)$.
\end{itemize}
The amplitudes are subjected to the potential energy constraint $(d\theta-nb+qA)_\lk \in 2\pi\Z$, and are evaluated by the path integral \eqref{theory_U1U1} (with \eqref{closedness_Lagrange_multiplier} understood) over $M$. The precise path integral measure will be given in \eqref{measure_U1U1} later; the state thus constructed may or may not be normalized. The amplitudes satisfy a Gauss's constraint as described before, which is a reminiscent of the gauge equivalence of $b$; such superposition is the defining character of long range, intrinsic topological order \cite{Wen:2016ddy}. Inner product in the Hilbert space is performed by glueing two discretized manifolds along a shared discretized boundary, which is manifestly equivalent to performing the path integral  \eqref{theory_U1U1} over the closed manifold thereby formed (recall that the summation over $l$ on the boundary links is equivalent to having the said potential energy constraint). Therefore our lattice path integral defines a consistent (topological) quantum field theory.

Among the quantum states built by a path integral, the ground states are of particular importance. The fundamental example is the $n^2$ degenerate ground states on a two-torus space.
\footnote{The two-torus space is fundamental because states on spaces of other topologies can be obtained from the two-torus space via Dehn surgery \cite{Witten:1988hf}, a remarkable property of Chern-Simons theory.}
Let $M$ be a solid torus whose $\partial M$ is a two-torus; we set $A=0$ and consider $W=wC$, $V=vC$ inserted around a non-contractible loop $C$ inside $M$ \cite{Verlinde:1988sn}. The $n^2$ ground states are built by having the charges $w, v$ taking values in $\{0, 1, \cdots, n-1\}$. (Inserting additional contractible Wilson loops in $M$ does not change the states on $\partial M$, up to overall phases. Changing $v$ by $n$ can be absorbed by an $l$ loop around $C_f$, and changing $w$ by $n$ along with $v$ by $k$ can be absorbed by an $s$ loop around $C$.)

It is worth noting that the full Hilbert space is not unique, if we change some perspectives. We may, say, integrate out $l$ first in the path integral \eqref{theory_U1U1}, before we consider the presence of spacetime boundary. This leads to a $\Z_n$ gauge field in place of $b$. More particularly, we let the $\Z_n$ variable $\bar{b}$ be the constrained version of $(-d\theta+nb-qA)/(2\pi)$; the invariance of the theory under a shift of $\bar{b}$ by $n\Z$ is preserved by the corresponding $1$-form $\Z$ transformation of $s$. The theory becomes
\begin{align}
Z[W, V, A] & = \int [Da]^{U(1)} \: [D\bar{b}]^{\Z_n} \: [Ds]^\Z_{ds=0} \ e^{iS[W,V, A]}, \nonumber \\[.2cm]
S[W, V, A] & = \int a \cdot \left( d\bar{b} - ns - W\right) + \frac{\pi k}{n^2} \int \left( \bar{b} \cp d\bar{b} - \bar{b}\cp ns - ns \cp \bar{b} \right) - \frac{2\pi}{n}\int \bar{b} \cp V \nonumber\\[.2cm]
& \phantom{=} + \frac{kq}{2n^2} \int \left( A \cp (d\bar{b}-ns) + (d\bar{b}-ns) \cp A \right) - \frac{p}{n} \int \left(d\bar{b}-ns\right) \cp A \nonumber\\[.2cm]
& \phantom{=} - \frac{q}{n} \int A \cp V + \left(\frac{kq^2}{n^2} - \frac{2pq}{n}\right) \frac{1}{4\pi} \int A \cp dA
\label{theory_U1Zn}
\end{align}
which is still local. In this formulation of the theory, the observable $A$ \emph{has to} appear in this complicated way in order for the observables to respect the background $1$-form $\Z$ gauge invariance \eqref{electromagnetic_string_absorbed} and \eqref{EMU1_compactness}. Now, if we consider to build a quantum state on a boundary $\partial M$, the local product Hilbert space is described by the local degrees of freedom
\begin{itemize}
\item
$\bar{b}\in C^1(\partial M; \{0, 1, \cdots, n-1\})$;
\item
$s\in C^2(\partial M; \Z)$.
\end{itemize}
This is the Hilbert space of the charging Hamiltonian in \cite{Levin:2011hq}, which corresponds to $k=0$, $q=0$, $p=2$. (The charging Hamiltonian in \cite{Levin:2011hq} has $A$ coupled to $s$ only but not to $\bar{b}$, which breaks the gauge invariance of the dynamical gauge field. This is not a problem because in this Hamiltonian the dynamical gauge invariance is considered \emph{emergent} rather than \emph{fundamental}; the physics in these two views however can be mapped to each other. We will focus on this in Section \ref{sect_Hamiltonian}.)

By the same idea, we may also further integrate out $s$ before we consider the presence of spacetime boundary. The theory becomes
\begin{align}
Z[W, V, A] & = \int [D\bar{a}]^{\Z_n} \: [D\bar{b}]^{\Z_n} \ e^{iS[W,V, A]}, \nonumber \\[.2cm]
S[W, V, A] & = \frac{2\pi}{n}\int \bar{a} \cdot \left( d\bar{b} - W\right) + \frac{\pi k}{n^2} \int \left( -\bar{b} \cp d\bar{b} +\bar{b}\cp W + W \cp \bar{b} \right) - \frac{2\pi}{n}\int \bar{b} \cp V \nonumber\\[.2cm]
& \phantom{=} + \frac{kq}{2n^2} \int \left( A \cp W + W \cp A \right) - \frac{p}{n} \int W \cp A - \frac{q}{n} \int A \cp V \nonumber\\[.2cm]
& \phantom{=} + \left(\frac{kq^2}{n^2} - \frac{2pq}{n}\right) \frac{1}{4\pi} \int A \cp dA.
\label{theory_ZnZn}
\end{align}
The dynamical $\Z_n$ gauge invariances does not require any accompanied $1$-form $\Z$ invariance. If we set $A=0$ or equivalently $p=q=0$, this is just the $\Z_n$ DW model \cite{Dijkgraaf:1989pz} with the $k$ term a representation of the DW term. The $A$ field does not directly couple to any dynamical variable, however it still \emph{has to} appear in this complicated way in order for the observables to respect the background $1$-form $\Z$ gauge invariance \eqref{electromagnetic_string_absorbed} and \eqref{EMU1_compactness}. The local product Hilbert space on $\partial M$ is described by the local degrees of freedom $\bar{b}\in H^1(\partial M; \Z_n)$, as is in the usual presentation of DW model \cite{Dijkgraaf:1989pz, hu2013twisted, mesaros2013classification, Lin:2014aca}.

The DW form \eqref{theory_ZnZn} is convenient for figuring out the correct normalization of the path integral measure for $\bar{a}, \bar{b}$; this in turn makes it easy to find the correct normalization for $a, b$ in the original theory \eqref{theory_U1U1}. The measure in \eqref{theory_ZnZn} is precisely
\begin{align}
\int [D\bar{a}]^{\Z_n} \: [D\bar{b}]^{\Z_n} \ &\equiv \ \left(\prod_{\mathrm{plaq.}\: \pl} \ \frac{1}{n} \sum_{\bar{a}_\pl \in \Z_n} \right) \left(\prod_{\mathrm{links}\: \lk} \ \sum_{\bar{b}_\lk \in \Z_n} \right) \left(\prod_{\mathrm{vert.}} \frac{1}{n} \right) \nonumber \\[.25cm]
&= \ \left(\prod_{\mathrm{plaq.}\: \pl} \ \sum_{\bar{a}_\pl \in \Z_n} \right) \left(\prod_{\mathrm{links}\: \lk} \frac{1}{n} \ \sum_{\bar{b}_\lk \in \Z_n} \right) \left(\prod_{\mathrm{tetra.}} \ \frac{1}{n} \right)
\label{measure_ZnZn}
\end{align}
(the same holds for cubic lattice, as long as $\mathrm{tetra.}$ is understood as cubes) where the second equality is because for oriented odd-dimensional manifold the Euler characteristic vanishes due to Poincar\'{e} duality. Note that the $1/n$ factors are a product of local factors, i.e. they are local counter terms in the action, receiving no input knowledge about the global topology, as is required for a locally defined quantum field theory. To verify the normalization, we may compute the partition function on the three-torus, which should yield $n^2$, the trace over the ground state subspace. The $1/n$ average over $\bar{a}$ on each plaquette yields $1$ if $d\bar{b}$ vanishes on this plaquette, and $0$ otherwise. The $\bar{b}$ configurations with $d\bar{b}=0$ are, up to gauge redundancy, specified by the $\Z_n$ holonomies around the three non-contractible loops, yielding a factor of $n^3$. There is an independent $\Z_n$ gauge redundancy on each vertex, except for a global $\Z_n$ transformation that leaves $\bar{b}$ unchanged. This gauge redundancy factor is extensive in the number of vertices, and it is removed by the product of $1/n$ over the vertices, leaving the $1/n$ global symmetry factor. Thus, we indeed obtain $n^2$ for the partition function, as desired. Obviously, for more general $H^2(M; \Z)=\mathcal{F}=\Z^{B_1}$, the partition function is $n^{B_1-B_0}$. For the most general $H^2(M; \Z)=\mathcal{F}\oplus\mathcal{T}$, there is the additional factor \eqref{U1U1_torsion_factor} which is generically complex, so the partition function is
\begin{align}
Z \ = \ n^{B_1-B_0} \ \sum_{\tau\in \mathcal{T}, \ \mathrm{p}_\tau | n} \ \exp\left( -i\pi k\mathrm{L}(s_\tau, s_\tau) \right).
\label{PF_U1U1}
\end{align}
The $n^{-B_0}$ factor counts the global symmetry (we usually assume $B_0=1$), on the other hand, the factors $n^{B_1}$ and \eqref{U1U1_torsion_factor} arise from the summation over flat gauge configurations \cite{Dijkgraaf:1989pz}.

Tracing back along how we obtained \eqref{theory_ZnZn} from \eqref{theory_U1U1}, we can determine the measure in the theory \eqref{theory_U1U1} (recall \eqref{closedness_Lagrange_multiplier}):
\begin{align}
& \ \int [Da]^{U(1)} \: [Db]^{U(1)} \: [Dl]^\Z_{\partial l =0} \: [Ds]^\Z_{ds=0} \nonumber \\[.25cm]
\equiv& \ \left(\prod_{\mathrm{plaq.}\: \pl} \int_0^{2\pi} \frac{da_\pl}{2\pi} \ \sum_{s_\pl\in\Z} \right) \left(\prod_{\mathrm{links}\: \lk} n \int_0^{2\pi} \frac{db_\lk}{2\pi} \ \sum_{l_\lk\in\Z} \right) \nonumber \\[.2cm]
& \  \left(\prod_{\mathrm{vert.}\: \vt} \frac{1}{n} \int_0^{2\pi} \frac{d\theta_\vt}{2\pi} \ e^{i\theta_{\vt}(\partial l)_\vt}\right) \left(\prod_{\mathrm{tetra.} \: \td} \int_0^{2\pi} \frac{d\varpi_\td}{2\pi} \ e^{i\varpi_{\td}(ds)_\td}\right).
\label{measure_U1U1}
\end{align}
Again we may simultaneously move the $n$ from $\prod_\lk$ to $\prod_\pl$ and the $1/n$ from $\prod_\vt$ to $\prod_\td$ since the Euler characteristic vanishes. One can easily verify the measure \eqref{measure_U1U1} produces the correctly normalized partition functions, and can be reduced to \eqref{measure_ZnZn} (constant $\theta$ and constant $\varpi$ do not provide closedness constraints for $l$ and $s$, but the integral over these global degrees of freedom yield $1$ due to the $2\pi$ denominators).

We have thus completed the discussion of the physical observables, the Hilbert space and the lattice path integral measure of the doubled $U(1)$ bosonic Chern-Simons theory on spacetime lattice. We may view it as an extension to the usual $\Z_n$ DW theory incorporating electromagnetic coupling, guided by the $1$-form $\Z$ invariance due to the indistinguishability between a narrow $2\pi$ flux tube and certain Wilson loops. This lattice path integral construction can be naturally retrieved from the continuum path integral, as we will elaborate on in Section \ref{sect_DB}. The generalization to \eqref{doubledU1CS_cont} beyond $m=1$ is obvious.

\subsection{A $U(1)\times \R$ Chern-Simons}
\label{ssect_U1RCS}

The method of gauging $1$-form $\Z$ symmetries can also be applied to one instead of both of the doubled $\R$ gauge fields, yielding $U(1)\times \R$ CS theories. Here we focus on one which has its $U(1)$ sector observables and its complex phase of the partition function the same as those for a single $U(1)$ CS.

In the continuum, a single $U(1)$ CS of level $-k$ is given by
\begin{align}
S = -\frac{k}{4\pi} \int_\M a\, \d a.
\end{align}
with $k\in \Z$, and in particular $k\in 2\Z$ for bosonic phases. As we mentioned in Section \ref{ssect_RRCS}, we cannot simply realize the local aspects of the theory by a cup product on the lattice, and the difficulty is profoundly related to the chiral central charge \cite{Kitaev:2006lla}. One may however consider adding an $\R$ CS with the opposite chiral central charge, and thus the non-chiral theory may be realizable on the lattice. On the other hand, the $\R$ sector CS, unlike the $U(1)$ sector, does not have topologically non-trivial gauge configurations, and cannot naturally couple to a $U(1)$ background. By a trivial rescaling of the $\R$ gauge field, we can always normalize it as $(k/4\pi) \int_\M b\, \d b$, then we shift $b$ to $b+a$, the theory appears as
\begin{align}
S = \frac{k}{2\pi} \int_\M a\, \d b + \frac{k}{4\pi} b\, \d b.
\end{align}
The $U(1)\times \R$ CS thus appears as a Hubbard-Stratonovich transformation of the single $U(1)$ CS, which is sometimes used in perturbative CS theory \cite{Guadagnini:2014mja, BarNatan:1991rn}. This appears similar to a doubled $U(1)$ CS with $n=k$, but we have to keep in mind that $b+a$ is an $\R$ rather than $U(1)$ gauge field, and hence $-a$ and $b$ are understood to share the same set of Dirac string variables (see Section \ref{sect_DB} for details); similar care has to be taken on the flat holonomies and the identification of the global symmetry, as we shall see later.

We will focus on the observables coupled to the original $U(1)$ sector only, which are
\begin{align}
-\oint_W a - \frac{q}{2\pi} \int_\M a\, \d A.
\end{align}
Again, a narrow $2\pi$ electromagnetic flux tube is indistinguishable from a charge $q$ anyon Wilson loop. On the other hand, the $\R$ sector couples to $\R$-valued closed observables; since there is no natural identification of a subset of $\Z$-valued observables, the narrow $2\pi$ flux tube and hence the $A$ field also does not have a natural coupling to the $\R$ sector, we therefore assume $A$ only couples to the $U(1)$ sector.

\subsubsection{Observables of the $U(1)$ Sector}

The construction of the lattice theory is straightforward. It appears as a doubled $U(1)$ CS theory on the lattice with $n=k, p=0$, but in addition with the Dirac string variables $l$ and $-s$ identified because $b+a$ is an $\R$ gauge field. More precisely, since $l$ lives on links while $s$ lives on plaquettes, we choose to identify $l$ with $-s\capp \mathrm{M}$. The lattice theory is therefore
\begin{align}
Z[W, A] & = \int [Da]^{\R} \: [Db]^{U(1)} \: \: [Ds]^\Z_{ds=0} \ e^{iS[W, A]}, \nonumber \\[.2cm]
S[W, A] & = \int a \cdot \left( \frac{k}{2\pi} db - ks - W - \frac{q}{2\pi} dA\right) + \int s\cp \left( kb - qA \right) \nonumber \\[.2cm] 
& \phantom{=} + \frac{k}{4\pi} \int \left( b \cp db - b\cp 2\pi s - 2\pi s \cp b \right).
\label{theory_U1R}
\end{align}
There is only one $1$-form $\Z$ gauge invariance as opposed to two, which acts as
\begin{align}
a\ \rightarrow a - 2\pi \beta\capp \mathrm{M}, \ \ \ \ \ b \ \rightarrow \ b + 2\pi \beta, \ \ \ \ \ s \ \rightarrow \ s + 2\pi d\beta, \ \ \ \ \ \ \beta\in C^1(M; \Z).
\end{align}
As it acts on both $a$ and $b$, by fixing its gauge we may reduce either $a$ or $b$ from $\R$-valued to $U(1)$-valued. It turns out to be technically more convenient to fix $b$ into $U(1)$-valued, as we have denoted in the path integral measure; the observables $W$ and $A$ coupled to $a$ are however still recognized as the ``$U(1)$ sector observables'', as this should not depend on the gauge fixing. As a result of this convention, the $\R$-valued $a$ respects a $0$-form $\R$ gauge invariance while the $U(1)$-valued $b$ respects a $0$-form $U(1)$ gauge invariance.

The observables are easily computed. Let's set $A=0$ first and compute the Wilson loop observables. The most crucial difference to the doubled $U(1)$ case is that there is no $l$ to be summed over which would restrict $b$ to be discrete. It is again convenient to apply a central extension by $\Z$ as in \eqref{central_extension}, except now $\wt{b}$ is valued in $\R$ rather than $\Z$. The theory becomes
\begin{align}
Z[W] & \propto \int [Da]^{\R} \: [D\wt{b}]^{\R} \: \sum_{\gamma\in H^2(M;\Z)} \ e^{iS[W]}, \nonumber \\[.2cm]
S[W] & = \int a \cdot \left(d\wt{b} - k s_\gamma - W\right) + \frac{\pi}{k} \int \left( \wt{b} \cp d\wt{b} -\wt{b}\cp ks_\gamma + ks_\gamma \cp \wt{b} \right)
\label{theory_U1R_no_A_b_s}
\end{align}
in analogy to \eqref{theory_U1U1_no_A_b_s}, but note the sign of the $ks_\gamma \cp \wt{b}$ term. Alternatively, we may note the similarity with the doubled $\R$ CS \eqref{theory_RR}, but including a summation over $\gamma\in H^2(M; \Z)$ which represents the non-trivial $U(1)$ bundles, and a change of the path integral normalization as we will study later. Integrating out the $\R$-valued $a$ yields
\begin{align}
d\wt{b} - k s_\gamma = W.
\end{align}
Since $\wt{b}$ is $\R$-valued here, we cannot map this into $H^2(M; \Z)$ as we did for \eqref{a_EoM}. But we may still map this into $\mathcal{F}=\mathrm{Hom}(H_2(M; \Z), \Z)$, i.e. integrate it over non-contractible $2$-cycles, and get the constraint $[[W]]=-k [\gamma]$, which means the total number of anyons on any closed $2$-cycle must be a multiple of $k$, as is expected for a single $U(1)$ CS. Again, this constraint uniquely determines $[\gamma]\in \mathcal{F}$ for a given $W$.

When the spacetime has torsion, there are some notable differences with the previous doubled $U(1)$. We may set $W=0$ first. The equation $d\wt{b}=k s_\gamma$ has solution for \emph{any} $\gamma=\tau \in\mathcal{T}$ (i.e. $[\gamma]=0$). Thus, the partition function $Z[0]$ contains a complex factor 
\begin{align}
\sum_{\tau\in \mathcal{T}} \ \exp\left( i \pi k \mathrm{L}(s_\tau, s_\tau) \right) 
\label{U1R_torsion_factor}
\end{align}
which has the condition on the period $\mathrm{p}_\tau$ removed compared to \eqref{U1U1_torsion_factor}, and has a sign difference. The same factor appears in a single $U(1)$ CS of level $-k$.

When $W\neq 0$, because $\wt{b}$ is $\R$-valued, there is no requirement for $[W]$ in addition to $[[W]]\in k\mathcal{F}$, unlike the previous doubled $U(1)$ CS in which $\wt{b}$ is $\Z$-valued. More precisely, we may let $\gamma_W$ be a representative such that $-k[\gamma_W]=[[W]]$, but there might not exist one such that $-k \gamma_W = [W]$, so $\mathcal{W}\equiv W+ks_{\gamma_W}$ may be a non-trivial torsion element; moreover, any such choice of $\gamma_W$, which may differ by a torsion element, works equally well. The solution to $\wt{b}$ is, up to gauge transformation,
\begin{align}
& \wt{b} = \wt{b}^{sol.} + \wt{b}_\tau + k \wt{b}^{aux.} + \wt{b}^{flat}, \nonumber \\[.2cm]
& d\wt{b}^{sol.} = W + ks_{\gamma_W} \equiv \mathcal{W}, \ \ \ \ \ d\wt{b}_\tau = ks_\tau, \ \ \ \ \ d\wt{b}^{aux.}_\tau = s_{\gamma_W + \tau} - (s_{\gamma_W} + s_\tau), \ \ \ \ \ d\wt{b}^{flat}=0.
\end{align}
We find the Wilson loop expectation value by substitution into the last term of \eqref{theory_U1R_no_A_b_s}. Again $\wt{b}^{aux.}$ will decouple as a reminiscence of gauge invariance. $\wt{b}^{flat}$ also decouples due to the signs of $\wt{b}\cp s_\gamma$ and $s_\gamma\cp \wt{b}$ in \eqref{theory_U1R_no_A_b_s}. The expectation value is
\begin{align}
\left\langle e^{-i\int a\cdot W} \right\rangle = \frac{Z[W]}{Z[0]} \ & = \ \exp\left( i\frac{\pi}{k} \mathrm{L}(\mathcal{W}, \mathcal{W}_f) - i\pi\left(\mathrm{L}(\mathcal{W}, {s_{\gamma_W}}_f) - \mathrm{L}(s_{\gamma_W}, \mathcal{W}_f) \right) \right) \nonumber\\[.2cm]
& \phantom{= \ } \cdot \frac{\sum_{\tau\in \mathcal{T}} \ \exp\left( i\pi k \mathrm{L}(s_\tau, s_\tau) + i 2\pi \mathrm{L}(s_\tau, \mathcal{W}) \right)}{\sum_{{\tau'}\in \mathcal{T}} \ \exp\left( i\pi k \mathrm{L}(s_{\tau'}, s_{\tau'}) \right) }
\end{align}
with the constraint $[[W]] = -k[\gamma_W] \in k\mathcal{F}$. Note the expectation almost depends on $\mathcal{W}$ only (which must have trivial free part), except for a $\pi$ phase which depends on $\gamma_W=-[[W]]/n$. This is indeed the Wilson loop expectation value for a single $U(1)$ CS of level $-k$.

Next we let $A\neq 0$. Upon shifting $\wt{b}$ by $qA/2\pi$, we find $S[W, A]$ is given by
\begin{align}
(\mbox{the action \eqref{theory_U1R_no_A_b_s} without $A$}) + \frac{q}{2k} \int \left( A \cp W + W \cp A \right) + \frac{q^2}{k} \frac{1}{4\pi} \int A \cp dA
\label{theory_U1R_shifted_A}
\end{align}
which captures the anyon charge $q/k$ and the Hall conductivity $q^2/2\pi k$. Again they are the same as a single $U(1)$ CS as expected.

\subsubsection{Partition Function}
\label{sssect_U1R_partition}

It remains to specify the path integral measure for the $\R$ gauge field, a problem left over since Section \ref{ssect_RRCS}. Let's first consider the path integral measure of the doubled $\R$ theory \eqref{theory_RR}. Schematically, the measure can still be expressed in a manner of \eqref{measure_ZnZn} of \eqref{measure_U1U1}, but now the variables are $\R$ rather than $\Z_n$ or $U(1)$, so one can imagine the $\R$ gauge redundancy on each lattice vertex has infinite measure, and we need to divide it by an infinite counter term factor. A precise prescription is needed to make sense of such ``infinite divided by infinite'' on each vertex. This leads to the standard Faddeev-Popov method, which couples the gauge redundant degree of freedom on each vertex to an auxiliary field and integrate out the auxiliary fields. We will denote the standard Faddeev-Popov measure as  $[D(a/2\pi)]^{\R}_{FP} \: [D(b/2\pi)]^{\R}_{FP}$. The partition function of the doubled $\R$ theory \eqref{theory_RR} is given by
\begin{align}
Z = \int \left[D\frac{a}{2\pi}\right]^{\R}_{FP} \: \left[D\frac{b}{2\pi}\right]^{\R}_{FP} \ \mathcal{N} \ \exp\left( i\frac{n}{2\pi}\int a \cdot db \right)
\end{align}
where we have removed the $k$ term via a shift of $a$ by $-(k/2n)(b\capp\mathrm{M}+\mathrm{M}\capp b)$, and $\mathcal{N}$ is a locally finite extra counter term factor that will be specified soon. The remaining integral is just that of a doubled $\R$ BF theory \cite{Adams:1996yf}. We review its computation in details.

The Lagrange multiplier integral of $a$ and $b$ gives a factor of $1/|{\det}'(n d_1)|$, where $d_1$ is $d$ acting on $C^1$ elements, and ${\det}'$ means determinant with zero modes excluded; note the determinant of $d_1: C^1(\R)\rightarrow C^2(\R)$ is well-defined, because given a simplicial complex, there is a natural measure of $C^n(\R)$ inherited from each $n$-simplex. The total number of non-zero modes of $d_1$ is just the total number of linearly independent $db$ degrees of freedom, which is equal to $(N_2-B_2)-(N_3-B_3)-B_2 = (N_1-B_1)-(N_0-B_0)$, where $N_i$ is the number of $n$-simplices.
\footnote{There are $N_2$ plaquettes for $db$ to live on. Each of the $N_3$ tetrahedra gives an independent closedness constraint, except for one in each connected component of the manifold ($B_3=B_0$ many in total). Moreover, there are $B_2=B_1$ many closed, non-exact real degrees of freedom that cannot be written as $db$.}
So the $n$ can be taken out from the ${\det}'$ and gives an overall factor of $n^{B_1-B_0} n^{N_0-N_1}$. The $n^{N_0-N_1}$ factor is extensive, but it is a local product, so it can be removed by the local counter term factor $\mathcal{N}$, if we define it to be $\mathcal{N}\equiv \left(\prod_{\mathrm{links}} n\right)\left(\prod_{\mathrm{vert.}} 1/n \right)$ as in \eqref{measure_U1U1}. We are left with the $n^{B_1-B_0}$ topological factor.

There are also zero modes of the lattice BF term. The zero modes include the exact part, i.e. the gauge redundancies, which are taken care of by the Faddeev-Popov measure, as well as the closed, non-exact part, $H^1(\R)$ for $b$ and $H_2(\R)$ for $a$. Both parts have extra contributions to the partition function.

Consider the closed, non-exact contributions first. The flat $a, b$ configurations in $H_2(\R)\cong H^1(\R) \cong \R^{B_1}$ are unconstraint in the path integral, so they give a diverging factor of $\delta(0)^{B_1+B_2}=\delta(0)^{2B_1}$. To make sense of the diverging factor, we may consider inserting a closed $W, V$ loop around each of the $B_1$ free non-contractible directions, normalized as $-i\int a\cdot W - i\int b\cp V$ (recall that in doubled $\R$ theory the $W, V$ charges can be real valued), and the argument ``$0$'' in $\delta(0)^{2B_1}$ means the constrained real charge for each loop.

Finally consider the Faddeev-Popov measure contributions. The $[Db]^\R_{FP}$ measure, removing the gauge redundancy $\delta b = d\kappa$, gives a factor of $|{\det}'(d_0)|$. Note that $d_0$ in turn has a zero mode, constant $\kappa\in H^0(\R)$, which is a global symmetry rather than gauge redundancy, and this leads to an extra global symmetry factor of $\delta(0)^{-B_0}$. The factor is formally $0$, and it should be understood such that, if the spacetime time contains an extra scalar field that transforms as $X\rightarrow X+\kappa$ and is dynamically constrained to $dX=0$, then integrating it out will precisely cancel the $\delta(0)^{-B_0}$. Likewise, the $[Da]_{FP}$ measure, removing the gauge redundancy $\delta a = \partial\varphi$, gives a factor of $|{\det}'(\partial_3)|=|{\det}'(\partial_3^T)|=|{\det}'(d_2)|$. The zero mode of $\partial_3$, i.e. constant $\varphi\in H_3(\R)$, gives an extra global symmetry factor $\delta(0)^{-B_3}=\delta(0)^{-B_0}$.

Piecing up the above, we arrive at the partition function for doubled $\R$ theory:
\begin{align}
Z \ = \ \left(\sqrt{n} \: \delta(0)\right)^{2(B_1-B_0)} \ |{\det}'(d_2)| \: |{\det}'(d_1)|^{-1} \: |{\det}'(d_0)|.
\end{align}
The product of the determinants with alternating power is exactly one way to define the R-torsion $\tau_R$  \cite{muller1993analytic, Adams:1996yf} with $\R$ coefficient, a topological invariant of the manifold independent of the triangulation.
\footnote{The R-torsion is originally defined on simplicial complex, as we did, and known as the Reidemeister torsion. Later, the Ray-Singer analytic torsion was defined in the continuum. The two has been shown to be equal \cite{cheeger1979analytic, muller1978analytic, muller1993analytic}. Hence we will not differentiate them and just call it the R-torsion.}
In fact $\tau_R$ with $\R$ coefficient has a simple expression, $\tau_R(\R)=|\mathcal{T}|^{-1}=(\prod_i \mathrm{p}_i)^{-1}$ \cite{Witten:2003ya}, the inverse of the size of the torsion subgroup $\mathcal{T}$. Therefore, the partition function for doubled $\R$ theory is, finally,
\begin{align}
Z \ = \ \left(\sqrt{n} \: \delta(0)\right)^{2(B_1-B_0)} \ |\mathcal{T}|^{-1} \ = \ \left(\sqrt{n} \: \delta(0)\right)^{-B_3+B_2+B_1-B_0} \ |\mathcal{T}|^{-1}.
\label{PF_RR}
\end{align}
More precisely, $\tau_R(\R)$ is not well-defined unless an extra measure is prescribed to a set of generators of the cohomology groups $H^i(M; \R)=\R^{B_i}$, and our definition of $\delta(0)$ explained in the above is such a measure that arises naturally from the path integral, completing the definition of $\tau_R(\R)$. Note that by rescaling the fields $a$ and $b$ by $1/\sqrt{n}$ we can absorb the $\sqrt{n}$ factor into the redefined $\delta(0)$, as expected.
\footnote{In \cite{Adams:1996yf} the $\delta(0)$ factors are dropped. This is what would happen if the closed, non-exact holonomies and the global symmetries associated with $a$ and $b$ were $U(1)$-valued instead of $\R$-valued, as we will discuss soon in \eqref{PF_U1U1_alt}. Since in \cite{Adams:1996yf} this dropping of $\delta(0)$ (viewed as a choice of measure) had to be artificially imposed on the non-local degrees of freedom -- the flat holonomies and global symmetries, we still refer to \cite{Adams:1996yf} as a doubled $\R$ BF theory. Indeed, the model in \cite{Adams:1996yf} does not have the level quantization condition and the self-linking number factor in \eqref{PF_U1U1_alt}.}
The partition function is equal to the absolute value square of that of a single $\R$ CS \cite{Schwarz:1978cn}.
\footnote{The partition function of a single $\R$ CS is the square root of the above, $\left(\sqrt{n} \: \delta(0)\right)^{B_1-B_0} \ |\mathcal{T}|^{-1/2}$, times a framing anomaly complex phase which can be set to $1$ in the ``canonical framing'' \cite{Witten:1988hf}. Historically, \cite{Schwarz:1978cn} developed the method relating perturbative CS path integral to Ray-Singer analytic torsion, but misidentified the partition function as $\tau_R(\R)$. The need for the framing anomaly complex phase and the $H^i(\M; \R)$ measures was rectified in \cite{Witten:1988hf}, and the need to take the square root of $\tau_R(\R)$ as rectified in \cite{Freed:1991wd}.}

The reduction to our $U(1)\times \R$ CS is now straightforward. In \eqref{theory_U1R} (where $n=k$), we define the measure as
\begin{align}
& \int [Da]^{\R} \: [Db]^{U(1)} \: \: [Ds]^\Z_{ds=0} \nonumber \\[.2cm] 
\equiv & \ \int\left[D\frac{a}{2\pi}\right]^{\R}_{FP} \left(\prod_{\mathrm{plaq.}\: \pl} \sum_{s_\pl\in\Z} \right) \left(\prod_{\mathrm{links}\: \lk} |k| \int_0^{2\pi} \frac{db_\lk}{2\pi}\right) \left(\prod_{\mathrm{vert.}\: \vt} \frac{1}{|k|} \right) \left(\prod_{\mathrm{tetra.} \: \td} \int_0^{2\pi} \frac{d\varpi_\td}{2\pi} \ e^{i\varpi_{\td}(ds)_\td}\right).
\end{align}
Exploiting the $1$-form $\Z$ symmetry between $b$ and $s$, we may redefine
\begin{align}
& s = s_\gamma - dy, \ \ [s_\gamma]=[s]=\gamma\in H^2(M; \Z), \ \ \ y\in C^1(M; \Z) \nonumber \\[.1cm]
& b' \equiv b + 2\pi y \in C^1(M; \R)
\end{align}
(the $\wt{b}$ in \eqref{theory_U1R_no_A_b_s} is $k/2\pi$ times this $\R$-valued $b'$), and use the Faddeev-Popov measure for $b'$. The partition function then appears as that of a doubled $\R$ CS, with an extra factor \eqref{U1R_torsion_factor} arising from the summation over $\gamma$. But there is an extra caveat: The closed, non-exact holonomy of $y$ does not contribute to $s$ and hence should not be taken as part of the original theory, but it becomes the closed, non-exact holonomies of $b'$, so we need to remove it, which amounts to replacing the associated $\delta(0)^{B_1}$ with $1^{B_1}$ in our normalization. Moreover, the Faddeev-Popov measure of $b'$ contains an $\R$-valued global symmetry, which should have been $U(1)$-valued, and this amounts to replacing the associated $\delta(0)^{-B_0}$ with $1^{B_0}$ in our normalization. Therefore, the partition function for our $U(1)\times \R$ theory is
\begin{align}
Z \ = \ \left(|k| \delta(0)\right)^{B_1-B_0} \  |\mathcal{T}|^{-1} \ \sum_{\tau\in \mathcal{T}} \ \exp\left( i \pi k \mathrm{L}(s_\tau, s_\tau) \right).
\label{PF_U1R}
\end{align}
This is indeed the product of partition functions of a single $U(1)$ CS and a single $\R$ CS of opposite chirality, as we originally motivated it.
\footnote{In the Hamiltonian perspective, the diverging $\delta(0)$ factors are associated with the real valued observables that couple to the $\R$ sector gauge field $b+a$. We will not go into the details here.}

Further carrying out the procedure of gauging $1$-form $\Z$ symmetry for $a$, we can provide an alternative perspective to the doubled $U(1)$ partition function \eqref{PF_U1U1}. Similar to $b$ and $s$, we replace $\partial a - 2\pi l$ with $\partial a' - 2\pi l_{\gamma'}$ so that $\gamma'\in H_1(M; \Z)$ and $a'$ becomes real-valued. The partition function we thus obtain is
\begin{align}
Z \ = \ n^{B_1-B_0} \  |\mathcal{T}|^{-1} \ \sum_{\tau, \tau'\in \mathcal{T}} \ \exp\left( i2\pi n \mathrm{L}(s_\tau, l_{\tau'}) - i \pi k \mathrm{L}(s_\tau, s_\tau) \right).
\label{PF_U1U1_alt}
\end{align}
The summation over $\tau'$ is non-vanishing only when $ns_\tau$ is contractible, i.e. $\mathrm{p}_\tau | n$, in which case $n \mathrm{L}(s_\tau, l_{\tau'}) \in \Z$ and leads to a factor $\sum_{\tau'\in\mathcal{T}} =|\mathcal{T}|$ that cancels the $|\mathcal{T}|^{-1}$ prefactor \cite{Witten:2003ya}, resulting in \eqref{PF_U1U1}.

\section{Fermionic Phases}
\label{sect_fermionic}

When the doubled $U(1)$ CS theory \eqref{doubledU1CS_cont} contains any odd diagonal element $k_{ii}$, the theory is a fermionic topological order. A theory with fermionic topological order can only be defined on the lattice if local fermionic degrees of freedom are available on the lattice \cite{Gu:2012ib, Gu:2013gma, Tarantino:2016qfy, Gaiotto:2015zta, Bhardwaj:2016clt}. By combining with the spacetime formalism developed in the recent years \cite{Gu:2012ib, Gaiotto:2015zta, Bhardwaj:2016clt}, we can easily extend our construction, include the electromagnetic coupling, to the fermionic theories. Again it suffices to consider $m=1$ in \eqref{doubledU1CS_cont}, as generalization to $m>1$ is straightforward.

The reason we demanded $k$ to be even in the bosonic theories is for the $1$-form $\Z$ gauge transformation of the action, \eqref{deltaS_U1U1}, to be a multiple of $2\pi$. When $k$ is odd, there is a $\pi$ phase ambiguity in \eqref{deltaS_U1U1} under the gauge transformation \eqref{s_gauge_transf}. In order for the odd $k$ theory to be well-defined, we must introduce new contributions to the action that absorb this gauge ambiguity.

A more illuminating perspective towards the problem is to use the $4D$ formulation \eqref{theory_U1U1_4D}, as is usually done in the continuum \cite{Dijkgraaf:1989pz, Witten:2003ya, Seiberg:2016rsg}. We want the theory to be independent of the choice of the $4D$ extension $N$. For odd $k$, evidently the last term of \eqref{theory_U1U1_4D}
\begin{align}
\pi k \int_N s \cp s
\label{theory_U1U1_f_4D}
\end{align}
is a $\pi$ phase that cannot be dropped. But it can be made independent of $N$ if we impose an additional restriction on the allowed choices of $N$: We not only require $\partial N=M$, but also pick a spin structure on $M$ (details given below) and require it to be extendable into $N$. If we have different extensions $N$ and $N'$, the ambiguity of the action is given by $\pi k \int_{N''} s\cp s$ where $N''$ is the closed simplicial complex formed by glueing $N$ and $-N'$ along $M$. Notably, with the restriction above imposed on $N$ and $N'$, the $N''$ they form must be a spin manifold, on which $\int_{N''} s\cp s$ is always \emph{even} \cite{Dijkgraaf:1989pz}, and hence the ambiguity can be dropped. 

Although the $N$ dependent ambiguity can be dropped, the term \eqref{theory_U1U1_f_4D} itself may be $0$ or $\pi \mod 2\pi$, depending on $M$ and the spin structure we pick on $M$. The task becomes whether \eqref{theory_U1U1_f_4D} with the desired spin structure dependence can be evaluated as a  new, local contribution within $M$, without reference to some $4D$ $N$. This cannot be done with local bosonic degrees of freedom only,
\footnote{We cannot simply express the closed $s$ as some $dy$: First, there are closed, non-exact $s$; second, $y$ itself would have extra flat holonomy (see \eqref{central_extension}).}
but it can be done if we allow local fermionic variables on $M$ \cite{Gu:2012ib, Gaiotto:2015zta, Bhardwaj:2016clt}:
\begin{align}
e^{i\pi \int_N s \cp s} = e^{i\pi \int_M \Sigma \cdot s} \ \mathcal{Z}_\psi[s].
\label{4D_to_3D_f}
\end{align}
Here $\mathcal{Z}_\psi[s]$ is a fermionic path integral over $M$ that yields $\pm 1$ depending on the $\mod 2$ reduction of $s$, and $\Sigma\in C_2(M; \Z_2)$ is a background data such that $\partial\Sigma$ is fixed by the vertex ordering. The details of their definitions will be presented later. For now, we only need to note that while $\partial\Sigma$ is fixed by the vertex ordering, we may still consider $\Sigma$ and $\Sigma'$ that differ up to a flat holonomy:
\begin{align}
\Sigma'-\Sigma=\Xi, \ \ \ \ \Xi \in Z_2(M; \Z_2); \ \ \ \ \ \ \Xi \sim \Xi + \partial\Pi \ \mbox{for} \ \Pi\in C_3(M; \Z_2).
\label{spin_structure_choice}
\end{align}
An exact change of $\Xi$ by $\partial\Pi$ has trivial effect on the theory because $s$ is closed. Thus, the different legitimate choices of $\Sigma$ are classified by their differences $[\Xi]\in H_2(M; \Z_2) \cong H^1(M; \Z_2)$, which corresponds to different choices of spin structure \cite{Gaiotto:2015zta} (see details later; all oriented three manifolds admit spin structures).
\footnote{The identity element of $H^1(M; \Z_2)$ has no canonical identification with a particular choice of spin structure, since $[\Xi]\in H_2(M; \Z_2) \cong H^1(M; \Z_2)$ classifies the difference between the spin structures only.}

The doubled $U(1)$ CS theory on spacetime lattice is therefore
\begin{align}
Z[W, V, A] & = \int [Da]^{U(1)} \: [Db]^{U(1)} \: [Dl]^\Z_{\partial l =0} \: [Ds]^\Z_{ds=0} \ \ e^{iS[W,V, A]} \ \ e^{i\pi k \int_M \Sigma \cdot s} \ \ \left(\mathcal{Z}_\psi \right)^k, \nonumber \\[.2cm]
S[W, V, A] & = \int a \cdot \left( \frac{n}{2\pi} db - ns - W - \frac{q}{2\pi} dA\right) - \int l \cdot \left( nb - qA \right) \nonumber \\[.2cm] 
& \phantom{=} + \frac{k}{4\pi} \int \left( b \cp db - b\cp 2\pi s - 2\pi s \cp b \right) - \int b \cp V - \frac{p}{2\pi} \int \left(db-2\pi s\right) \cp A
\label{theory_U1U1_f}
\end{align}
with $n, k, q, p \in \Z$; when $k$ is even, the above reduces back to the bosonic \eqref{theory_U1U1}. The bosonic path integral measure is given by \eqref{measure_U1U1}, while the definitions of $\mathcal{Z}_\psi$ and $\Sigma$ will be given later.

The physical effect of the choice of spin structures \eqref{spin_structure_choice} is obvious. First of all, for odd $k$, by inspecting the exchange statistics of the anyon $W=ns$ (which braids trivially with $V$), an odd / even valued $s$ loop describes a fermion / boson. Under \eqref{spin_structure_choice}, the weight of the partition function changes by a $e^{i\pi \int \Xi\cdot s}$, which maps $[s]$ to a sign $\pm 1$. More particularly, let $s$ be a generator of $H^2(M; \Z)$; a generator is odd valued and hence describes a fermion traveling around a non-contractible loop. A change of spin structure by $\Xi$ may change this fermion loop by weight $-1$, which can be interpreted as a change of the fermion boundary condition around that non-contractible loop. Note that $[\Xi]\in H_2(M; Z_2) \cong H^1(M; \Z_2) = \mathrm{Hom}(H_1(M; \Z), \Z_2) = \left(\Z_2\right)^{B_1} \oplus \bigoplus_{j \ (\mathrm{p}_j \: \mathrm{even})} \Z_2$ (see Appendix \ref{app_alg_topo}), i.e. only free generators and torsion generators with even periods admit a choice of fermion boundary condition / spin structure, while for consistency reason torsion generators with odd periods does not.

If the background $\Sigma\in C_2(M; \Z_2)$ is promoted to dynamical, only even valued $s$ will survive its fluctuation; as a result, $\mathcal{Z}_\psi[s]=1$. Upon proper redefinitions of $a$ and $b$, one can show the resulting theory is a bosonic one with $n_b=2n$ and $k_b=4k$. This theory is known as the ``bosonic shadow'' of the original fermionic theory, and the promotion of $\Sigma$ to dynamical corresponds to gauging the fermion parity \cite{Bhardwaj:2016clt}. 

Sometimes the change of fermion boundary condition can be equivalently achieved with the electromagnetic field $A$. This happens when the ``spin-charge'' condition is satisfied:
\begin{align}
q = 0 \mod 2, \ \ \ \ p = k \mod 2.
\end{align}
\footnote{This condition is called ``spin-charge'' because under this condition, a bosonic / fermionic anyon must have even / odd electric charge.}
Under this condition, the simultaneous shift of the background fields
\begin{align}
\Sigma \rightarrow \Sigma + \Xi, \ \ \ \ \ A \rightarrow A - \pi \, \mathrm{M} \capp \Xi
\label{spin-c}
\end{align}
by $[\mathrm{M}\capp\Xi]\in H^1(M; \Z_2)$ leaves the theory \eqref{theory_U1U1_f} invariant. The shift $-\pi \, \mathrm{M}\capp\Xi$ of $A$ is a flat, non-exact $\pi$ holonomy. The background data $\Sigma$ and $A$ together, modulo the equivalence relation \eqref{spin-c}, constitutes a spin-c structure in three spacetime dimensions.
\footnote{In general dimensions, a spin-c structure requires $A$ to be \emph{half} Dirac quantized over the $2$-cycles on which the second Stiefel-Whitney class $w_2$ does not vanish (and hence does not admit a spin structure) \cite{Seiberg:2016rsg}. In three dimensions or lower, $w_2$ is always trivial so a spin structure is always admitted, hence the spin-c structure reduces to the equivalence relation \eqref{spin-c}. However, if we extend $(\Sigma, A)$ to some $4D$ $N$ and $(\Sigma+\Xi, A-\pi \mathrm{M}\capp\Xi)$ to some $4D$ $N'$, and glue $N$ and $N'$ along the boundary $M$ to form $N''$, we will indeed encounter a $w_2$ and a half Dirac quantized gauge configuration on $N''$, due to the mismatching $\Xi$.}
Thus, given $\partial\Sigma$ (which depends on vertex ordering as we will describe below), we can fix a choice of $\Sigma$, and let the flat $\pi$ holonomies of $A$ control the fermion boundary conditions. Such electromagnetic field $A$ is referred to as a spin-c connection.
\footnote{This notion of spin-c connection agrees with the continuum presentation. If the lattice theory is viewed as a microscopic model, a more stringent definition of spin-c structure may be given. We will disuss this at the end of the section. In this paper, we our lattice description as a deep IR effective theory rather than microscopic theory.}

The $U(1)\times \R$ CS introduced in Section \ref{ssect_U1RCS} can also be extended to fermionic theory with odd $k$. The same fermionic factor $e^{i\pi k \int_M \Sigma \cdot s}  \ \left(\mathcal{Z}_\psi \right)^k$ is included into the partition function \eqref{theory_U1R}. Since $b$ and $-a$ share the same Dirac string variable $s$, the spin-charge relation becomes $-q = p-q = k\mod 2$ (since $p=0$).

It remains to review the details of $\mathcal{Z}_\psi$ and $\Sigma$, defined by the formalism of \cite{Gu:2012ib, Gaiotto:2015zta}. Originally, they are defined on the simplicial complex. In order to provide some intuition behind the formalism, here we will first introduce a construction, of the same spirit, on the cubic lattice. Then we move to the original construction on the simplicial complex.

On the cubic lattice, the fermionic degrees of freedom are:
\begin{itemize}
\item
On each plaquette $\pl$ there are two Grassmann variables $\psi_\pl$ and $\bar\psi_\pl$. Consider the two cubes on the two sides of $\pl$. If $\pl$ faces the $x$-direction, we associate $\bar\psi_\pl$ to the cube centered at $\pl+\hat{x}/2$, and $\psi_\pl$ to the cube at $\pl-\hat{x}/2$. Likewise if $\pl$ faces the $y$-direction. However, if $\pl$ faces the $z$-direction, we associate $\bar\psi_\pl$ to the cube centered at $\pl-\hat{z}/2$, and $\psi_\pl$ to the cube at $\pl+\hat{z}/2$.
\end{itemize}
We treat the $z$-direction differently to simplify some computations below; we may say the ``default'' fermion flow direction through a plaquette is $+x, +y$ or $-z$. Alternatively, the Grassmann variables can be \emph{equivalently} introduced as
\begin{itemize}
\item
In each cube $\cb$, there are six Grassmann variables, associated with the six faces of the cube respectively. They are $\bar\psi_{\cb-\hat{x}/2}, \: \psi_{\cb+\hat{x}/2}$, and $\bar\psi_{\cb-\hat{y}/2}, \: \psi_{\cb+\hat{y}/2}$, and $\psi_{\cb-\hat{z}/2}, \: \bar\psi_{\cb+\hat{z}/2}$.
\end{itemize}
Given the Grassmann variables, the fermionic path integral is defined as
\begin{align}
\mathcal{Z}_\psi[s] \equiv \left(\prod_{\mathrm{plaq.} \: \pl} \int d\psi_\pl \, d\bar{\psi}_\pl \right) \left( \prod_{\mathrm{cubes} \: \cb} h_\cb [s] \right)  \left( \prod_{\mathrm{plaq.} \: \pl} \left(1+\bar\psi_\pl \psi_\pl\right) \right).
\label{Z_psi_def}
\end{align}
Here, $h_\cb [s]$ describes the hopping of fermions along the $\mod 2$ reduction of $s$ across a cube $\cb$:
\begin{align}
h_\cb[s] = \psi_{\cb+\hat{x}/2}^{\mathtt{s}_{\cb+\hat{x}/2}} \ \psi_{\cb-\hat{z}/2}^{\mathtt{s}_{\cb-\hat{z}/2}} \ \psi_{\cb+\hat{y}/2}^{\mathtt{s}_{\cb+\hat{y}/2}} \ \bar\psi_{\cb-\hat{x}/2}^{\mathtt{s}_{\cb-\hat{x}/2}} \ \bar\psi_{\cb+\hat{z}/2}^{\mathtt{s}_{\cb+\hat{z}/2}} \ \bar\psi_{\cb-\hat{y}/2}^{\mathtt{s}_{\cb-\hat{y}/2}}
\end{align}
where the power $\mathtt{s}_\pl = s_\pl \mod 2$ may be $0$ or $1$, and the order of the variables is important.
\footnote{The factor $\bar\psi_\pl^{\mathtt{s}_\pl}$ may be alternatively presented as e.g. $(1/2) \sum_{\bar{u}_\pl=\pm 1} \bar{u}_\pl^{s_\pl} \left( 1 + \bar{u}_\pl \bar\psi_\pl \right)$, and likewise $\psi_\pl^{\mathtt{s}_\pl}$ as $(1/2) \sum_{u_\pl=\pm 1} u_\pl^{s_\pl} \left( 1 + u_\pl \psi_\pl \right)$.}
\footnote{The idea behind our design of $h_{\cb}$ is the following. If $\beta$ is $1$ on, say, the link $\cb-\hat{x}/2-\hat{y}/2$, it creates $d\beta$ on the two plaquettes $\cb-\hat{x}/2$ and $\cb-\hat{y}/2$, which are placed at the fourth and the sixth position in $h_\cb$. Placed between them, the fifth position, is the plaquette $\cb+\hat{z}/2$, which is associated to the link $\cb-\hat{x}/2-\hat{y}/2$ in the definition Figure \ref{f2-1_bdb} of $\cp$. Likewise for $\beta$ on other links. This design is crucial to the property \eqref{Z_psi_transf}.}
Since $s$ is closed, $h_\cb$ always involves an even power of Grassmann variables, though not necessarily equal numbers of $\psi$ and $\bar\psi$. On the other hand, $\left(1+\bar\psi_\pl \psi_\pl\right) = e^{\bar\psi_\pl \psi_\pl}$ is the mass factor on a plaquette $\pl$. If $s$ is even on a plaquette $\pl$, $\bar\psi_\pl$ and $\psi_\pl$ would not appear in any $h_\cb$, the Grassmann integral $d\psi_\pl \, d\bar{\psi}_\pl$ will be absorbed by the $\bar\psi_\pl \psi_\pl$ in the mass factor, yielding a trivial factor $1$; if $s$ is odd on $\pl$, $\bar\psi_\pl$ and $\psi_\pl$ appear in the adjacent $h_{\cb=\pl\pm\hat{\mu}/2}$ already, so the $1$ in the mass factor will be picked up. Thus, the sign $\mathcal{Z}_\psi[s]$ only depends on the Grassmann integral of $h_\cb$ along the loops specified by $\mathtt{s}$, the $\mod 2$ reduction of $s$.
\footnote{In the original presentation \cite{Gu:2012ib, Gaiotto:2015zta}, the Grassmann variables only exist on these loops, so the mass term is not needed. Obviously, for fixed $s$, our presentation is equivalent to the original, but we do not need to redefine the fermionic path integral when $s$ changes.}

To evaluate $\mathcal{Z}_\psi[s]$, it is crucial to observe the property \cite{Gu:2012ib, Gaiotto:2015zta}
\begin{align}
\mathcal{Z}_\psi[s + d\beta] = \mathcal{Z}_\psi[s] \: \mathcal{Z}_\psi[d\beta] \ e^{i\pi \int (\beta \cp s - s\cp \beta)}.
\label{Z_psi_transf}
\end{align}
One may verify this from the definition;
\footnote{The relatively tedious verification procedure can be organized as the following \cite{Gaiotto:2015zta}. Consider the product $\mathcal{Z}_\psi[s] \: \mathcal{Z}_\psi[d\beta]$; it suffices to assume $s$ and $d\beta$ are odd valued. On the path of $s$ we have $d\psi_\pl \: d\bar\psi_\pl$ (see the previous footnote), and on the path of $d\beta$ we have $d\chi_\pl \: d\bar\chi_\pl$. If $d\beta$ pass through $\pl$ but $s$ does not, then we can relabel $\chi_\pl, \bar\chi_\pl$ as $\psi_\pl, \bar\psi_\pl$ which then enter the evaluation of $\mathcal{Z}_\psi[s+d\beta]$. But on the plaquettes which both $s$ and $d\beta$ pass through (hence this plaquette should drop out in $\mathcal{Z}_\psi[s+d\beta]$), we have $d\psi_\pl \: d\bar\psi_\pl \ d\chi_\pl \: d\bar\chi_\pl$ that requires extra care. Suppose $s$ and $d\beta$ intersects in $\cb$, then in $h_\cb[s]\: h_\cb[d\beta]$, we move every $\chi_\pl$ to the immediate right of $\psi_\pl$ (so the $\psi_\pl \chi_\pl$ pair is bosonic), and every $\bar\chi_\pl$ to the immediate right of $\bar\psi_\pl$. This introduces some sign. Moreover, $\int d\psi_\pl \: d\bar\psi_\pl \ d\chi_\pl \: d\bar\chi_\pl \ \psi_\pl \chi_\pl \: \bar\psi_\pl \bar\chi_\pl = -1$. Combining these signs, we find $(-1)^{\int(\beta \cp s - s\cp \beta)}$, due to our deliberate design of $h_\cb$.}
the sign factor on the right-hand-side arises on the cubes where $d\beta$ intersects $s$. Letting $s$ be exact leads to the general solution
\begin{align}
\mathcal{Z}_\psi[d\beta] = e^{i\pi \int \beta \cp d\beta + i\pi \int \Lambda \cdot \beta},
\label{Z_psi_exact}
\end{align}
where we cannot rule out a linear ansatz term involving some background $\Lambda\in C_1(M; \Z_2)$. We will find $\Lambda$ soon. Combining \eqref{Z_psi_transf} and \eqref{Z_psi_exact} we can readily see that under the $1$-form $\Z$ transformation \eqref{s_gauge_transf}, the transformation of $\mathcal{Z}_\psi$ cancels the $k\pi$ (with odd $k$) term in transformation \eqref{deltaS_U1U1} of the bosonic action, as desired, but at the same time introduces an extra sign $e^{i\pi \int \Lambda\cdot \beta}$. We can cancel this extra sign if the theory contains an extra factor $e^{i\pi \int \Sigma\cdot s}$, as in \eqref{4D_to_3D_f}, where $\Lambda=\partial\Sigma$ for some $\Sigma\in C_2(M; \Z_2)$.

But is $\Lambda$ exact? One may find $\Lambda$ on each link by evaluating $\mathcal{Z}_\psi[d\beta]$ with $\beta$ being $1$ on that link and $0$ elsewhere. For instance, consider $\beta_{\lk^z}=1$ on a $z$-direction link $\lk^z$, which creates a $d\beta$ going counter-clockwise around it. In evaluating $\mathcal{Z}_\psi[d\beta]$, a $(-1)$ factor arises when the default fermion ordering on a neighboring plaquette or cube appears clockwise (hence going against $d\beta$) -- these include the plaquettes $\lk^z-\hat{x}/2$, $\lk^z+\hat{y}/2$, and the cube $\lk^z-\hat{x}/2+\hat{y}/2$; moreover, there is an overall $(-1)$ from the fermion loop. Thus, $\mathcal{Z}_\psi[d\beta]=(-1)^4=1$ for $d\beta$ around a $z$-direction link.
\footnote{The result would be the same if $\beta_{\lk^z}=-1$ and $d\beta$ goes clockwise, because $h_\cb[s]$ only depends on the $\mod 2$ reduction of $s$. In this ``default fermion direction'' picture, reversing the direction of $s$ (assume it takes the shape of a loop without self-intersection) leads to an extra $(-1)$ factor on each plaquette and each cube that $s$ passes through; since the numbers of these plaquettes and these cubes must be equal, the total extra factor is always $1$.}
We can show the same is true for $d\beta$ around an $x$- or $y$-direction link as well. Thus, we find $\Lambda=0$, and $\Sigma\in Z_2(M; \Z_2)$.

In order to relate $\Sigma$ to a spin structure / fermion boundary condition, we let the cubic lattice form a three torus. Consider a straight line $s$ threading around the $\mu$-direction, with the frame $s_f$ always at $-\hat{x}/2-\hat{y}/2-\hat{z}/2$ relative to $s$ (Figure \ref{f2-1_framing}). We may interpret this as that the transition function of ``rotation'' is trivial, so the sign of the straight non-contractible fermion loop $s$ should not receive any $(-1)$ from any local $2\pi$ rotation. On the other hand, $e^{i\pi\int \Sigma\cdot s} \: \mathcal{Z}_\psi[s]$ may still give a sign, hence this sign must be interpreted as the fermion boundary condition. It is easy to see for a straight non-contractible fermion loop $s$ in the $\mu$-direction, $\mathcal{Z}_\psi[s]=-1$ regardless of $\mu$. Therefore, for $[\Sigma]\in H_2(T^3; \Z_2)=\Z_2^3$, we may say its $\mu$-component being $0$ or $1$ corresponds to periodic or anti-periodic fermion boundary condition around the $\mu$-direction. (In this case there seems to be a simple canonical identification between the class $[\Sigma]\in H_2(M; \Z_2)$ with a particular spin structure. But this actually relied on our particular choice of $\cp$ and our detailed design of the ``default fermion directions'' on the plaquettes and cubes.
\footnote{For instance, suppose we had associated $\bar\psi_{\pl^z}$ and $\psi_{\pl^z}$ on the $z$-direction plaquette $\pl^z$ to the cubes $\cb=\pl\pm \hat{z}/2$ respectively, as opposed to the previous $\cb=\pl\mp \hat{z}/2$, then the default flow on $\pl^z$ is $+z$, but that in $h_\cb$ is still $-z$. This change does not violate the crucial property \eqref{Z_psi_transf} (and hence \eqref{Z_psi_exact}), and still $\Lambda=0$. But for a straight non-contractible loop around the $z$-direction, $\mathcal{Z}_\psi[s]$ will be $\mp 1$ depending on whether the number of cubes in the $z$-direction is even or odd. The identification of $[\Sigma]$ with a particular spin structure must change correspondingly. The fact that local modifications of $\mathcal{Z}_\psi$ -- as long as it does not violate the crucial property \eqref{Z_psi_transf} -- can be absorbed by a reinterpretation of $\Sigma$ is an important point emphasized in \cite{Gaiotto:2015zta}.}
As we will later see on the simplicial complex, there is in general no such simple identification; in fact, $\Sigma$ may not even be closed.)

\begin{figure}
\centering
\includegraphics[width=.3\textwidth]{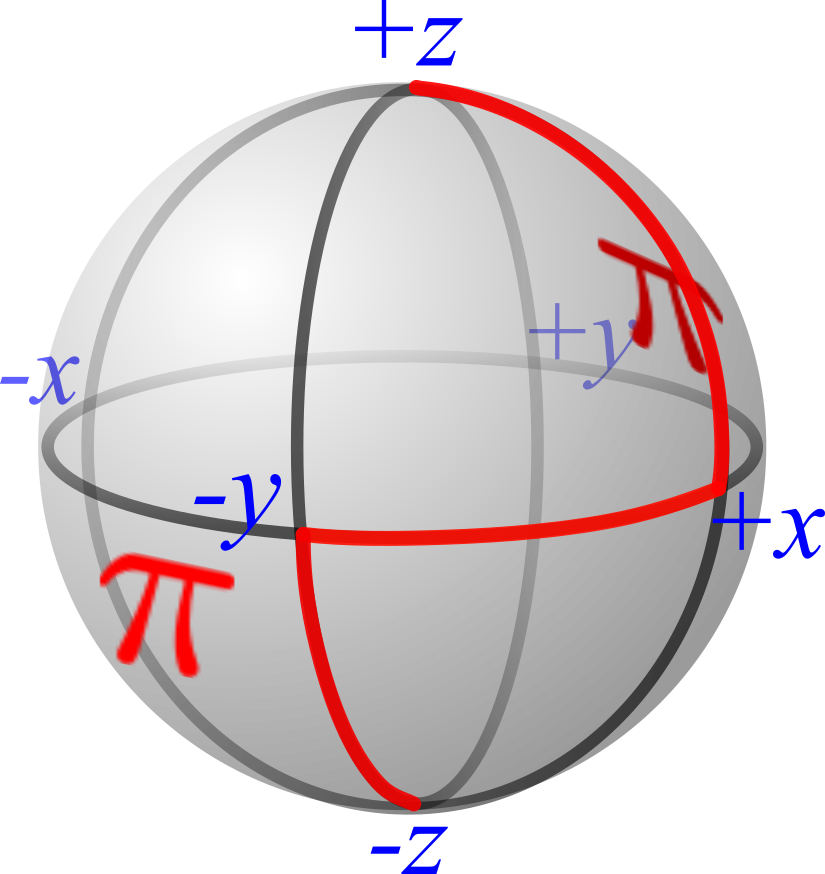}
\caption{$d\beta$ may travel in the $\pm x, \pm y, \pm z$ directions, indicated by the $6$ discrete points on the ``Bloch sphere''. The $12$ arcs connecting these points correspond to the $24$ ways that $d\beta$ may make a turn. The $3$ red arcs correspond to the $6$ ways of turning that will contribute a $(-1)$ factor to $\mathcal{Z}_\psi$, while the remaining $9$ black arcs are $18$ ways of turning that will contribute a $(+1)$ factor. We may interpret the $(\pm 1)$ factor as the exponentiation of the integration of a Berry connection along the arc. Thus, there is a $\pi$ Berry phase around the $1/8$-sphere cornered at $(+x, +y, +z)$ and a $\pi$ Berry phase around the $1/8$-sphere cornered at $(-x, -y, -z)$, and $0$ Berry phase around the other six $1/8$ spheres.}
\label{f3-1_Bloch}
\end{figure}

To gain more intuition about our design of $h_\cb$, we provide an alternative derivation of $\mathcal{Z}_\psi[d\beta]=e^{i\pi \int \beta\cp d\beta}$, making connection to Berry phase on the Bloch sphere. For simplicity, we assume $d\beta$ takes the shape of a loop of magnitude $1$ without self-intersection, so that each $h_\cb$ on the path of $d\beta$ has two Grassmann variables invoked. Again, the fermion loop has an overall $(-1)$ sign, and moreover, if the default fermion ordering on a plaquette or a cube on the path of $d\beta$ is against the direction of $d\beta$, it contributes a $(-1)$ factor. It is convenient to absorb the sign associated with a plaquette into the cube that $d\beta$ is heading towards, e.g. if $d\beta$ passes through a $z$-direction plaquette $\pl$ in the $+z$ direction, we absorb the associated $(-1)$ factor (since the default plaquette directions are $+x, +y, -z$) into the cube $\cb=\pl+\hat{z}/2$. Thus, aside from the overall $(-1)$ of the fermion loop, each cube on the path of $d\beta$ contributes a sign factor, given by the following. There are $30$ ways the loop $d\beta$ enters and exits $\cb$, including $6$ ways of straight through and $24$ ways of making a turn. It is easy to see the sign is $1$ for all $6$ straight through cases. One the other hand, when $d\beta$ makes a turn, a non-trivial sign factor may arise. Notably, one can check that exchanging the initial and final directions of a turn does not change the sign factor. Thus, the sign arising from a turn can be presented on the (discrete) ``Bloch sphere'', as pictured in Figure \ref{f3-1_Bloch}. The $\pi$ Berry phases around the $1/8$-sphere cornered at $(+x, +y, +z)$ and the $1/8$-sphere cornered at $(-x, -y, -z)$, along with the overall $(-1)$ fermion loop sign, measures the self-linking number $\mod 2$, which is equivalent to $e^{i\pi \int \beta\cp d\beta}$.
\footnote{The Berry connection associated with turning can be imposed as a local bosonic rule on each $h_\cb$, however the $-1$ factor of each fermion loop is non-local. This is another way to see why fermionic degrees of freedom are required if we want the theory to be local on the lattice.}
For examples, consider the plain loop Figure \ref{f2-1_framing} and the coil Figure \ref{f2-1_coil}. This Berry phase interpretation is motivated by Polyakov's geometrical regularization \cite{Polyakov:1988md} (see footnote \ref{geo_framing}). There, the integer self-linking number is replaced by the ``writhe'' -- a non-quantized, geometry dependent self-linking number -- in flat metric, and the Berry phase configuration of Figure \ref{f3-1_Bloch} is replaced by a uniform distribution of Berry curvature on continuous Bloch sphere (of strength $1/2$ per solid angle). 

Now we move on to simplicial complex, on which this formalism is originally defined \cite{Gu:2012ib, Gaiotto:2015zta}.
\footnote{In \cite{Gu:2012ib}, the $\mathtt{s}$ loop (the $\mod 2$ reduction of $s$) is a background variable rather than dynamical.}
The fermionic degrees of freedom are
\begin{itemize}
\item
On each plaquette $\pl$ there are two Grassmann variables $\psi_\pl$ and $\bar\psi_\pl$. Consider the two tetrahedra on the two sides of $\pl$. We associate $\bar\psi_\pl$ to the tetrahedron on the side that $\pl$ orients towards, and $\psi_\pl$ to the tetrahedron on the other side.
\end{itemize}
Here, the orientation of $\pl$ is determined by the vertex ordering: for $\pl=abc$ with vertices $a<b<c$, the right-hand-rule determines the direction that $\pl$ faces. See Figure \ref{f3-1_Grassmann} for example. An equivalent way to introduce the fermionic degrees of freedom is
\begin{itemize}
\item
In each tetrahedron $\td=abcd$ with $a<b<c<d$ appearing in the right-handed order (right-handed tetrahedron), there are Grassmann variables $\psi_{bcd}$, $\bar\psi_{acd}$, $\psi_{abd}$ and $\bar\psi_{abc}$.

\begin{figure}
\centering
\includegraphics[width=.3\textwidth]{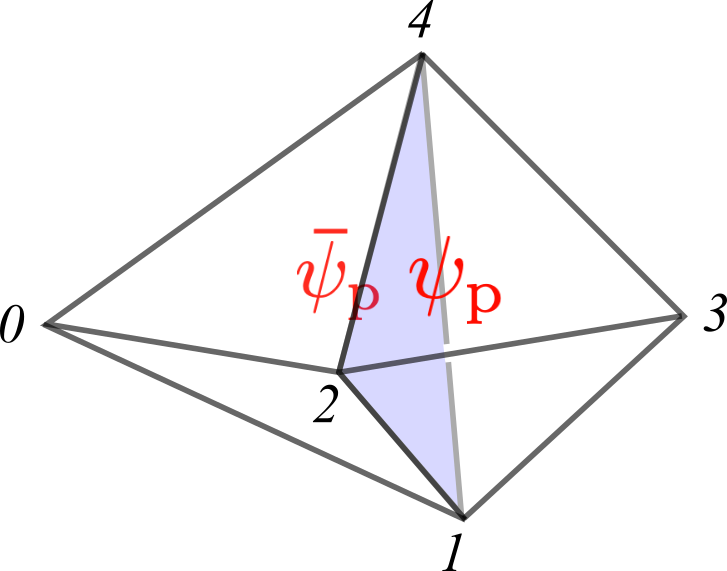}
\caption{The plaquette $\pl=124$ faces towards the left. Hence $\bar{\psi}_{124}$ is associated to the tetrahedron $0124$, while $\psi_{124}$ is associated to the tetrahedron $1234$.}
\label{f3-1_Grassmann}
\end{figure}

In each tetrahedron $\td=abcd$ with $a<b<c<d$ appearing in the left-handed order  (left-handed tetrahedron), there are Grassmann variables $\bar\psi_{bcd}$, $\psi_{acd}$, $\bar\psi_{abd}$ and $\psi_{abc}$.
\end{itemize}
Given the Grassmann variables, the fermionic path integral is defined as \eqref{Z_psi_def}, but with the hopping through tetrahedron $h_\td[s]$ in place of the hopping through cube $h_\cb[s]$. For a tetrahedron $\td=abcd$ with $a<b<c<d$, the hopping factor is given by
\begin{align}
h_\td[s] = \left\{
\begin{array}{ll}
\psi_{bcd}^{\mathtt{s}_{bcd}} \ \psi_{abd}^{\mathtt{s}_{abd}} \ \bar\psi_{acd}^{\mathtt{s}_{acd}} \ \bar\psi_{abc}^{\mathtt{s}_{abc}}, & \ \ \ \ \ \mbox{ if $\td$ is right-handed} \\[.3cm]
\psi_{abc}^{\mathtt{s}_{abc}} \ \psi_{acd}^{\mathtt{s}_{acd}} \ \bar\psi_{abd}^{\mathtt{s}_{abd}} \ \bar\psi_{bcd}^{\mathtt{s}_{bcd}}, & \ \ \ \ \ \mbox{ if $\td$ is left-handed} 
\end{array}
\right.
\end{align}
where again the power $\mathtt{s}_\pl= s_\pl \mod 2$ takes value $0$ or $1$.

\begin{figure}
\centering
\includegraphics[width=.8\textwidth]{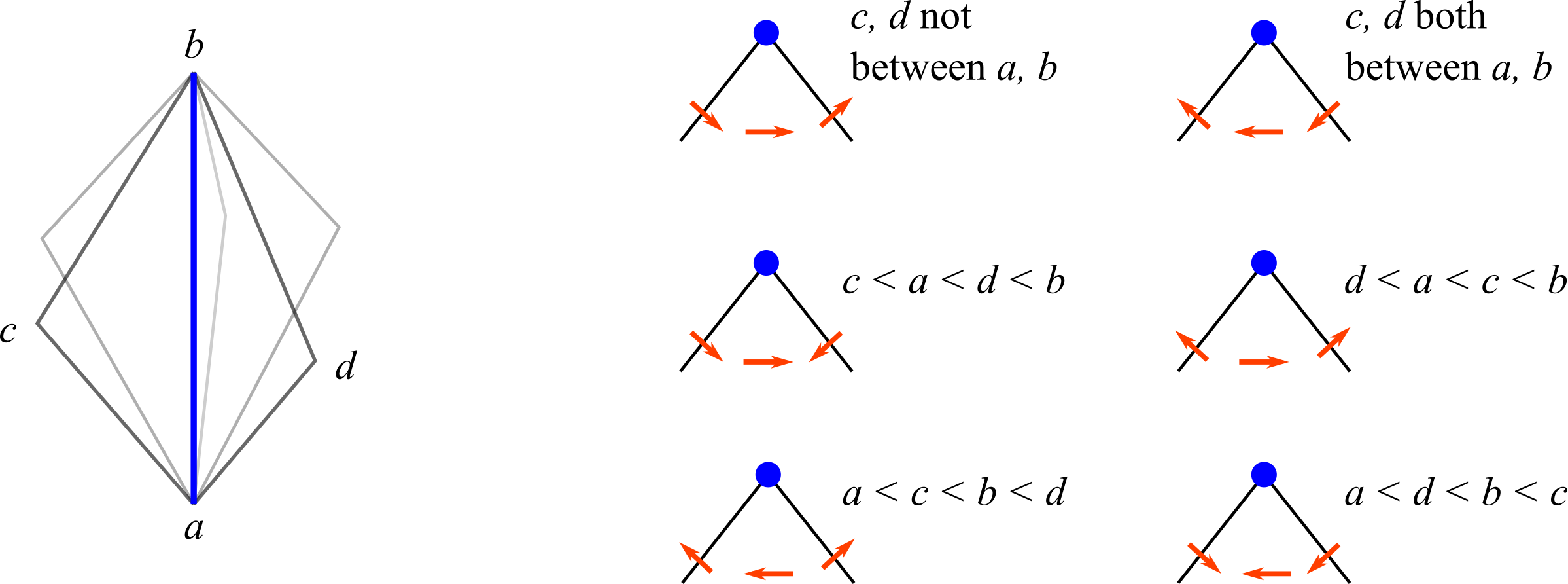}
\caption{Left: Let $\beta$ be $1$ on only one link $\lk=ab \ (a<b)$ and $0$ elsewhere. The loop $d\beta$ goes through the plaquettes around $ab$. \ Right: We focus on a wedge formed by $ab$ and the vertices $c$, $d$, and view it from the top. Depending on the relations between $c, d$ and $a, b$, we mark the default direction of fermion flow (encoded in the definition of $\mathcal{Z}_\psi$) through the plaquettes $abc$, $abd$ and the tetrahedron $abcd$. If a default flow arrow is counter-clockwise, it agrees with $d\beta$ and gives $1$; if a default flow arrow is clockwise, it disagrees with $d\beta$ and gives $-1$. Finally, the fermion loop gives an overall $-1$. This explains the first expression of $\Lambda$ in \eqref{SW_rep}. To explain the second expression, we absorb the sign from each plaquette into the adjacent tetrahedron on its counter-clockwise side (in the pictured wedge, absorb the sign from the plaquette $abc$ into the tetrahedron $abcd$.)}
\label{f3-1_sign}
\end{figure}

One can again verify from the definition that the crucial property \eqref{Z_psi_transf} and hence \eqref{Z_psi_exact} are satisfied \cite{Gu:2012ib, Gaiotto:2015zta}, with some $\Lambda\in C_1(M; \Z_2)$ to be determined. Again, we may find $\Lambda$ on each link by evaluating $\mathcal{Z}_\psi[d\beta]$ with $\beta$ being $1$ on that link only and $0$ elsewhere. The evaluation is illustrated by Figure \ref{f3-1_sign}. The result is
\begin{align}
\Lambda_{ab} \ = \ 1 & + \mbox{(number of plaquettes $acb$ with $a<c<b$)} \nonumber \\[.2cm]
& + \mbox{(number of tetrahedra $acbd$ with $a<c<b<d$)} \nonumber \\[.2cm]
& + \mbox{(number of tetrahedra $acdb$ with $a<c<d<b$)} \ \ \ \mod 2 \nonumber \\[.2cm]
= \ 1 & +  \mbox{(number of right-handed tetrahedra $adbc$ with $a<d<b<c$)} \nonumber \\[.2cm]
& + \mbox{(number of left-handed tetrahedra $dacb$ with $d<a<c<b$)} \ \ \ \mod 2
\label{SW_rep}
\end{align} 
for any link $\lk=ab \ (a<b)$.

It turns out the seemly tedious $\Lambda$ constructed as the above is (dual to) a representative of the second Stiefel-Whitney class of the manifold \cite{Gaiotto:2015zta}, $[\Lambda \capp \mathrm{M}] = w_2 \in H^2(M; \Z_2)$. In three dimensions or lower, $w_2$ is always trivial, so $\Lambda$ must be exact, i.e. a $\Sigma$ such that $\partial\Sigma=\Lambda$ always exists. Since $w_2$ is the obstruction to admitting a spin structure, it is natural to interpret $\Sigma$ as the spin structure data, with $\partial\Sigma=\Lambda$ fixed by the vertex ordering through \eqref{SW_rep}, and the different choices \eqref{spin_structure_choice} of the flat holonomy representing the choices of spin structure in $H^1(M; \Z_2)$. Note that not only the transformations of the two sides of \eqref{4D_to_3D_f} agree. For a fixed generator $s$ of $H^2(M; \Z)$, both sides are linear ($\mod 2\pi$ phase) under $s\rightarrow ms$ for integer $m$. Hence, by a proper identification of the choice of $\Sigma$ with the spin structure (the identification depends on the vertex ordering, which corresponds to local rotations), the two sides of \eqref{4D_to_3D_f} are identified.

We end this section by a final comment on the notion of spin-c structure, after we have introduced the details of $\Sigma$. We said that in $3D$ the equivalence relation \eqref{spin-c} amounts to a spin-c structure, in agreement with the notion in the continuum. However, if the lattice theory is viewed as a microscopic lattice model, a more stringent lattice definition may be given: that in addition, a change of $\partial\Sigma$ can also be compensated by a $\pi$ flux in $dA$, i.e. the entire $\Sigma$ term can be absorbed into $A$. This is expected for a microscopic model because the loop around an individual link (left panel of Figure \ref{f3-1_sign}) is the smallest possible loop of hopping, hence ``non-contractible'' in some sense. There can be a fermion boundary condition around this ``non-contractible'' small loop, specified by $\Lambda=\partial\Sigma$. Under the spin-charge relation, we can specify this fermion boundary condition by the $A$ holonomy, which is indeed a $0$ or $\pi$ flux of $dA$.
\footnote{There is a small technical issue, that $\Lambda$ lives on the links, while the $dA$ flux lives on the plaquettes, and the cap product $\mathrm{M}\capp dA$ may not have support on every individual link. This issue is not crucial. In a microscopic model we could have defined $A\in C_2(M; U(1))$ (on the links of dual lattice) instead of $A\in C^1(M; U(1))$.}
However, this definition of spin-c structure becomes inconsistent as we view the lattice theory as a deep IR effective theory rather than microscopic theory, because a narrow $2\pi$ flux of $dA$ is \emph{not} invisible, though indistinguishable from certain Wilson loop (see the discussion at the beginning of Section \ref{sssect_EM_Hall}). Therefore, in the main context of this paper, \eqref{spin-c} is the consistent notion of spin-c structure. In Section \ref{sect_Hamiltonian} we will discuss how the construction of this paper motivates microscopic Hamiltonian; there we will use the more stringent lattice definition of spin-c structure.

\section{Connection to Continuum Path Integral}
\label{sect_DB}

In the previous sections we constructed the lattice theories by proposing a doubled $\R$ CS theory on lattice and gauging 1-form global $\Z$ symmetries. The physical and mathematical properties of these resulting abelian topological lattice models indeed agree with their doubled $U(1)$ CS descriptions in the continuum. A natural question would then be whether such equivalence between the lattice and the continuum descriptions can be lifted to the level of the action and the path integral. For usual field theories this is certainly impossible, because the renormalization flow from the UV lattice to the IR continuum has dropped out irrelevant UV details so the flow cannot go backwards. For topological field theories under consideration, however, such equivalence is expected because being ``topological'' implies there is no irrelevant UV information to begin with. The purpose of this section is to manifest how our lattice theories can be retrieved from the doubled CS path integral in the continuum.  The abelian CS action in the continuum belongs to a mathematical structure known as the Deligne-Beilinson double cohomology, which is related to the BRST descent equation. We will introduce the structure through an elementary, explicit presentation, and show how this structure is naturally related to the lattice theory.

In this section we use $\d$ for exterior derivative acting on differential forms in the continuum, in order to distinguish with the lattice coboundary $d$ used in the previous sections. They will be related as we proceed. Also, to distinguish the continuum degrees of freedom from the corresponding lattice ones, in this section we will include a ``$\lat$'' superscript for the lattice variables. For instance the lattice variable that appeared as $a$ in the previous Sections \ref{sect_bosonic} and \ref{sect_fermionic} will become $a^\lat$ in this section, while $a$ is now reserved for the continuum gauge field which $a^\lat$ is reduced from.

\subsection{Review of Deligne-Beilinson Description}
\label{ssect_DBrev}

We first review the Deligne-Beilinson (DB) description of abelian CS in elementary terms. We may start with a single $U(1)$ CS theory $(k/4\pi) \int_\M a\d a$ for simplicity, and generalize towards other $K$-matrices later. Rigorously speaking, this expression of action is not well-defined because the gauge field $a$ may have discontinuities for topologically non-trivial gauge configurations. The CS action may be rigorously expressed if we let our $3D$ physical spacetime manifold $\M$ be the boundary $\M=\partial\N$ of a $4D$ oriented manifold $\N$, and extend the gauge field on $\M$ into $\N$, which is always possible for $U(1)$ \cite{Dijkgraaf:1989pz}. The unambiguous action is
\begin{align}
S = 2\pi\, \frac{k}{2} \int_\N \frac{\d a}{2\pi} \frac{\d a}{2\pi} \ \ \ \ \mod 2\pi.
\end{align}
This expression should be independent of the choice of $\N$. This is the case if $k$ is even. A straightforward explanation \cite{Dijkgraaf:1989pz, Witten:2003ya, Seiberg:2016rsg} (similar to the lattice version we introduced before) is that if we have two choices $\N$ and $\mathcal{N'}$, the difference  $S_{\N}-S_{\N'}$ is equal to the action $S_{\N''}$ over a closed oriented manifold $\N''$ obtained by gluing $\N$ and $-\N'$ along their common boundary $\M$. Due to Dirac quantization, the difference $S_{\N''}=S_{\N}-S_{\N'}$ is an integer multiple of $2\pi k/2$, so it is always trivial for even $k$. This explains the case of bosonic CS. For fermionic CS where $k$ is odd, the $\pi$ ambiguity in the action can be removed if $\M$ has a given spin structure and we demand it to be extended into $\N$. This restricts the possibilities of $\N$. If we have two choices $\N$ and $\mathcal{N'}$ under this restriction, the corresponding $\N''$ would be a spin manifold, and it is known that for such $\mathcal{N''}$, the difference $S_{\N''}=S_{\N}-S_{\N'}$ is an \emph{even} multiple of $2\pi k/2$, therefore $k$ can be odd.

When $a$ has no discontinuities, one may use the Stoke's theorem to retain the usual form $(k/4\pi) \int_\M a\d a$. When the gauge configuration is topologically non-trivial so that $a$ has discontinuities, the Stoke's theorem can still be applied, but extra care is required. We will elaborate on this extra care which leads to the DB cohomology structure. Through this procedure, we also gain a more explicit understanding of the quantization condition of $k$, in straightforward connection to the lattice picture.

We proceed in three steps. First we introduce a ``good polyhedral patch system'' on $\N$ and on $\M=\partial\N$. Second we specify the data defining the $U(1)$ bundle. In the last step we use this data to carefully apply the Stoke's theorem to the CS action.

\subsubsection{Good Polyhedral Decomposition}

\begin{figure}
\centering
\includegraphics[width=.95\textwidth]{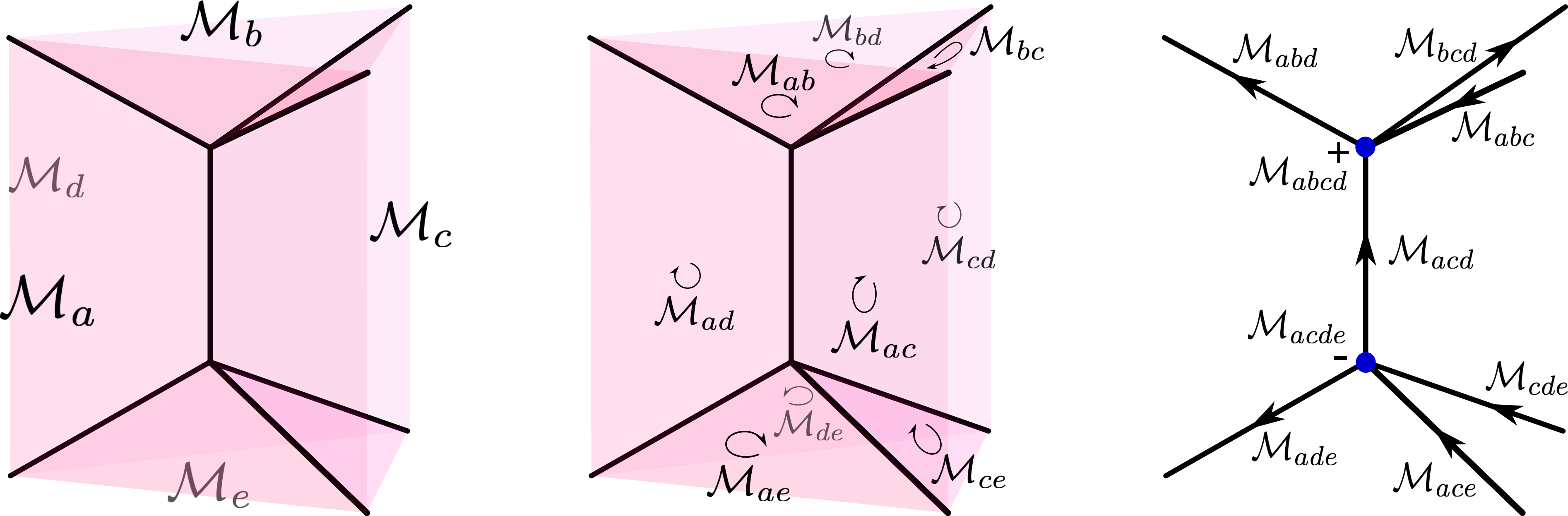}
\caption{Illustration of a few patches on the $3D$ manifold $\M$, and their intersections with orientations indicted. From left to right: some $3D$ (codimension-$0$) patches, their codimension-$1$ intersections, and their codimension-$2$ and codimension-$3$ intersections. The idea is similar in $\N$ though we cannot draw $4D$.}
\label{f4-1_patches}
\end{figure}

\begin{figure}
\centering
\includegraphics[width=.5\textwidth]{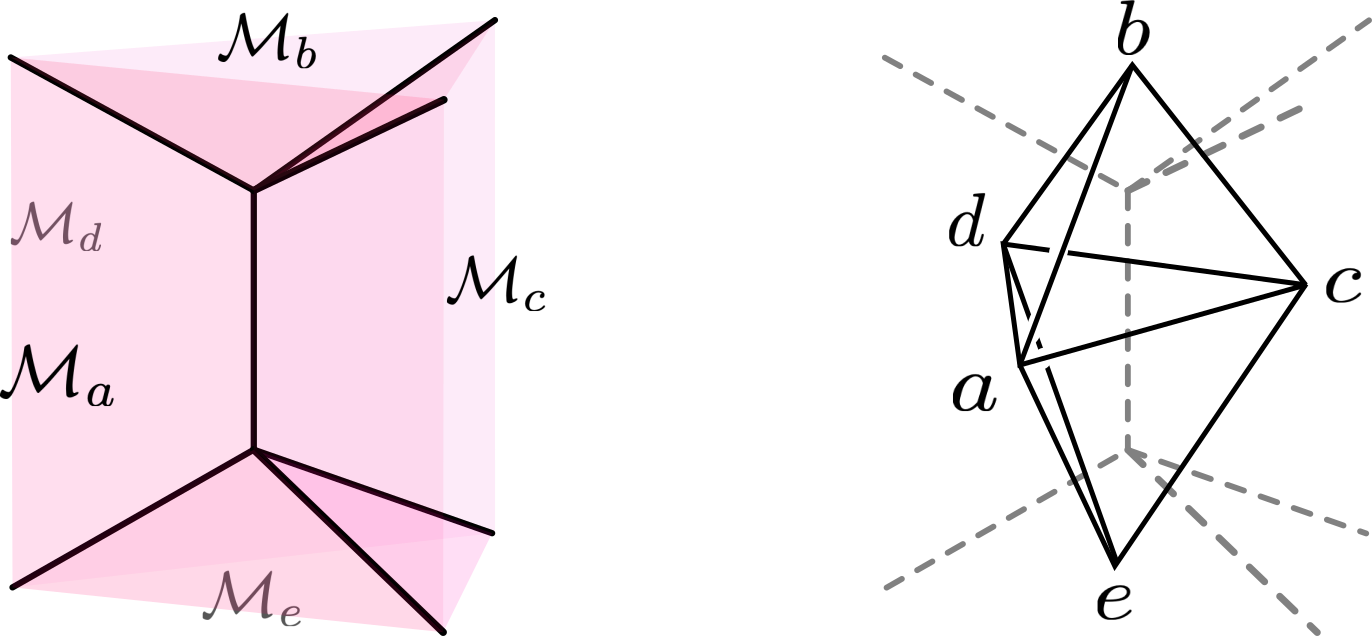}
\caption{Dualizing the patches and their intersections to the nerve simplicial complex. Our choice of orientation is consistent with the usual convention on the simplicial complex.}
\label{f4-1_nerve}
\end{figure}

Consider a ``good polyhedral decomposition'' of the $4D$ manifold $\N$. That is, we divide $\N$ into patches $\N_\a$ labeled by ``$\a$'', such that each patch $\N_{\a}$ is codimension-$0$ ($4D$), two patches $\N_\a, \N_\b$ may intersect at a codimension-$1$ ($3D$) submanifold $\N_{\a\b}$, three patches may intersect at a codimension-$2$ ($2D$) submanifold $\N_{\a\b\c}$, and so on, until that there is no intersection of six patches or more; moreover, all the patches and intersections must be topologically trivial. 
\footnote{Mathematically, this is equivalent to starting with a good open cover, take its nerve which is a simplicial complex, and then take a dual complex of the nerve. Alternatively, starting with a good open cover $\{ U_\a \}$, we let $\N_1$ be the closure of $U_1$, $\N_2$ be the closure of $U_2\backslash U_1$, $\N_3$ be the closure of $U_3\backslash (U_1\cup U_2)$, and so on.}
The orientation is understood that $\N_{\a\b}=-\N_{\b\a}$, and more generally, an odd / even permutation of the patch labels gives rise to a minus / plus sign. The nerve (or dual) of such a polyhedral patch system is a simplicial triangulation of $\N$, with ``$\a$'' labeling the vertices, ``$\a\b$'' labeling the links, and so on. The polyhedral patch system in $\N$ induces one on the physical spacetime $\M=\partial\N$. We label the patches of $\M$ by $a$, where $\{ a\}$ is a subset of $\{\a\}$ whose $\N_{\a}$ has a face on $\M$, being $\M_{a=\a}$. We have
\begin{align}
& \partial \N_{\a} = \sum_{\b} \N_{\a\b} + \M_{a=\a}, \ \ \ \partial \N_{\a\b} = \sum_{\c} \N_{\a\b\c} - \M_{ab=\a\b}, \ \ \ \partial \N_{\a\b\c} = \sum_{\d} \N_{\a\b\c\d} + \M_{abc=\a\b\c} \nonumber \\
& \partial \N_{\a\b\c\d} = \sum_{\e} \N_{\a\b\c\d\e} - \M_{abcd=\a\b\c\d}, \ \ \ \partial \N_{\a\b\c\d\e} = 0.
\end{align}
Note that $\M_a$ is codimension-$0$ which is now $3D$, $\M_{ab}$ is codimension-$1$ which is $2D$, and so on. The signs above indicate our choice of orientation, chosen so that
\begin{align}
& \partial \M_a = \sum_{b} \M_{ab}, \ \ \ \partial\M_{ab} = \sum_{c} \M_{abc}, \ \ \ \partial \M_{abc} = \sum_{d} \M_{abcd}, \ \ \ \partial \M_{abcd} = 0.
\end{align}
Clearly our convention is consistent with $\partial\partial=0$.
\footnote{Another way to motivate our convention of orientation is to think of $\N$ as a subregion of a larger $4D$ manifold. View the ``outside'' of $\N$ as another patch labeled by ``$\mathrm{o}$''. Then we let $\M_{a=\a} \equiv -\N_{\mathrm{o} \a}, \: \M_{ab=\a\b} \equiv -\N_{\mathrm{o} \a\b}, \: \M_{abc=\a\b\c} \equiv -\N_{\mathrm{o} \a\b\c}, \: \M_{abcd=\a\b\c\d} \equiv -\N_{\mathrm{o} \a\b\c\d}$.}
See Figure \ref{f4-1_patches} for an illustration of the orientations. When dualized to the nerve, the orientation is also consistent with the usual convention on the simplicial complex, see Figure \ref{f4-1_nerve}.

\subsubsection{$U(1)$ Bundle}
\label{sssect_U1Bundle}

The $U(1)$ gauge connection can be specified by the following data:
\begin{itemize}
\item
Real valued $1$-form $a_{\a}$ on each codimension-$0$ $\N_{\a}$.
\item
Real valued $0$-form (function) $\lambda_{\a\b}$ on each codimension-$1$ $\N_{\a\b}$, such that
\begin{align}
\delta a_{\a\b} \equiv a_{\b}-a_{\a} = 2\pi\d\lambda_{\a\b}.
\label{a_lambda_relation}
\end{align}
\item
Integer valued $(-1)$-form (constant number) $l_{\a\b\c}$ on each codimension-$2$ $\N_{\a\b\c}$, such that
\begin{align}
\delta \lambda_{\a\b\c} \equiv \lambda_{\b\c} - \lambda_{\a\c} + \lambda_{\a\b} = \d_{-1} l_{\a\b\c}.
\label{lambda_l_relation}
\end{align}
Here $\d_{-1}$ is simply to map a constant number to a constant function, hence $\d\d_{-1}=0$; this ensures the consistency of \eqref{a_lambda_relation} on $\N_{\a\b\c}$. For convenience, we will denote this integer valued constant function on $\N_{\a\b\c}$ as $\mathrm{l}_{\a\b\c} \equiv \d_{-1} l_{\a\b\c}$. Consistency of \eqref{lambda_l_relation} on each codimension-$3$ $\N_{\a\b\c\d}$ requires
\begin{align}
\delta l_{\a\b\c\d} \equiv l_{\b\c\d} - l_{\a\c\d} + l_{\a\b\d} - l_{\a\b\c} = 0.
\label{L_consistency_relation}
\end{align}
\end{itemize}
We may view $e^{i2\pi \lambda_{\a\b}}$ as the $U(1)$ transition function between patches. The integrity of $l_{\a\b\c}$ ensures the compatibility of the $U(1)$ transition functions:
\begin{align}
e^{i 2\pi \lambda_{\b\c}} e^{-i 2\pi \lambda_{\a\c}} e^{i 2\pi \lambda_{\a\b}} = 1.
\end{align}
The local curvature of the $U(1)$ bundle, i.e the $2$-form $\d a = \d a_{\a}$, is continuous across the patches because
\begin{align}
\d a_{\b} - \d a_{\a} = \d^2 \lambda_{\a\b}=0
\label{da_continuity}
\end{align}
on each intersection $\N_{\a\b}$.

The $U(1)$ gauge equivalence is given by:
\begin{itemize}
\item
Real valued $0$-form gauge transformation $\varphi_{\a}$ on each codimension-$0$ $\N_{\a}$, such that
\begin{align}
a_{\a} \rightarrow a_{\a} + \d\varphi_{\a}, \ \ \ 2\pi \lambda_{\a\b} \rightarrow 2\pi \lambda_{\a\b} + \delta\varphi_{\a\b} \equiv 2\pi\lambda_{\a\b} + \left( \varphi_{\b} - \varphi_{\a} \right).
\label{a_lambda_gauge_transf}
\end{align}
The case of constant $\varphi$, i.e. one satisfying $\d\varphi_\a=0$, $\delta \varphi_{\a\b}$ everywhere, requires extra care. The $2\pi\Z$ part of the constant $\varphi$ is viewed as gauge equivalence, while the $\mod 2\pi$ part of it is a $U(1)$ global symmetry.
\item
Integer valued $(-1)$-form gauge transformation $m_{\a\b}$ on each codimension-$1$ $\N_{\a\b}$, such that
\begin{align}
\lambda_{\a\b} \rightarrow \lambda_{\a} + \d_{-1} m_{\a\b}, \ \ \ l_{\a\b\c} \rightarrow l_{\a\b\c} + \delta m_{\a\b\c} \equiv l_{\a\b\c} + m_{\b\c} - m_{\a\c} + m_{\a\b}.
\label{lambda_l_gauge_transf}
\end{align}
\end{itemize}
Note however that the parametrization of gauge transformations by $\varphi$ and $m$ in turn has its own redundancy:
\begin{itemize}
\item
Integer valued $(-1)$-form ambiguity $n_{\a}$ on each codimension-$0$ $\N_\a$, such that
\begin{align}
\varphi_{\a} \rightarrow \varphi_{\a} + 2\pi \d_{-1} n_{\a} \ \ \Longleftrightarrow \ \ m_{\a\b} \rightarrow m_{\a\b} + \delta n_{\a\b} \equiv m_{\a\b} + n_{\b} - n_{\a}.
\label{varphi_m_equiv}
\end{align}
\end{itemize}
Gauge transformations that are equivalent in this manner are considered the same $U(1)$ gauge transformation.

The topology of the $U(1)$ bundle
\footnote{This discussion of the topology of the $U(1)$ bundle applies to base manifolds of general dimensions, not limited to our $3D$ spacetime $\M$.}
on $\M$ is specified by $l_{abc}$ up to gauge equivalence, hence its classification is $H^2(\M; \Z)$ by inspecting \eqref{L_consistency_relation} and \eqref{lambda_l_gauge_transf}. The universal coefficient theorem says $H^2(\M; \Z) = \mathcal{F} \oplus \mathcal{T}$ (see Appendix \ref{app_alg_topo}). The free part $\mathcal{F}=\mathrm{Hom}(H_2(\M; \Z), \Z)$ classifies the non-trivial flux configurations, i.e. it counts the number of $\d a$ fluxes (in units of $2\pi$) through the non-contractible $2$-cycles in $\M$, with the $\mathrm{l}_{abc}$ that lives on the codimension-$2$ $\M_{abc}$ being the Dirac strings. To understand this interpretation, recall that $\mathcal{F}$ is the image of  $H^2(\M; \Z)$ when mapped to $H^2(\M; \R)$. This means its non-trivial elements are the $l_{abc}$ configurations that \emph{cannot} (in any gauge) be produced by $\lambda_{ab}$ that takes constant value on each $\M_{ab}$ -- otherwise \eqref{lambda_l_relation} says $l_{abc}$ is a coboundary on the nerve of the patch system. This in turn means the $a_a$ cannot be everywhere flat -- otherwise we can set it to $0$ in each patch, in contradiction to \eqref{a_lambda_relation} and the fact that $\lambda_{ab}$ must not be constants. This explains how the non-exact $\d a$ fluxes are related to $\mathrm{l}$. On the other hand, the torsion part $\mathcal{T}$ counts the cocycles on the nerve that are coboundaries when valued in $\R$ but not in $\Z$. Its non-trivial elements correspond to the $l_{abc}$ configurations that \emph{can} be produced by $\lambda_{ab}$ that takes constant value on each $\M_{ab}$, but such that their constant values cannot be simultaneously set to integers -- otherwise they would be removed by \eqref{lambda_l_gauge_transf}. The curvature $\d a$ is globally exact (and maybe flat) if a configuration trivial in $\mathcal{F}$ but non-trivial in $\mathcal{T}$.

If we fix $l_{abc}$, the gauge transformation \eqref{lambda_l_gauge_transf} would be limited to $\delta m_{abc}=0$. Together with \eqref{varphi_m_equiv}, we learn that such gauge transformations are classified by $H^1(\M; \Z) \cong \mathcal{F}$, whose non-trivial elements are familiarly known as the large gauge transformations.

\subsubsection{Chern-Simons Action}
\label{sssect_CS_cont}

Now we look back at the CS action, given by
\begin{align}
S = 2\pi\, \frac{k}{2} \sum_{\a} \int_\mathcal{N_\a} \frac{\d a_\a}{2\pi} \frac{\d a_\a}{2\pi} \ \ \ \ \mod 2\pi
\end{align}
which is manifestly gauge invariant and thus well-defined thanks to \eqref{da_continuity}. We want to reduce it to an expression on the physical spacetime $\M$. We can use the Stoke's Theorem to write
\begin{align}
\sum_{\a} \int_\mathcal{N_\a} \frac{\d a_\a}{2\pi} \frac{\d a_\a}{2\pi} 
= + \sum_a \int_{\M_a} \frac{a_a}{2\pi} \frac{\d a_a}{2\pi} + \sum_{\a<\b}  \int_{\N_{\a\b}} \frac{-\delta a_{\a\b}}{2\pi} \frac{\d a_\b}{2\pi}
\end{align}
where in the last term we used \eqref{da_continuity}. The familiar $3D$ term on $\M$ appears. The $3D$ term in $\N$ can be further treated with Stoke's Theorem if we use \eqref{a_lambda_relation}:
\begin{align}
\sum_{\a<\b}  \int_{\N_{\a\b}} \frac{-\delta a_{\a\b}}{2\pi} \frac{\d a_\b}{2\pi} &= -\sum_{\a<\b}  \int_{\N_{\a\b}} \d \lambda_{\a\b} \frac{\d a_\b}{2\pi} \nonumber \\[.1cm]
&= - \left( -\sum_{a<b}  \int_{\M_{ab}} \lambda_{ab} \frac{\d a_b}{2\pi} + \sum_{\a<\b<\c} \int_{\N_{\a\b\c}} (+\delta\lambda_{\a\b\c}) \frac{\d a_\c}{2\pi} \right)
\end{align}
where in the last term we used \eqref{da_continuity}. A new $2D$ term on $\M$ involving the transition function appeared. The $2D$ term in $\N$ can again be treated with Stoke's Theorem if we use \eqref{lambda_l_relation}:
\begin{align}
-\sum_{\a<\b<\c} \int_{\N_{\a\b\c}} \delta\lambda_{\a\b\c} \frac{\d a_\c}{2\pi} &= -\sum_{\a<\b<\c} \int_{\N_{\a\b\c}} \mathrm{l}_{\a\b\c} \frac{\d a_\c}{2\pi} \nonumber \\[.1cm]
&= - \left( + \sum_{a<b<c} \int_{\M_{abc}} \mathrm{l}_{abc} \frac{a_c}{2\pi} + \sum_{\a<\b<\c<\d} \int_{\N_{\a\b\c\d}} \mathrm{l}_{\a\b\c} \frac{-\delta a_{\c\d}}{2\pi} \right)
\end{align}
(recall that $\mathrm{l}_{\a\b\c}\equiv \d_{-1} l_{\a\b\c}$ is the constant function taking value $l_{\a\b\c}$ on $\N_{abc}$) where in the last term we used \eqref{L_consistency_relation}. A new $1D$ term on $\M$ appeared. The $1D$ term in $\N$ can again be treated with Stoke's Theorem if we use \eqref{a_lambda_relation}:
\begin{align}
-\sum_{\a<\b<\c<\d} \int_{\N_{\a\b\c\d}} \mathrm{l}_{\a\b\c} \frac{-\delta a_{\c\d}}{2\pi} &= \sum_{\a<\b<\c<\d} \int_{\N_{\a\b\c\d}} \mathrm{l}_{\a\b\c} \d\lambda_{\c\d} \nonumber \\[.1cm]
&= - \sum_{a<b<c<d} \int_{\M_{abcd}} \mathrm{l}_{abc} \lambda_{cd} + \sum_{\a<\b<\c<\d<\e} \int_{\N_{\a\b\c\d\e}} \mathrm{l}_{\a\b\c} (+\delta\lambda_{\c\d\e})
\end{align}
where in the last term we used \eqref{L_consistency_relation}. A new $0D$ term on $\M$ appeared. The $0D$ term in $\N$ can be finally expressed, using \eqref{lambda_l_relation}, as
\begin{align}
\sum_{\a<\b<\c<\d<\e} \int_{\N_{\a\b\c\d\e}} \mathrm{l}_{\a\b\c} \delta\lambda_{\c\d\e} = \sum_{\a<\b<\c<\d<\e} \int_{\N_{\a\b\c\d\e}} \mathrm{l}_{\a\b\c} \mathrm{l}_{\c\d\e}
\end{align}
which cannot be further treated with Stoke's Theorem anymore. This term is however an integer, and is equal to $l\cup l$ on the nerve of $\N$. Piecing up the above, the CS action is equal to
\begin{align}
S &= 2\pi\, \frac{k}{2} \left( \sum_a \int_{\M_a} \frac{a_a}{2\pi} \frac{\d a_a}{2\pi} + \sum_{a<b}  \int_{\M_{ab}} \lambda_{ab} \frac{\d a_b}{2\pi} - \sum_{a<b<c} \int_{\M_{abc}} \mathrm{l}_{abc} \frac{a_c}{2\pi} - \sum_{a<b<c<d} \int_{\M_{abcd}} \mathrm{l}_{abc} \lambda_{cd} \right) \nonumber \\[.1cm]
& \ \ \ + 2\pi\, \frac{k}{2} \sum_{\a<\b<\c<\d<\e} \int_{\N_{\a\b\c\d\e}} \mathrm{l}_{\a\b\c} \mathrm{l}_{\c\d\e} \ \ \ \ \mod 2\pi.
\label{action_DB_class}
\end{align}
The first line are terms on the physical spacetime $\M$. The second line involves $\N$. When $k$ is an even integer, the second line can be dropped as a $2\pi$ phase, hence $S$ is manifestly expressed entirely on $\M$. This is the action explained in \cite{Guadagnini:2014mja} (but whose $k$ is our $k/2$). When $k$ is an odd integer, the second line gives a $\pi$ phase, which is independent of the particular choice of $\N$ as long as we demand $\N$ to admit an extension of a given spin structure on $\M$. Note that under \eqref{lambda_l_gauge_transf}, the first and the second line might simultaneously change by a $\pi$ phase. The second line with such transformation property cannot be realized with bosonic variables solely on $\M$ (as opposed to using $\N$), but it can be so if we allow fermionic variables, as we have seen in Section \ref{sect_fermionic}. This explicitly explains why $3D$ CS only allows integer level $k$, with the need of spin structure and fermionic variables if $k$ is odd.

Let's briefly mention the underlying mathematical context \cite{Carey:2004xt, Bauer:2004nh}. The relations between $a, \lambda$ and $l$ maybe compactly written in the BRST-like notation $\delta a/2\pi + \d(-\lambda)=0, \ \delta(-\lambda) + \d l = 0, \ \delta l = 0$,
\footnote{In our notation, $\delta \d = \d \delta$. To convert to the usual BRST convention $\delta \d = -\d \delta$, one may redefine some signs of $\delta$, depending on the degree of the differential form being acted on. We will not do so since our convention is geometrically more intuitive when dealing with a good patch system or its nerve (a simplicial complex).}
where $a$ is a $1$-form with ghost number $0$, $\lambda$ is a $0$-form with ghost number $1$, $l$ is an integer valued $(-1)$-form with ghost number $2$. Here the ``ghost number'' refers to the codimension of the submanifold on which the field lives. The presence of two coboundary operators, the de Rham one $\d$ and the \v{C}ech one $\delta$, leads to the notion of Delign-Beilinson (DB) double cochain complex. The ordered triplet $(a/2\pi, \lambda, l)$ related in such a BRST manner with integer valued $l$ is known as a $1$-hypercocycle in the complex with $\Z$ coefficient, where the degree $1$ of the hypercocycle refers to the sum of the differential form degree and the ghost number. More precisely, the triplet is a $1$-hypercohomology class because of gauge equivalence. The parenthesis of \eqref{action_DB_class} is in turn a $3$-hypercohomology class, formed by the so-called DB pairing of $(a/2\pi, \lambda, l)$ with itself \cite{Guadagnini:2008bh, Guadagnini:2014mja}. One may notice that the terms in the parenthesis of \eqref{action_DB_class} are in correspondence to the BRST descent equation \cite{Manes:1985df}.
\footnote{There is a geometrical picture for BRST ghost and descent equation \cite{ThierryMieg:1979kh,Manes:1985df}, in which a ghost is ``a direction of exploring the possible gauge transformations''. In our case the gauge transformation is not a ``exploration'' but rather a transition function located between the patches.}
More particularly, each term in the parenthesis of \eqref{action_DB_class} can be obtained by applying $\delta$ to the previous term, removing a total $\d$ derivative, and multiplying an extra $\mp 1$ depending on whether the codimension of the submanifold is odd or even.

We would like to note that the expression \eqref{action_DB_class} is not unique. When applying the Stoke's Theorem to derive \eqref{action_DB_class}, we had other ways to ``pull out a total derivative'', for instance $\d\lambda_{\a\b} \d a_\b$ can be equally well expressed as $\d(\lambda_{\a\b} \d a_\b)$ (as we originally did) or as $-\d (\d \lambda_{\a\b} a_\b)$. Thus, one may arrive at other equivalent expressions of $S$, such as
\begin{align}
& 2\pi\, \frac{k}{2} \left( \sum_a \int_{\M_a} \frac{a_a}{2\pi} \frac{\d a_a}{2\pi} - \sum_{a<b}  \int_{\M_{ab}} \d\lambda_{ab} \frac{a_b}{2\pi} + \sum_{a<b<c} \int_{\M_{abc}} \d\lambda_{ab} \lambda_{bc} - \sum_{a<b<c<d} \int_{\M_{abcd}} \lambda_{ab} \mathrm{l}_{bcd} \right) \nonumber \\[.1cm]
& + 2\pi\, \frac{k}{2} \sum_{\a<\b<\c<\d<\e} \int_{\N_{\a\b\c\d\e}} \mathrm{l}_{\a\b\c} \mathrm{l}_{\c\d\e} \ \ \ \ \mod 2\pi.
\label{action_DB_class_alt}
\end{align}
Alternatively, \eqref{action_DB_class_alt} and other equivalent expressions can be obtained from \eqref{action_DB_class} via some integration by parts. The expression \eqref{action_DB_class_alt} is particularly convenient for those $U(1)$ bundles trivial in $\mathcal{F}$ (but maybe non-trivial in $\mathcal{T}$), because their $\lambda_{ab}$ can be made constant on $\M_{ab}$, so that the two $\d\lambda$ terms drop out.

The phase of a Wilson loop of charge $w$ along path $W$ is 
\begin{align}
-w\oint_W a \equiv -w\sum_{a} \int_{W\bigcap \M_a} a_a - 2\pi w\sum_{a<b} \int_{W\bigcap \M_{ab}} \lambda_{ab}.
\label{WL_cont}
\end{align}
\footnote{In the previous sections on the lattice, we absorbed the charge $w$ into the path $W$. In the continuum it is more customary not to do so.}
When the loop $W$ is contractible, this agrees with the integral of the curvature $\d a$ over a surface bounded by $W$. In order for this phase to be invariant under \eqref{lambda_l_gauge_transf} up to $2\pi$, the charge $w$ must be an integer.
\footnote{We cannot avoid this quantization by claiming the loop lies entirely within one patch $\M_a$, because the theory should be independent of how we decompose the manifold into patches.}

It is straightforward to generalize to CS with multiple $U(1)$ gauge groups. The action is
\begin{align}
S &= 2\pi\, \sum_{I, J} \frac{K_{IJ}}{2} \int_\N \frac{\d a^I}{2\pi} \frac{\d a^J}{2\pi} \ \ \ \ \mod 2\pi \nonumber \\[.1cm]
&= 2\pi\, \sum_{I} \frac{K_{II}}{2} \int_\N \frac{\d a^I}{2\pi} \frac{\d a^I}{2\pi} + 2\pi\, \sum_{I<J} K_{IJ} \int_\N \frac{\d a^I}{2\pi} \frac{\d a^J}{2\pi}  \ \ \ \ \mod 2\pi
\end{align}
where $K_{IJ}$ is symmetric. Repeating the procedure above, we easily arrive at the generalization of \eqref{action_DB_class}:
\begin{align}
S &= 2\pi\, \sum_I \frac{K_{II}}{2} \left( \sum_a \int_{\M_a} \frac{a^I_a}{2\pi} \frac{\d a^I_a}{2\pi} + \sum_{a<b}  \int_{\M_{ab}} \lambda^I_{ab} \frac{\d a^I_b}{2\pi} - \sum_{a<b<c} \int_{\M_{abc}} \mathrm{l}^I_{abc} \frac{a^I_c}{2\pi} \right. \nonumber \\[.1cm]
& \hspace{4cm} \left. - \sum_{a<b<c<d} \int_{\M_{abcd}} \mathrm{l}^I_{abc} \lambda^I_{bcd} \right) + 2\pi\, \sum_I \frac{K_{II}}{2} \sum_{\a<\b<\c<\d<\e} \int_{\N_{\a\b\c\d\e}} \mathrm{l}^I_{\a\b\c} \mathrm{l}^I_{\c\d\e} \nonumber \\[.2cm]
& \ \ \ + 2\pi\, \sum_{I<J} K_{IJ} \left( \sum_a \int_{\M_a} \frac{a^I_a}{2\pi} \frac{\d a^J_a}{2\pi} + \sum_{a<b}  \int_{\M_{ab}} \lambda^I_{ab} \frac{\d a^J_b}{2\pi} - \sum_{a<b<c} \int_{\M_{abc}} \mathrm{l}^I_{abc} \frac{a^J_c}{2\pi} \right. \nonumber \\[.1cm]
& \hspace{4cm} \left. - \sum_{a<b<c<d} \int_{\M_{abcd}} \mathrm{l}^I_{ab} \lambda^J_{bcd}  \right) + \ 2\pi\, \sum_{I<J} K_{IJ} \sum_{\a<\b<\c<\d<\e} \int_{\N_{\a\b\c\d\e}} \mathrm{l}^I_{\a\b\c} \mathrm{l}^J_{\c\d\e} \nonumber \\[.2cm] &
\ \ \ \ \mod 2\pi.
\label{action_DB_class_multiple}
\end{align}
For the action to be independent of the choice of $\N$, we need $K_{IJ}$ to be integers, and in particular $K_{II}$ be even integers. We can allow $K_{II}$ to be odd if a spin structure and fermionic variables are available on $\M$.

\subsection{Retrieve of Lattice Theory for Doubled $U(1)$ Chern-Simons}
\label{ssect_retrieveU1U1}

Through our elementary and explicit introduction to the DB description of the abelian CS in the continuum, one can envision that the nerve of the continuum patch system is going to serve as the simplicial complex on which our lattice theory is defined.
\footnote{This aspect is the same as in the ``Swiss cheese'' method which leads to the Turaev-Viro lattice constructions for more general topological field theories \cite{Bhardwaj:2016clt}. The Turaev-Viro construction applies to more general theories with given spherical fusion category data; our abelian theories under consideration are special cases within the formalism. However, the Dirac strings are considered invisible in the spherical fusion category data, so the theory cannot couple to a background electromagnetic field. To include this coupling is one of the main purposes of our present work.}
In the below we will present the detailed mapping from the continuum to the lattice. We consider the doubled $U(1)$ CS first; it is straightforward to reduce to the $U(1)\times \R$ CS by identifying some Dirac strings.

We start with the continuum action \eqref{action_DB_class_multiple} with three gauge fields -- two dynamical ones $a, b$ and a background one $A$. The $K$-matrix is given by
\begin{align}
K_{ab}= n, \ \ \ \ \ K_{aA}= -q, \ \ \ \ \ K_{bA}= -p, \ \ \ \ \ K_{aa}=0, \ \ \ \ \ K_{bb}=k, \ \ \ \ \ K_{AA}=0
\label{U1U1_K_matrix}
\end{align}
More precisely, the dynamical gauge fields are the triplets $(a/2\pi, \lambda, l)$ and $(b/2\pi, \sigma, s)$, while the background one can be chosen as $(A/2\pi, 0 ,0)$. We can ignore the Dirac string of the background $A$ field because it is essentially equivalent to a Wilson loop with charges $w=q$, $v=p$ under the $a$ and $b$ fields, as we mentioned below \eqref{electromagnetic_string_absorbed}.
\footnote{Though we will ignore the Dirac strings $L$ associated with the $A$ field because of this equivalence, if one wants to one can easily include $L_A$ and verify the equivalence we claimed. It is easy to do so because the way we will treat the $L$ Dirac strings in the $K_{aA}$, $K_{bA}$ terms is essentially the same as the way we treat the $s$ Dirac strings in the $K_{ab}$ terms in \eqref{cont_action_general_topo}.}

We start with trivial $H^2(\M; \Z)$ to introduce the local aspects of the mapping first. Then we consider $H^2(\M; \Z)$ with torsion only, and finally general $H^2(\M; \Z)$.

\subsubsection{On Trivial Topology}

On trivial topology, we can always choose a gauge so that the transition functions $\lambda_{ab}$ and $\sigma_{ab}$ vanish, and hence so do $l_{abc}$ and $s_{abc}$; we are left with $a_a$ and $b_a$ in the patches. Doing so completely fixes the \eqref{lambda_l_gauge_transf} gauge (up to the ambiguity \eqref{varphi_m_equiv}) and also enforces \eqref{a_lambda_gauge_transf} to satisfy $\delta \varphi_{ab}=0$. The remaining gauge redundancy in \eqref{a_lambda_gauge_transf} will then be fixed by the Faddeev-Popov measures for $a$ and $b$. A caveat regarding the global symmetry (introduced below \eqref{a_lambda_gauge_transf}) will be addressed after \eqref{recovered_lattice_action_trivial_topo} and \eqref{b_a_lat_redef}. \footnote{On general topology, there is the extra caveat of large gauge transformations (introduced at the end of Section \ref{sssect_U1Bundle}), which we will address after \eqref{recovered_lattice_action_general_topo}.}

The continuum action \eqref{action_DB_class_multiple}, upon the inclusion of a Wilson loop of charge $w$ under $a$ (the generalization to multiple Wilson loops is obvious) and one with charge $v$ under $b$, reads
\begin{align}
\sum_a \int_{\M_a} a_a \left( \frac{n}{2\pi}\, \d b_a - w \delta^2_W - \frac{q}{2\pi} \d A_a \right) + \sum_a \int_{\M_a} \frac{k}{4\pi} b_a \d a_a - \sum_a \int_{\M_a} b_a \left( v \delta^2_V + \frac{p}{2\pi} \d A_a \right)
\end{align}
where $\delta^2_W$ is a two-form distribution concentrated on the Wilson loop $W$ and likewise for $\delta^2_V$; more explicitly $\left(\delta^2_W\right)_{\mu\nu}(x) \equiv \epsilon_{\mu\nu\lambda} \oint_W \d y^\lambda \delta^3(x-y)$.

To proceed, let's first turn off the background $A$ field. Consider the scenario that the Wilson loop $W$ runs along the codimension-$2$ ($1D$) intersections $\M_{abc}$. This is always possible since we can choose how to decompose $\M$ into patches. The role of the $a$ field is to be the Lagrange multiplier enforcing $\d b$ to be a narrow flux satisfying
\begin{align}
\frac{n}{2\pi}\, \d b = w \delta^2_W
\label{cont_flux_attachment}
\end{align}
(we dropped the patch label on $\d b$ because of \eqref{da_continuity}). Since the $\d b$ flux is a narrow distribution concentrated on the curves $\M_{abc}$, we can in turn let the $b$ gauge field to be a narrow distribution concentrated on the surfaces $\M_{ab}$; note that the strength of the distribution is constant over the surface.

Viewed from the nerve simplicial complex, the constant strength of the $b$ field distribution over $\M_{ab}$ can be associated to a lattice variable $b^\lat_{ab}$ residing on the nerve link $ab$, and the strength of the $\d b$ flux along $\M_{abc}$ can be associated to $(db^\lat)_{abc}$ on the nerve plaquette $abc$. The mode of the Lagrange multiplier $a$ field that fixes the value of $db^\lat_{abc}$ is given by the integral of $a$ along $\M_{abc}$, which we denote as $a^\lat_{abc}$, a degree of freedom on the nerve plaquette $abc$. All the other fluctuating modes of $a$ fixes the other modes of $b$ to be gauge equivalent to $0$. In other words, we integrate out those other modes first, since the Wilson loop placed along $\M_{abc}$ will never invoke those modes anyways, so we are left with the modes associated to $a^\lat_{abc}$ and $b^\lat_{ab}$ lattice gauge degrees of freedom.
\footnote{Notably, the continuum path integral measure must ensure this procedure of ``integrating out the other modes constrained by the Lagrange multiplier'' to yield a numerical factor $1$ to the partition function (or at most yield a product of \emph{local} factors insensitive to the spacetime topology, so that this factor can be removed by local counter terms). Intuitively this is because the patches are topologically trivial, within which the Lagrange multiplier $adb$ integral (this is the importance of having \emph{doubled} CS) should be trivial up to local numerical factor. We hope this point can be made more rigorous at the functional analysis level, which we will not pursue in this work. In the end, this point can be \emph{a posteri} justified since our construction produces the right partition function normalization.}
The lattice degrees of freedom are coupled as $a^\lat_{abc} (db^\lat)_{abc}$, which is the $a^\lat \cdot db^\lat$ term in our lattice gauge theory.

We are almost reaching a lattice gauge theory description, yet we still need to show how the $b \d b$ term in the continuum is related to the $b^\lat d b^\lat$ term on the lattice. It is well-known that the Wilson loop in continuum CS requires a point-splitting regularization \cite{Witten:1988hf}.
\footnote{As mentioned in footnote \ref{geo_framing}, there is another regularization, Polyakov's geometrical regularization, which requires a metric and is not under our consideration.}
Practically, this means we artificially prescribe a framing loop $W_f$ that stays close to $W$, so that $W$ and $W_f$ look like the two edges of a ribbon without self-intersection (the statement does not depend on the details of the metric). We then regularize the term $b \d b$ as $b_f \d b$ where $b$ satisfies \eqref{cont_flux_attachment} while $b_f$ satisfies the same equation but with $W_f$ in place of $W$. In the below we incorporate this framing regularization, and see how the lattice cup product, which depends on the ordering of the nerve vertices (patch labels), arises from our artificial choice of splitting $W$ and $W_f$.

\begin{figure}
\centering
\includegraphics[width=.55\textwidth]{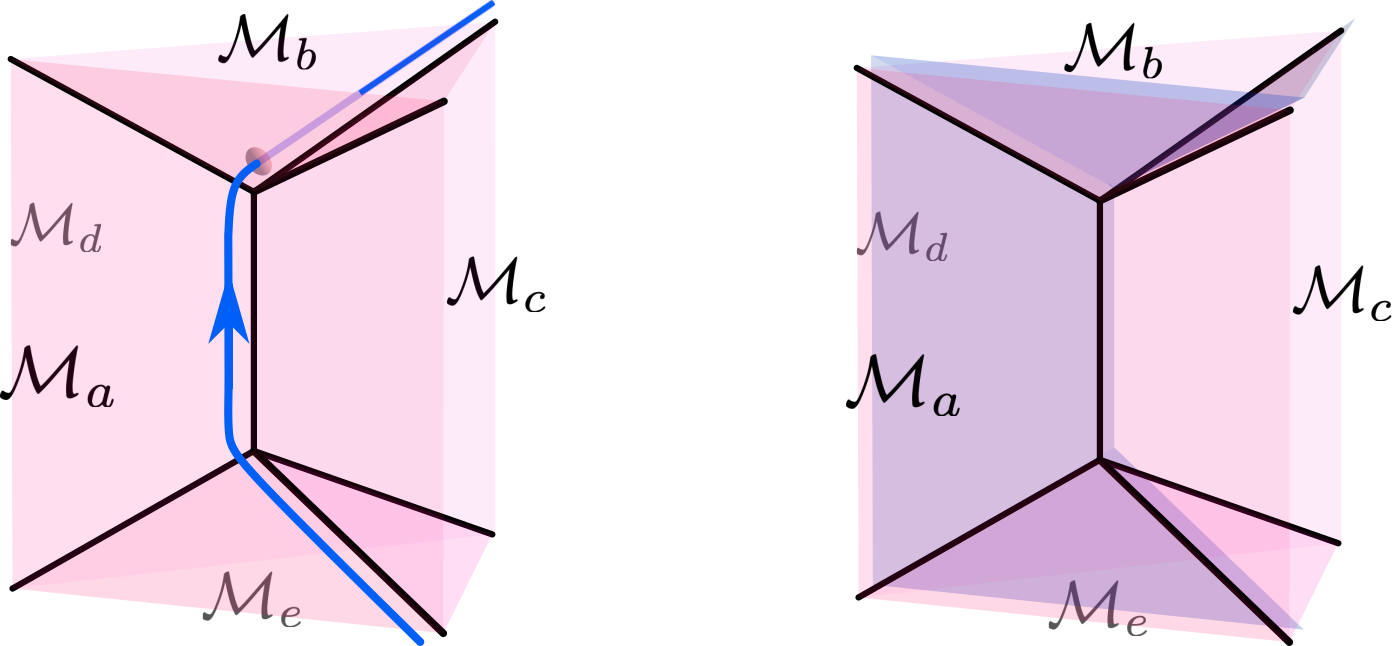}
\caption{Displacement of the Wilson loop $W \sim \d b$ (left), and displacement of the framed gauge field $b_f$ (right), assuming $a<b<c<d<e$. The $b\d b$ term, understood as $b_f \d b$, receives a contribution from the indicated point in the vicinity of the intersection point $\M_{abcd}$ (in reference to Figure \ref{f4-1_patches}). On the nerve simplicial complex, this contribution is given by the term $(b^\lat \cp db^\lat)_{abcd}$ on the $abcd$ tetrahedron. The fact that the indicated point occurs inside $\M_b$ in the continuum is related to the fact that $b$ is the repetitive vertex in the definition of the cup product on the simplicial complex.}
\label{f4-2_WL}
\end{figure}

First, let's displace the ``magnetic'' Wilson loop $W$ slightly away from $\M_{abc}$. Assuming $a<b<c$, we choose the displacement so that the segment of $W$ along $\M_{abc}$ is slightly displaced into the patch $\M_a$; see the left panel of Figure \ref{f4-2_WL}. This displacement prescription does not depend on the details of the metric. Denote this displaced image of $\M_{abc}$ as $\M_{a(bc)} \subset \M_a$. In each patch $\M_a$, \eqref{cont_flux_attachment} implies
\begin{align}
\d b_a = \sum_{b,c \: (a<b<c)} (db^\lat)_{abc} \: \delta^2_{\M_{a(bc)}}, \ \ \ (db^\lat)_{abc} = \pm \frac{2\pi w}{n} \mbox{ \ if $W$ runs near $\pm \M_{abc}$}.
\label{W_displacement}
\end{align}
Accordingly, while we previously defined $a^\lat_{abc}$ to be the integral of the $a$ field along $\mathcal{M}_{abc}$, now we adjust its definition to be the integral of $a_a$ along $\mathcal{M}_{a(bc)}$. On the other hand, we slightly displace the framed Wilson loop $W_f$ into the patch $\M_c$ assuming $a<b<c$, and denote the displaced image as $\M_{(ab)c} \subset \M_c$. Thus
\begin{align}
\d (b_f)_c = \sum_{a,b \: (a<b<c)} (db^\lat)_{abc} \: \delta^2_{\M_{(ab)c}}, \ \ \ (db^\lat)_{abc} = \pm \frac{2\pi w}{n} \mbox{ \ if $W$ runs near $\pm \M_{abc}$}.
\label{W_f_displacement}
\end{align}
The framed gauge field $b_f$ which originally distributes over the surface $\M_{ab}$ is then slightly displaced into the patch $\M_b$ assuming $a<b$; see the right panel of Figure \ref{f4-2_WL}. Denote this displaced image of $\M_{ab}$ as $\M_{(a)b}\subset \M_b$, and let $\delta^1_{\M_{(a)b}}$ is the $1$-form distribution concentrated on it, we have
\begin{align}
(b_f)_b = \sum_{a\: (a<b)} b^\lat_{ab} \: \delta^1_{\M_{(a)b}}.
\label{b_f_displacement}
\end{align}
Piecing up the above, we find the framed $b\d b$ term is equal to
\begin{align}
\sum_b \int_{\M_b} (b_f)_b \: \d b_b &= \sum_b \ \sum_{a\: (a<b)} \ \sum_{c, d \: (b<c<d)} \ b^\lat_{ab} \ (db^\lat)_{bcd} \int_{\M_b} \delta^1_{\M_{(a)b}} \delta^2_{\M_{b(cd)}} \nonumber \\[.1cm]
&= \sum_{a<b<c<d} b^\lat_{ab} \ (db^\lat)_{bcd} \ \sgn(\M_{abcd})
\end{align}
where $\sgn(\M_{abcd})$ equals $\pm 1$ if the orientation of the point $\M_{abcd}$ is $\pm 1$ (on the nerve, this is equivalent to whether the tetrahedron has its four vertices $a<b<c<d$ appearing in the right- or left-handed-rule configuration), and equals $0$ if the point $\M_{abcd}$ does not exist. This is nothing but our lattice cup product summation $\int b^\lat \cp db^\lat$ defined in Section \ref{sect_bosonic}. The computation is illustrated in Figure \ref{f4-2_WL}. Similar to the ``magnetic'' Wilson loop $W$, we also place the ``electric'' Wilson loop $V$ on $\M_{a(bc)} \subset \M_a$ in the vicinity of $\M_{abc}$, and we understand $b \delta^2_V$ as $b_f \delta^2_V$.

We have thus arrived at a lattice action
\begin{align}
\int a^\lat  \cdot \left( \frac{n}{2\pi} db^\lat - W^\lat \right) + \frac{k}{4\pi}\int b^\lat \cp db^\lat - \int b^\lat \cp V^\lat
\label{recovered_lattice_action_trivial_topo}
\end{align}
where $W^\lat_{abc}$ takes value $\pm w$ if $W$ runs along / against the vicinity of $\M_{abc}$,
\footnote{In the lattice theories in Section \ref{sect_bosonic}, we introduced the framed Wilson loop on the lattice, $W^\lat_f$ in \eqref{frame_capp}. In turns out that geometrically, the relative separation between $W^\lat_f$ and $W^\lat$ resembles the separation between $W$ and $W_f$, rather than that between $W_f$ and $W$, according to our continuum definition \eqref{W_displacement} and \eqref{W_f_displacement}. This issue is not substantial, because if one wants to one may either switch the definitions of $W^\lat$ and $W^\lat_f$ on the lattice, or of $W$ and $W_f$ in the continuum; any linking number stays the same anyways. We will not do so for book-keeping, because either switch will significantly increase the use of the ``$f$'' subscript.}
and $V^\lat_{abc}$ takes value $\pm v$ if $V$ runs along / against the vicinity of $\M_{abc}$. The lattice variables $b^\lat_{ab}$ and $a^\lat_{abc}$ take real values. In order to compare this to our doubled $U(1)$ CS lattice action \eqref{theory_U1U1_no_A}, we separate the real valued $b^\lat_{ab}$ to its $2\pi\Z$ part, denoted as $2\pi y_{ab}$, and its $\mod 2\pi$ part, still denoted as $b^\lat_{ab}$; likewise, we separate the real valued $a^\lat_{abc}$ to its $2\pi\Z$ part, denoted as $2\pi z_{abc}$, and its $\mod 2\pi$ part, still denoted as $a^\lat_{abc}$:
\begin{align}
& b^\lat_{ab} \in \R \ \ \longrightarrow \ \ b^\lat_{ab} + 2\pi y^\lat_{ab} \ \mbox{ with } \ y^\lat_{ab}\in \Z, \ \ b^\lat_{ab} \in [0, 2\pi), \nonumber \\[.1cm]
& a^\lat_{abc} \in \R \ \ \longrightarrow \ \ a^\lat_{abc} + 2\pi z^\lat_{abc} \ \mbox{ with } \ z^\lat_{abc}\in \Z, \ \ a^\lat_{abc} \in [0, 2\pi).
\label{b_a_lat_redef}
\end{align}
On trivial topology, any $s^\lat\in Z^2(M; \Z)$ can be written as $s^\lat=-dy^\lat$, and any $l^\lat\in Z_1(M; \Z)$ can be written as $l^\lat = -d^\star z^\lat$, therefore we have indeed recovered \eqref{theory_U1U1_no_A} without $A^\lat$.

We comment on a subtlety related to the path integral measure. In \eqref{recovered_lattice_action_trivial_topo}, before we perform \eqref{b_a_lat_redef}, both $b^\lat_{ab}$ and $a^\lat_{abc}$ take real values, so naively, on trivial topology, the theory seems like a doubled $\R$ CS. In fact, there is still a difference between our doubled $U(1)$ CS and a doubled $\R$ CS, even on trivial topology. In the doubled $U(1)$ CS under consideration, in the Faddeev-Popov measure we remove gauge redundancy but keep a $U(1)\times U(1)$ global symmetry, while in doubled $\R$ CS we keep an $\R\times \R$ global symmetry. Their difference can be understood through the discussion below \eqref{a_lambda_gauge_transf}, yielding different normalizations of the partition function, $n^{-B_0}$ versus $(\sqrt{n} \:\delta(0))^{-2B_0}$ as seen in Section \ref{sssect_U1R_partition}. After performing \eqref{b_a_lat_redef}, the global symmetry being $U(1)\times U(1)$ rather than $\R\times \R$ is manifestly implemented, because we would then regard $s^\lat$ and $l^\lat$ as the dynamical integer degrees of freedom, manifestly getting rid of the integer ``gauge fields'' $y^\lat$ and $z^\lat$.

It remains to include the electromagnetic background $A$. In order to have exact mapping from the continuum to the lattice, we need $\d A$ to form narrow flux tubes, so that we can treat them as we treated the Wilson loops, i.e. we place the narrow $\d A$ flux on $\M_{a(bc)} \subset \M_a$ in the vicinity of $\M_{abc}$.
\footnote{Just as we interpreted $b\d b$ as $b_f \d b$ and $b \delta^2_V$ as $b_f \delta^2_V$, now we interprete $b \d A$ as $b_f \d A$.}
We can in turn choose to make $A$ a thin distribution of strength $A^\lat$ on $\mathcal{M}_{a(b)}$; on the nerve simplicial complex, the $A$ field is then associated to the links, just as the $b$ field is. As a result, in \eqref{recovered_lattice_action_trivial_topo},
\begin{align}
W^\lat \rightarrow W^\lat + \frac{q}{2\pi} dA^\lat, \ \ \ \ V^\lat \rightarrow V^\lat + \frac{p}{2\pi} dA^\lat.
\end{align}
When we perform \eqref{b_a_lat_redef}, since $dA^\lat$ might not be quantized, unlike $W^\lat$ and $V^\lat$, we have the extra terms $s^\lat \cp A^\lat$ and $l^\lat \cp A^\lat$ seen in \eqref{theory_U1U1}. What if the $\d A$ flux smears over the continuum instead of forming flux tubes? Then we just invoke the usual assumption underlying almost any use of lattice gauge theories, that as long as the lattice is fine enough, the physics of a smearing background gauge field can be approximated by fluxes on a discretized spacetime (which form narrow flux tubes on the dual lattice).

\subsubsection{On Topology with Torsion Only}

Let's then move to non-trivial spacetime topology. In trivial topology we only needed to integrate over gauge fields with trivial transition functions. In general topology, we must include a summation over $H^2(\M; \Z)$ which classifies the $U(1)$ bundles (see Section \ref{sssect_U1Bundle}). We take a representative gauge configuration $(b_\gamma/2\pi, \sigma_\gamma, s_\gamma)$ for each class $\gamma\in H^2(\M; \Z)$; any other gauge configuration in the class $\gamma$ can be written as the sum of the representative plus a non-topological gauge configuration $(b_{\mathrm{n.t.}}/2\pi, 0, 0)$ which we then integrate over. Likewise for $(a/2\pi, \lambda, l)$. It has been shown that such prescription of $U(1)$ path integral measure is gauge invariant \cite{Bauer:2004nh, Guadagnini:2008bh}. Let's first consider topologies with torsion only, i.e. $H^2(\M; \Z)=\mathcal{T}$ with the free part $\mathcal{F}$ trivial. These cases are simpler for two reasons: First, the representative gauge configurations can be chosen flat, as mentioned in Section \ref{sssect_U1Bundle}. On contrary, configurations with non-trivial free part necessarily have non-trivial (non-exact) flux. Second, as long as we have fixed $\sigma$ and $\lambda$ in the representative gauge configurations, the \eqref{lambda_l_gauge_transf} gauge is still completely fixed (up to the ambiguity \eqref{varphi_m_equiv}); there is no large gauge transformation.

More exactly, for each class $\tau \in H^2(\M; \Z)=\mathcal{T}$, we can choose some representative Dirac string configuration $(s_\tau)_{abc}$ such that $[s_\tau] =\tau$. When viewed from the nerve simplicial complex (which we denote as $M$), the Dirac string configuration $s_\tau\in Z^2(M; \Z)$ must be expressible as $d \sigma^\lat_{\tau'}$ for some $\sigma^\lat_{\tau'} \in C^1(M; \R)$, because the free part $\mathcal{F}$ -- the image of $H^2(\M; \Z)$ in $H^2(\M; \R)$ -- is trivial. In particular, if $j_\tau \tau$ is a trivial class for some integer $j_\tau$, then $j_\tau s_\tau\in B^2(M; \Z)$, which means we can choose $\sigma^\lat_\tau \in C^1(M; (1/j)\Z)$ \cite{Guadagnini:2014mja}. When viewed from the continuum patches, according to \eqref{lambda_l_relation}, we can simply let the transition function $(\sigma_\tau)_{ab}$ on $\M_{ab}$ to be a function taking constant value $(\sigma^\lat_\tau)_{ab}$; in turn, it is compatible with \eqref{a_lambda_relation} to make $b_\tau=0$ in the representative gauge configuration. Thus, the representative gauge configuration is chosen to be $(0, \sigma_\tau, s_\tau)$ for each $\tau\in\mathcal{T}$. Since $\sigma_\tau$ take constant non-integer values (except for $\tau=0$), they cannot be completely removed by \eqref{lambda_l_gauge_transf}. On top of the representative $U(1)$ gauge configuration $(0, \sigma_\tau, s_\tau)$, we add non-topological fluctuations $(b_{\mathrm{n.t.}}/2\pi, 0, 0)$. The same applies to $(a/2\pi, \lambda, l)$. Therefore, the degrees of freedom of the gauge configurations are
\begin{align}
& s_{abc} = (s_\tau)_{abc}, \ \ \ \ \sigma_{ab} = (\sigma_\tau)_{ab}, \ \ \ \ b_a = (b_{\mathrm{n.t.}})_a \nonumber \\[.2cm]
& l_{abc} = (l_{\tau'})_{abc}, \ \ \ \ \lambda_{ab} = (\lambda_{\tau'})_{ab}, \ \ \ \ a_a = (a_{\mathrm{n.t.}})_a.
\end{align}
Note that the classes $\tau$ and $\tau'$ are to be summed over separately in the path integral; when $\tau=\tau'$, it does not matter whether our gauge choices of the representatives $l_{\tau'}$ and $s_\tau$ are the same or different, so we may assume them to the be same for defniteness.

We substitute the ingredients above into the action \eqref{action_DB_class_multiple} with the K-matrix \eqref{U1U1_K_matrix}. For simplicity we first consider a bosonic theory with even $k$. We simply find
\footnote{The easiest way to derive this is to use (the multiple gauge field analogue of) the equivalent form \eqref{action_DB_class_alt}, for the reason explained below \eqref{action_DB_class_alt}.}
\begin{align}
&\phantom{+ \ } \frac{n}{2\pi} \left( \sum_a \int_{\M_a} (a_{\mathrm{n.t.}})_a \d (b_{\mathrm{n.t.}})_a - (2\pi)^2 \sum_{a<b<c<d} \int_{\M_{abcd}} (\lambda_{\tau'})_{ab} (\mathrm{s}_\tau)_{bcd} \right) \nonumber \\[.2cm]
& - \sum_a \int_{\M_a} (a_{\mathrm{n.t.}})_a \left( w \delta^2_W + \frac{q}{2\pi} \d A_a \right) - 2\pi w \sum_{a<b} \int_{\M_{ab}} (\lambda_{\tau'})_{ab} \delta^2_W  \nonumber \\[.2cm]
& + \frac{k}{4\pi} \left( \sum_a \int_{\M_a} (b_{\mathrm{n.t.}, \, f})_a \d (b_{\mathrm{n.t.}})_a - (2\pi)^2 \sum_{a<b<c<d} \int_{\M_{abcd}} (\sigma_\tau)_{ab} (\mathrm{s}_\tau)_{bcd} \right) \nonumber \\[.2cm]
& - \sum_a \int_{\M_a} (b_{\mathrm{n.t.}, \, f})_a \left( v \delta^2_V + \frac{p}{2\pi} \d A_a \right) - 2\pi v \sum_{a<b} \int_{\M_{ab}} (\sigma_\tau)_{ab} \delta^2_V
\label{cont_action_torision_only}
\end{align}
where the first two lines come from the $K_{ab}, K_{aA}$ entries, and the last two lines come from the $K_{bb}, K_{bA}$ entries; notice the transition functions also couple to the Wilson loop, see \eqref{WL_cont}. The non-topological fluctuations $a_{\mathrm{n.t.}}, b_{\mathrm{n.t.}}$ are then treated as we did for \eqref{recovered_lattice_action_trivial_topo} in trivial topology, so that they reduce to the lattice degrees of freedom $a^\lat, b^\lat$. Also recall that $(\sigma_\tau)_{ab}=\d_{-1} (\sigma^\lat_\tau)_{ab}, \ (\mathrm{s}_\tau)_{abc} = (\delta \sigma_\tau)_{abc} = \d_{-1} (d\sigma^\lat_\tau)_{abc}$ and likewise for $\lambda_{\tau'}$ and $\mathrm{l}_{\tau'}$. Moreover, we assume the background $\d A$ forms narrow flux tubes as before. We arrive at a lattice action
\begin{align}
&\phantom{+ \ } \int a^\lat \cdot \left( \frac{n}{2\pi} db^\lat - W^\lat - \frac{q}{2\pi} dA^\lat \right) - 2\pi n \int \lambda^\lat_{\tau'} \cp d\sigma^\lat_\tau - 2\pi \int \lambda^\lat_{\tau'} \cp W^\lat \nonumber \\[.2cm]
& + \frac{k}{4\pi}\int \left( b^\lat \cp db^\lat - (2\pi)^2 \sigma^\lat_\tau \cp d\sigma^\lat_\tau \right) - \int b^\lat \cp \left( V^\lat + \frac{p}{2\pi} dA^\lat \right) - 2\pi \int \sigma^\lat_\tau \cp V^\lat
\label{lat_action_torision_only}
\end{align}
We may get rid of $\lambda^\lat_{\tau'}, \sigma^\lat_\tau$ and only use the gauge independent $l_{\tau'} = d\lambda^\lat_{\tau'}, s_\tau=d\sigma^\lat_\tau$ if we shift
\begin{align}
b^\lat \ \rightarrow \ b^\lat - 2\pi \sigma^\lat_\tau, \ \ \ \ a^\lat \ \rightarrow \ a^\lat - 2\pi \lambda^\lat_{\tau'} \capp \mathrm{M}
\label{lat_dof_redef}
\end{align}
within in each pair of classes $\tau, \tau'$; here, $\mathrm{M}$ is the $3$-chain of the nerve of $\M$. We get (ignoring integer multiples of $2\pi$)
\begin{align}
& \phantom{+ \ } \int a^\lat \cdot \left( \frac{n}{2\pi} db^\lat - ns_\tau - W^\lat - \frac{q}{2\pi} dA^\lat \right) - \int l_{\tau'} \cp (nb^\lat - qA^\lat) \nonumber \\[.2cm]
& + \frac{k}{4\pi} \int \left( b^\lat \cp db^\lat - b^\lat\cp 2\pi s_\tau - 2\pi s_\tau \cp b^\lat \right) \nonumber \\[.2cm]
& - \int b^\lat \cp V^\lat - \frac{p}{2\pi} \int \left(db^\lat - 2\pi s_\tau\right) \cp A^\lat.
\label{recovered_lattice_action_torsion_only}
\end{align}
Finally, to recover our lattice action \eqref{theory_U1U1}, we perform \eqref{b_a_lat_redef}, and use the fact that when $H^2(\M; \Z)=\mathcal{T}$, any $s^\lat\in Z^2(M; \Z)$ can be written as $s^\lat=s_\tau-dy^\lat$ for some $\tau$, and any $l^\lat\in Z_1(M; \Z)$ can be written as $l^\lat = l_{\tau'}\capp \mathrm{M} -d^\star z^\lat$ for some $\tau'$.

The generalization to fermionic theory with odd $k$ is obvious -- the anomalous term in $\N$ with prefactor $\pi K_{bb}=\pi k$ in \eqref{action_DB_class_multiple} simply becomes \eqref{theory_U1U1_f_4D}.

\subsubsection{On General Topology}

We are ready to consider general topology $H^2(\M; \Z) = \mathcal{F}\oplus\mathcal{T}$. We need to choose a representative gauge configuration for each $\gamma\in H^2(\M; \Z)$. Let's recall from Appendix \ref{app_alg_topo} that any $\gamma \in H^2(\M; \Z)$ can be written as $\gamma = \gamma_\eta+\tau$ for some $\eta\in\mathcal{F}$ and $\tau\in \mathcal{T}$, where $\gamma_\eta$ is a representative so that $[\gamma_\eta]=\eta\in\mathcal{F}$.
\footnote{Recall that $[\gamma_\eta]=\eta\in\mathcal{F}$ through the natural map from $H^2(\M; \Z)$ to $H^2(\M; \R)$ whose image is $\mathcal{F}$, but for fixed $\eta$ there is no canonical choice of $\gamma_\eta$ because the map does not go backwards. On the other hand, by $\tau\in\mathcal{T}$ we actually mean the image of $\mathcal{T}$ under the natural map from $\mathcal{T}$ into  $H^2(\M; \Z)$.}
In turn, we can make the representative $s_\gamma \in Z^2(M;\ Z)$ so that $s_\gamma=s_{\gamma_\eta} + s_\tau$, with $[s_{\gamma_\eta}]=\gamma_\eta, \ [[s_{\gamma_\eta}]]=\eta\in\mathcal{F}$ and $[s_\tau]=\tau, \ [[s_\tau]]=0\in\mathcal{F}$. But there is a crucial distinction between $s_\tau$ and $s_{\gamma_\eta}$. Previously we have seen that the representative gauge configuration $(b_\tau/2\pi, \sigma_\tau, s_\tau)$ for $\tau$ can be conveniently chosen, in compatible with \eqref{a_lambda_relation} and \eqref{lambda_l_relation}, so that $b_\tau=0$ and $\sigma_\tau$ is constant over each $\M_{ab}$. For $\gamma_\eta$, however, this is impossible, for reasons explained near the end of Section \ref{sssect_U1Bundle} \cite{Guadagnini:2014mja}.

\begin{figure}
\centering
\includegraphics[width=.25\textwidth]{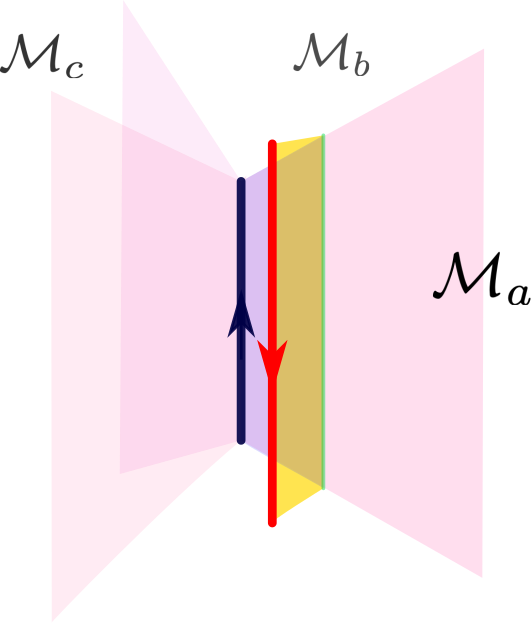}
\caption{Suppose $s_{\gamma_\eta}$ is non-zero (dark blue) along $\M_{abc}$. Assuming $a<b<c$, locally we choose $\d b_{\gamma_\eta}$ to be a narrow flux tube (red) in $\M_a$, in the vicinity of $\M_{abc}$, of strength $-2\pi (s_{\gamma_\eta})_{abc}$. We may choose a gauge so that $b_{\gamma_\eta}$ is a thin distribution on the yellow sheet in $\M_a$. In turn, $\sigma_{\gamma_\eta}$ takes value $(s_{\gamma_\eta})_{abc}$ on the purple subregion of $\M_{ab}$, and $\d\sigma_{\gamma_\eta}$ is a narrow distribution along the green line where the yellow and the purple sheets intersect. (The framed flux tube $\d b_{\gamma_\eta, \, f}$, on the other hand, is chosen to be placed in $\M_c$.)}
\label{f4-2_free_rep}
\end{figure}

For each $\eta$ with fixed choice of $\gamma_\eta$, we can choose the representative gauge configuration $(b_{\gamma_\eta}/2\pi, \sigma_{\gamma_\eta}, s_{\gamma_\eta})$ as the following. The same applies to $(a_{\gamma_{\eta'}}/2\pi, \lambda_{\gamma_{\eta'}}, l_{\gamma_{\eta'}})$.
\begin{itemize}
\item
Obviously $s_{\gamma_\eta}$ satisfies $[s_{\gamma_\eta}]=\gamma_\eta, [[s_{\gamma_\eta}]]=\eta$ as mentioned. Geometrically, we let the Dirac string $(\mathrm{s}_{\gamma_\eta})_{abc}$ which resides on $\M_{abc}$ form a closed loop going around some free non-contractible path specified by $\eta$.
\footnote{Think of the Poincar\'{e} duality, $\eta \in \mathcal{F} \subset H^2(\M; \R) \cong H_1(\M; \R)$.}
\item
The $\d b_{\gamma_\eta}$ flux must be non-zero. We choose the $\d b_{\gamma_\eta}$ flux to be a narrow flux tube residing near $\M_{abc}$ if $s_{\gamma_\eta}$ runs through $\M_{abc}$. In particular, the flux tube and its framed version (which we will use later) are displaced from $\M_{abc}$ in a way similar to \eqref{W_displacement} and \eqref{W_f_displacement}:
\begin{align}
\d (b_{\gamma_\eta})_a = \sum_{b,c \: (a<b<c)} -2\pi (s_{\gamma_\eta})_{abc} \: \delta^2_{\M_{a(bc)}}, \ \ \ \d (b_{\gamma_\eta, \, f})_c = \sum_{a,b \: (a<b<c)} -2\pi (s_{\gamma_\eta})_{abc} \: \delta^2_{\M_{(ab)c}}
\label{gamma_eta_flux}
\end{align}
The negative sign is because the Dirac string is opposite to the flux direction (as is obvious if one imagine the Dirac string as a thin, invisible solenoid).
\item
$b_{\gamma_\eta}$ and $\sigma_{\gamma_\eta}$ can be chosen in compatible with \eqref{a_lambda_relation}, \eqref{lambda_l_relation} and \eqref{gamma_eta_flux}. Consider a slightly displaced image of $\M_{abc}$ into $\M_{ab}$ assuming $a<b<c$; we denote the image as $\M_{ab(c)}$. On $\M_{ab}$, we let $(\sigma_{\gamma_\eta})_{ab}$ take value $(s_{\gamma_\eta})_{abc}$ in the narrow stripe region between $\M_{abc}$ and $\M_{ab(c)}$, and $0$ elsewhere, so that \eqref{lambda_l_relation} is satisfied. This results in a derivative $\d (\sigma_{\gamma_\eta})_{ab}$ that concentrates on $\M_{ab(c)}$. We let $(b_{\gamma_\eta})_a$ concentrate on a narrow two-dimensional stripe region in $\M_a$ between $\M_{a(bc)}$ and $\M_{ab(c)}$, so that \eqref{a_lambda_relation} and \eqref{gamma_eta_flux} are both satisfied. On the other hand, for $b_{\gamma_\eta, \, f}$ and the associated $\sigma_{\gamma_\eta, \, f}$, we displace them into $\M_c$ and $\M_{bc}$ respectively.
\end{itemize}
This completes the specification of the representative gauge configuration for $\gamma_\eta$. The idea is illustrated with Figure \ref{f4-2_free_rep}.

Piecing up the above, the degrees of freedom of the gauge configurations are
\begin{align}
& s_{abc} = (s_\tau)_{abc} + (s_{\gamma_\eta})_{abc}, \ \ \ \ \sigma_{ab} = (\sigma_\tau)_{ab} + (\sigma_{\gamma_\eta})_{ab}, \ \ \ \ b_a = (b_{\mathrm{n.t.}})_a + (b_{\gamma_\eta})_a, \nonumber \\[.2cm]
& l_{abc} = (l_{\tau'})_{abc} + (l_{\gamma_{\eta'}})_{abc}, \ \ \ \ \lambda_{ab} = (\lambda_{\tau'})_{ab} + (\lambda_{\gamma_{\eta'}})_{ab}, \ \ \ \ a_a = (a_{\mathrm{n.t.}})_a + (a_{\gamma_{\eta'}})_a.
\end{align}
Now we substitute these into the action \eqref{action_DB_class_multiple} with the K-matrix \eqref{U1U1_K_matrix}, first assuming the theory is bosonic, i.e. with $k$ even. We organize the terms into three sets: those without the $\gamma_\eta$ or $\gamma_{\eta'}$ representatives, those linear in the representatives, and those quadratic in the representatives. The first set of terms contribute \eqref{cont_action_torision_only} as before. The last set of terms can be ignored because for even $k$, their contribution is a multiple of $2\pi$. It remains to compute the set of terms that are linear in the $\gamma_\eta$ representatives. Due to our restrictions \eqref{W_displacement} and \eqref{b_f_displacement} (in which $b$ and $b_f$ should be understood as $b_{\mathrm{n.t.}}$ and $b_{\mathrm{n.t.}, \, f}$) and the conventions \eqref{gamma_eta_flux}, only some of these terms will ever be invoked.
\footnote{For instance, the term $\int_{\M_c} (b_{\gamma_\eta, \, f})_c \d(b_{\mathrm{n.t.}})_c$ will never make a contribution, because $(b_{\gamma_\eta, \, f})_c$ is only non-vanishing in the vicinity of some $\M_{abc}$ with $a<b<c$, while $\d(b_{\mathrm{n.t.}})_c$ is only non-vanishing in the vicinity of some $\M_{cde}$ with $c<d<e$, but five patches $\M_a, \cdots, \M_e$ cannot be all close to each other in a good patch system decomposing the four-dimensional $\M$. \label{reason_not_contribute}}
The resulting action is
\begin{align}
& \ \ \ \ \mbox{(All terms from \eqref{cont_action_torision_only})} \nonumber \\[.2cm]
& +\frac{n}{2\pi} \left( \sum_a \int_{\M_a} (a_{\mathrm{n.t.}})_a \d (b_{\gamma_\eta})_a + 2\pi \sum_{a<b} \int_{\M_{ab}} (\lambda_{\tau'})_{ab} \d (b_{\gamma_\eta})_b - 2\pi \sum_{a<b<c} \int_{\M_{abc}} (\mathrm{l}_{\gamma_{\eta'}})_{abc} (b_{\mathrm{n.t.}})_c \right. \nonumber \\[.2cm]
& \hspace{2cm} \left. - (2\pi)^2 \sum_{a<b<c<d} \int_{\M_{abcd}} (\mathrm{l}_{\gamma_{\eta'}})_{abc} (\sigma_\tau)_{cd} \right) \ - \ \frac{q}{2\pi}  (-2\pi) \sum_{a<b<c} \int_{\M_{abc}} (\mathrm{l}_{\gamma_{\eta'}})_{abc} A_c \nonumber \\[.2cm]
& + \frac{k}{4\pi} \left( \sum_a \int_{\M_a} (b_{\mathrm{n.t.}, \, f})_a \d (b_{\gamma_\eta})_a + 2\pi \sum_{a<b} \int_{\M_{ab}} (\sigma_\tau)_{ab} \d (b_{\gamma_\eta})_b - 2\pi \sum_{a<b<c} \int_{\M_{abc}} (\mathrm{s}_{\gamma_\eta})_{abc} (b_{\mathrm{n.t.}})_c \right. \nonumber \\[.2cm]
& \hspace{2cm} \left.  - (2\pi)^2 \sum_{a<b<c<d} \int_{\M_{abcd}} (\mathrm{s}_{\gamma_\eta})_{abc} (\sigma_\tau)_{cd} \right) \ - \ \frac{p}{2\pi}  (-2\pi) \sum_{a<b<c} \int_{\M_{abc}} (\mathrm{s}_{\gamma_{\eta'}})_{abc} A_c  \nonumber \\[.2cm]
&+ \mbox{(terms that never contribute under our restrictions for $W, V$ and $dA$)}.
\label{cont_action_general_topo}
\end{align}
Through the procedure similar to how we obtained \eqref{recovered_lattice_action_trivial_topo} and \eqref{lat_action_torision_only} before, now the above reduces to a lattice action on the nerve simplicial complex:
\begin{align}
& \ \ \ \ \mbox{(All terms from \eqref{lat_action_torision_only})} \nonumber \\[.2cm]
& + n \left( -\int a^\lat\cdot s_{\gamma_\eta} - 2\pi \int \lambda^\lat_{\tau'} \cp l_{\gamma_\eta} - \int l_{\gamma_{\eta'}} \cp \left( b^\lat + 2\pi \sigma^\lat_\tau \right) \right) + q \sum_{a<b<c} \int l_{\gamma_{\eta'}} \cp A  \nonumber \\[.2cm]
& +\frac{k}{2} \int \left( - \left(b^\lat + 2\pi \sigma^\lat_\tau \right) \cp s_{\gamma_\eta} - s_{\gamma_\eta} \cp \left(b^\lat + 2\pi \sigma^\lat_\tau \right) \right) + p \sum_{a<b<c} \int s_{\gamma_\eta} \cp A.
\end{align}
Performing the redefinition \eqref{lat_dof_redef} leads to
\begin{align}
& \phantom{+ \ } \int a^\lat \cdot \left( \frac{n}{2\pi} db^\lat - ns_\gamma - W^\lat - \frac{q}{2\pi} dA^\lat \right) - \int l_{\gamma'} \cp (nb^\lat - qA^\lat) \nonumber \\[.2cm]
& + \frac{k}{4\pi} \int \left( b^\lat \cp db^\lat - b^\lat\cp 2\pi s_\gamma - 2\pi s_\gamma \cp b^\lat \right) \nonumber \\[.2cm]
& - \int b^\lat \cp V^\lat - \frac{p}{2\pi} \int \left(db^\lat - 2\pi s_\gamma\right) \cp A^\lat.
\label{recovered_lattice_action_general_topo}
\end{align}
in which all topological degrees of freedom appear in $s_{\gamma}=s_{\gamma_{\eta}} + s_{\tau}$ and $l_{\gamma'}=l_{\gamma_{\eta'}} + l_{\tau'}$. Again, to recover our lattice action \eqref{theory_U1U1}, we perform \eqref{b_a_lat_redef}, and use the fact that any $s^\lat\in Z^2(M; \Z)$ can be written as $s^\lat=s_\gamma-dy^\lat$ for some $\gamma$, and any $l^\lat\in Z_1(M; \Z)$ can be written as $l^\lat = l_{\gamma'}\capp \mathrm{M} -d^\star z^\lat$ for some $\gamma'$.

When $\mathcal{F}$ is non-trivial, there is an extra subtlety in the gauge fixing. We have fixed the choices of $\sigma_\gamma$ and $\lambda_{\gamma'}$ in the representative gauge configurations by exploiting the gauge transformation \eqref{lambda_l_gauge_transf} as well as the $\delta\varphi$ part of \eqref{a_lambda_gauge_transf}. Is the gauge \eqref{lambda_l_gauge_transf} completely fixed, at most up to the ambiguity \eqref{varphi_m_equiv}? Clearly there is an unfixed part classified by $[m] \in H^1(\M; \Z)$, the large gauge transformations. We can choose $m$ in \eqref{lambda_l_gauge_transf} to be non-trivial in $H^1(\M; \Z)\cong\mathcal{F}$, and accompany it with $\varphi$ in \eqref{a_lambda_gauge_transf} so that $\lambda_{\gamma'}$ (and likewise $\sigma_\gamma$) remains unchanged, while $a_{\gamma'}$ (or $b_{\gamma}$) changes by flat $2\pi$ holonomies. In other words, the role of such $m$ configuration is to make the real valued $\varphi$ be effectively identified with $\varphi+2\pi$, which allows the familiar large gauge transformation. Therefore, in \eqref{recovered_lattice_action_general_topo}, before we perform \eqref{b_a_lat_redef}, the large gauge transformations are not fixed. But they are also not removed in the Faddeev-Popov measures for the real valued $b^\lat$ and $a^\lat$, because their $\varphi$'s are understood to be real there. This distinction has been seen in Section \ref{sssect_U1R_partition}, in the distinction between the factor $n^{B_1}$ versus $(\sqrt{n}\: \delta(0))^{2B_1}$. Hence we need to make an extra demand that the large gauge transformations be removed, just as that we needed to impose the summation over the $\gamma$ representatives, and that we needed to specify the global symmetry (discussed below \eqref{recovered_lattice_action_trivial_topo}). On the other hand, once we performed \eqref{b_a_lat_redef}, all these extra prescriptions are automatically implemented, because we would then regard $s^\lat$ and $l^\lat$ as the dynamical degrees of freedom rather than $y^\lat$ and $z^\lat$.

Finally, we shall generalize to fermionic theories with odd $k$. Clearly the anomalous term in $\N$ with prefactor $\pi K_{bb}=\pi k$ in \eqref{action_DB_class_multiple} becomes \eqref{theory_U1U1_f_4D}. But is there extra, undesired contribution? Above \eqref{cont_action_general_topo}, we mentioned that there are terms quadratic in the $\gamma_\eta$ or $\gamma_{\eta'}$ representatives, which only contribute integer multiples of $2\pi$ when $k$ is even, so one might worry there are multiples of $\pi k$ which cannot be dropped when $k$ is odd. A careful examination of these terms, however, shows that all of them in fact belong to the kind that never contribute given our restrictions of $W, V$ and $\d A$ (the last line of \eqref{cont_action_general_topo}), for reasons similar to the example in footnote \ref{reason_not_contribute}. Therefore, for odd $k$, the anomalous term in $\N$ with prefactor $\pi K_{bb}=\pi k$ is the only extra contribution, as desired.

\subsection{Retrieve of Lattice Theory for $U(1)\times \R$ Chern-Simons}
\label{ssect_retrieveU1R}

It is straightforward to carry out the same procedure to retrieve our lattice $U(1)\times \R$ CS introduced in Section \ref{ssect_U1RCS} from the continuum. In our convention, $(b+a)$ is a real gauge field, which does not have any topologically non-trivial configuration. Therefore, we have $\gamma'=-\gamma$, and we can set the representatives so that $(a_\gamma/2\pi, \lambda_\gamma, l_\gamma)=-(b_\gamma/2\pi, \sigma_\gamma, s_\gamma)$. Only the non-topological degrees of freedom $a_{\mathrm{n.t.}}$ and $b_{\mathrm{n.t.}}$ remain different. The $K$-matrix is
\begin{align}
K_{ab}= k, \ \ \ \ \ K_{aA}= -q, \ \ \ \ \ K_{bA}= 0, \ \ \ \ \ K_{aa}=0, \ \ \ \ \ K_{bb}=k, \ \ \ \ \ K_{AA}=0.
\label{U1R_K_matrix}
\end{align}
Moreover, we do not include the $V$ Wilson loop coupled to the $b$ field.

Upon imposing these changes, the lattice action \eqref{recovered_lattice_action_general_topo} that we retrieved from the continuum simply becomes
\begin{align}
& \phantom{+ \ } \int a^\lat \cdot \left( \frac{k}{2\pi} db^\lat - ks_\gamma - W^\lat - \frac{q}{2\pi} dA^\lat \right) + \int s_\gamma \cp (kb^\lat - qA^\lat) \nonumber \\[.2cm]
& + \frac{k}{4\pi} \int \left( b^\lat \cp db^\lat - b^\lat\cp 2\pi s_\gamma - 2\pi s_\gamma \cp b^\lat \right).
\label{recovered_lattice_action_general_topo_U1R}
\end{align}
Instead of \eqref{b_a_lat_redef}, we perform the following redefinition
\begin{align}
& b^\lat_{ab} \in \R \ \ \longrightarrow \ \ b^\lat_{ab} + 2\pi y^\lat_{ab} \ \mbox{ with } \ y^\lat_{ab}\in \Z, \ \ b^\lat_{ab} \in [0, 2\pi), \nonumber \\[.1cm]
& a^\lat_{abc} \in \R \ \ \longrightarrow \ \ a^\lat_{abc} + 2\pi (y^\lat\capp \mathrm{M})_{abc} \ \mbox{ with } \ a^\lat_{abc} \in \R.
\label{b_a_lat_redef_U1R}
\end{align}
so that $b^\lat$ becomes $U(1)$ valued while $a^\lat$ is still $\R$ valued, and we let $s^\lat \equiv s_\gamma-dy^\lat$. If this is a fermionic theory with odd $k$, then again there is the extra anomalous $\pi k$ term in $\mathcal{N}$. This recovers \eqref{theory_U1R} with \eqref{theory_U1U1_f_4D}. Moreover, similar to the discussions for the previous doubled $U(1)$ CS, now after we perform \eqref{b_a_lat_redef_U1R}, the global symmetry and the large gauge transformations of the $U(1)\times \R$ CS are correctly implemented, giving rise the the correct singular factors in \eqref{PF_U1R}.

\section{Gapless Boundary}
\label{sect_boundary}

The doubled abelian CS theory has been extensively studied partly for its admission of topological (gapped) boundary conditions \cite{Kapustin:2010hk, Wang:2012am, Levin:2013gaa, Barkeshli:2013yta, Kapustin:2013nva}. To complete our understanding of the doubled abelian CS theory on lattice, and to provide a consistency check, in this section we study the presence of a spatial boundary. We in particular focus on the gapless boundary condition, which may be protected by the conservation of electric charge on the boundary \cite{Levin:2013gaa}.

A gapless boundary is not topological, so the boundary physics will quantitatively depend on the details of the boundary condition. In the context of lattice gauge theory, different discretizations of the spacetime are not equally convenient for computing boundary properties. For definiteness, in this section we use cubic lattice, whose lattice translational invariance allows concrete calculations to be done in the momentum space. More particularly, we let the lattice vertices take integer coordinates $(x, y, t)$. The system exists for $y\geq 0$ and the spatial boundary is at $y=0$. On the other hand, the $x$- and $t$-directions may be infinite or periodic, which we will specify in the context.

We first study the long range correlation between the gapless boundary excitations \cite{Verlinde:1988sn, Elitzur:1989nr, Wen:1995qn}, featured by the anomalous scaling dimension in its magnitude, and, in its complex phase, a reminiscent of Kac-Moody algebra. Then we turn to the issue of (the vanishing of) chiral central charge. We will explicitly see the presence of two counter-propagating momenta modes on the boundary (or thermal Hall currents, if we switch the notions of $x$ and $t$), each carrying universal zero-point expectations that correspond to chiral central charges $\pm \sgn(k)$ respectively \cite{Verlinde:1988sn, Witten:1988hf, Elitzur:1989nr, Kane:1997fda}, as is expected for a doubled CS theory. Related to this, the system exhibits modular invariance in the $x, t$-directions.

\subsection{Anomalous Scaling Dimension}
\label{ssect_scaling}

The gapless boundary is primarily a local property, which can be largely analyzed in the ``parent'' doubled $\R$ CS theory. The reduction to doubled $U(1)$ CS and $U(1)\times \R$ CS will be discussed later; at that point we will see in what sense the gapless boundary may be protected by the conservation of electric charge on boundary.

We have encountered spacetime boundary before, when we discuss the Hilbert space in Section \ref{sssect_HS_PF_DW}. Each basis state is specified by the fixed values of $b$ on the boundary links, which in our spacetime geometry is the $y=0$ layer of square lattice. (In Section \ref{sssect_HS_PF_DW} there is also $s$ on the boundary plaquettes, but we consider doubled $\R$ CS for now, so there is only $b$.) When viewed as a spatial boundary, such fixed specification of all boundary variables corresponds to a topological, or gapped, boundary condition \cite{Kapustin:2010hk}. The interpretation is straightforward. Consider a $W$ observable insertion that may have ends on the boundary (recall $W$ can be real valued in double $\R$ CS). The constraint equation $(n/2\pi)db = W$ cannot be satisfied on the boundary and hence the expectation vanishes, except for a unique $W$ configuration that precisely matches with our specification of $b$ on the boundary. This is interpreted as having an infinite energy gap for states that are not the ground state required by the specified boundary condition (recall that in a topological theory, or deep IR topological limit of any gapped theory, any energy is either $0$ or infinite).
\footnote{Further including $V$ insertions which end on the boundary does not alter the boundary $b$ degrees of freedom, and at most creates bulk excitations.}

What if we allow the boundary $b$ variables to be dynamical and integrate them out? This would still be a topological boundary condition: the expectation vanishes unless $V+kW/2n$ vanishes on each boundary plaquette. This is most easily seen by noting that $\int b\cp db$ is not gauge invariant on the boundary, and consider the fluctuation of the ``gauge'' degree of freedom -- which is thereby physical on the boundary -- on each boundary vertex. One easy way (particularly easy when $k=0$) to think of why this is another topological boundary condition, is to imagine a fictitious layer of $x$- and $t$-facing plaquettes centered at $y=-1/2$, which now plays the role of the boundary, and we are essentially specifying the fixed value $a=0$ on this boundary. These two types of topological boundary conditions, and the more possibilities in more general doubled abelian CS, are classified by the notion of ``Lagrangian subgroups'' \cite{Kapustin:2010hk, Wang:2012am, Levin:2013gaa} (more precisely, the notion is applicable when the doubled $\R$ is reduced to doubled $U(1)$ as we will do later).

The first step to have a gapless boundary is to let some $b$ fields on the boundary be fixed and some fluctuate; this is required for single CS \cite{Elitzur:1989nr} (which does not have simple realization on the lattice), and an applicable choice for doubled CS.
\footnote{When we reduce the doubled $\R$ theory to doubled $U(1)$ or $U(1)\times \R$, conservation of to the electromagnetic field on boundary may require the gapless boundary condition to be imposed.}
In our spacetime geometry, we make the convenient choice
\begin{align}
b_{\lk^t} = 0 \ \ \mbox{for any $t$-direction link $\lk^t$ on the $y=0$ boundary}
\label{boundary_condition}
\end{align}
which resembles the $b_t=0$ ``temporal'' boundary condition commonly used in the continuum \cite{Elitzur:1989nr}.
\footnote{In applications to fractional quantum Hall effect \cite{Wen:1995qn}, the boundary condition is usually imposed as $b_t-vb_x = 0$ for some ``edge velocity'' $v$ determined by the microscopic details of the physical system. In the continuum theory, this boundary condition is equivalent to $a_t=0$ by a coordinate redefinition $x'=x-vt$. It is sometimes said that the sign of $v$ is fixed by the sign of $k$ \cite{Wen:1995qn}. In fact what really matters is the $i\epsilon$ prescription that we will introduce soon (which plays the role of the Kac-Moody algebra); the sign of $v$ just provides a convenient convention that relates the $i\epsilon$ prescription to the apparent Hamiltonian, as we will explain in footnote \ref{FQH_bc_2}. In this paper we simply use the convention $v=0$. \label{FQH_bc_1}}
The $b_{\lk^x}$ on the $x$-direction boundary link $\lk^x$, on the other hand, is dynamical; this corresponds to the $a_t=0$ boundary condition in the continuum, if we think of a fictitious layer of $t$-facing plaquettes at $y=-1/2$.

The remaining procedure to obtain the gapless boundary theory is essentially the same as in the continuum. Let's set $W, V=0$ in the doubled $\R$ theory \eqref{theory_RR} for now. We first integrate out the $t$-direction bulk variables $a_{\pl^t}$ and $b_{\lk^t}$, where $\pl^t, \lk^t$ is any bulk plaquette or link pointing in the $t$-direction. This leads to $b_{\lk^{x,y}}=(d\chi)_{\lk^{x,y}}$ on all links in the spatial $x, y$-directions, where $\chi$ is a real field on all lattice vertices, including the ones on the $y=0$ boundary. On the other hand, $a_{\pl^{x,y}}=(d^\star\phi)_{\pl^{x,y}}$ on all \emph{bulk} plaquettes facing the spatial directions, where $\phi$ is a real field on all lattice cubes. The $a$ on the boundary plaquettes facing the $y$-direction are unconstrained, however they can be absorbed into $\phi$ on the lattice cubes in the $y=1/2$ layer right beneath the boundary. Substituting these solutions back into the remaining action, we find most $\chi$ and $\phi$ degrees of freedom drop out, except for the $\chi$ on the $y=0$ boundary and $\phi$ on the $y=1/2$ cube layer beneath the boundary:
\begin{align}
S_{bd} \, = \, \sum_{\mathrm{boundary \: vert.} \: \vt} \left(\frac{k}{4\pi} \chi_\vt + \frac{n}{2\pi}\phi_\vt \right) \left( -\chi_{\vt+\hat{x}+\hat{t}} + \chi_{\vt+\hat{t}} + \chi_{\vt+\hat{x}} - \chi_{\vt}  \right)
\label{boundary_action}
\end{align}
where we have relabeled the $\phi$ field on the cube centered at $\cb=\vt+\hat{x}/2+\hat{y}/2+\hat{t}/2$ by $\phi_\vt$. The $\phi_\vt$ field is a Lagrange multiplier imposing the classical chiral equation of motion $-\chi_{\vt+\hat{x}+\hat{t}} + \chi_{\vt+\hat{t}} + \chi_{\vt+\hat{x}} - \chi_{\vt} = 0$, which clearly resembles $\partial_t \partial_x \phi = 0$ in the continuum. The variation of $\chi_\vt$ provides another equation of motion of opposite chirality. The overall boundary theory being non-chiral is why it can be presented within a finite depth into the $y$-direction (in our case, $y=0$ and $y=1/2$ only), making it essentially an intrinsically $2D$ system.

We have obtained the lattice realization of the classical equations of motion on the gapless boundary. An additional prescription is needed to quantize the theory, because the boundary theory is no longer topological and we need to make proper sense of ``ground state expectation value''. In the continuum, this is implemented by the $i\epsilon$ prescription terms $iS_{bd} \, \rightarrow \, iS_{bd} - \epsilon \int dt\, dx\, (\partial_x \phi)^2 - \epsilon' \int dt\, dx\, (\partial_x \chi)^2$.
\footnote{If one uses canonical quantization, the role of the $i\epsilon$ prescription is played by the identification of creation and annihilation modes of $\phi$ and $\chi$ and the subsequent normal ordering of composite operators such as $e^{\pm iw\phi}$.}
On the lattice we implement the same:
\begin{align}
iS_{bd} \ \ \longrightarrow \ \ iS_{bd} \ - \!\!\! \sum_{\mathrm{boundary \: vert.} \: \vt} \left(\epsilon \left( \phi_\vt - \phi_{\vt-\hat{x}} \right)^2 + \epsilon' \left( \chi_\vt - \chi_{\vt-\hat{x}} \right)^2 \right)
\label{iep_prescription_phi_chi}
\end{align}
(Although in \eqref{iep_prescription_phi_chi} we have taken the discretized second derivative at the lattice scale, it can be equally well taken at some larger scale smoothly, as we will see in Appendix \ref{app_scaling}. Moreover, we may also equally well add some term mixing the lattice $x$-derivatives of $\phi$ and $\chi$, as long as the total quadratic form stays positive definite; for simplicity we will not do so.) Tracing the origin of the $\phi$ variable, one may recognize $\left( \phi_\vt - \phi_{\vt-\hat{x}} \right)^2$ as a Maxwell term $\left( d^\star a \right)_{\lk^t}^2$ for the $d^\star a$ magnetic field on the boundary $t$-direction links. Similarly, $\left( \chi_\vt - \chi_{\vt-\hat{x}} \right)^2$ may be recognized as a Maxwell term for the $db$ magnetic field on the $t$-facing plaquette at the fictitious  $y=-1/2$ layer. The prescription can be written in the original $a$, $b$ variables as
\begin{align}
iS_{bd} \ \ \longrightarrow \ \ iS_{bd} \ - \!\!\! \sum_{\mathrm{boundary \: \lk^t}} \epsilon \left( d^\star a \right)_{\lk^t}^2 - \sum_{\mathrm{boundary \: \lk^x}} \epsilon' b_{\lk^x}^2.
\label{iep_prescription_a_b}
\end{align}
We should recognize these Maxwell terms as an infinitesimal Hamiltonian that projects the system to the ground state as $t\rightarrow \pm\infty$.
\footnote{As mentioned in footnote \ref{FQH_bc_1}, in applications to fractional quantum Hall effect \cite{Wen:1995qn}, the boundary condition is usually imposed as $b_t-vb_x = 0$. The boundary action then involves a term $-(kv/4\pi) \int dt\, dx (\partial_x \chi)^2$, which can be thought of as an apparent ``Hamiltonian'' term. This apparent ``Hamiltonian'' term matches with the Hamiltonian term used in the $i\epsilon$ prescription if $\sgn(v)=\sgn(k)$. This convenient matching is what is really meant by ``the sign of $v$ is fixed by $k$''. There is also a mixed term of $\partial_x \phi$ and $\partial_x \chi$, which is equivalent to the $(\partial_x \phi)^2$ term in the $i\epsilon$ prescription (in the sense of changing the details of the quadratic form but keeping it positive definite). \label{FQH_bc_2}}
While the infinitesimal Maxwell terms are negligible in the bulk,
\footnote{This statement is true for doubled CS. For single CS, the Maxwell term, though infinitesimal, is important in the bulk to regularize the theory \cite{Witten:1988hf, BarNatan:1991rn}, as there is no Lagrange multiplier; one of its consequences is the regularization in the definition of the $\eta$-invariant which leads to well-defined chiral central charge. In the lattice gauge theory perspective, a small Maxwell term is needed to remove the doubling problem we mentioned in Section \ref{ssect_RRCS} \cite{Berruto:2000dp}, making the single CS theory not exactly soluble.}
they play an important role on the gapless boundary (as will become manifest soon), due to the boundary's lack of gauge invariance and gapless nature. Why the suppression of the $db$ and $d^\star a$ magnetic fields in particular? In the bulk, these magnetic fields are constrained to zero by the equations of motion of the temporal components of $a$ and $b$. While on the gapless boundary these temporal components are set to zero (see discussion below the boundary condition \eqref{boundary_condition}), the reminiscent of their equations of motion should still specify the ground state in the infinite past and infinite future.

\begin{figure}
\centering
\includegraphics[width=.9\textwidth]{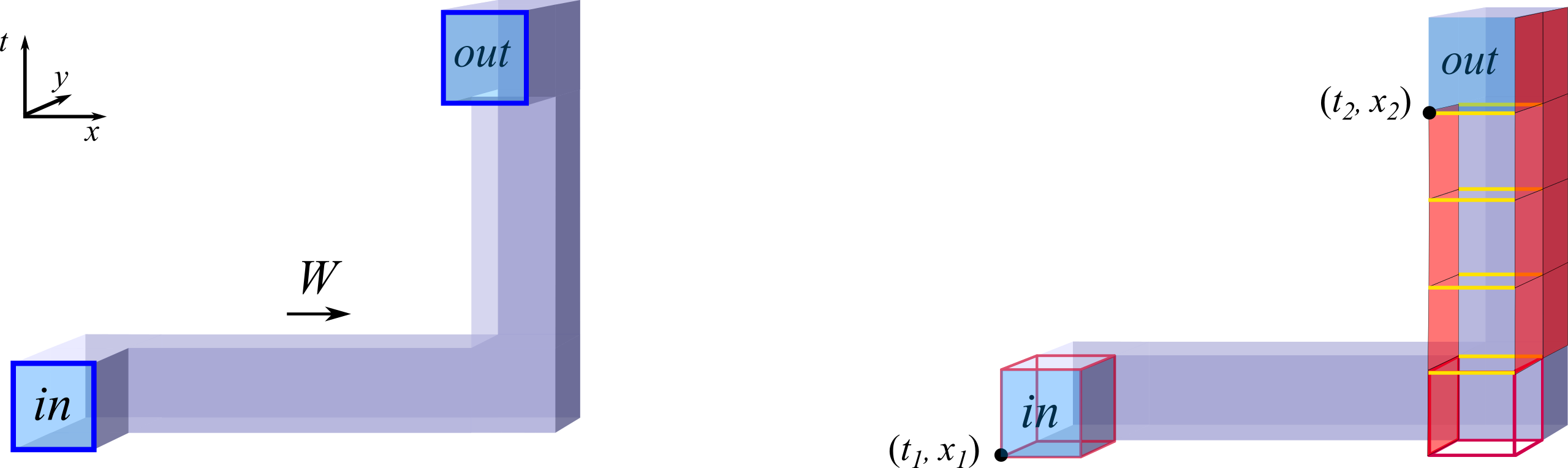}
\vspace{.3cm}
\caption{On the left is an example of a Wilson line starting and ending on the $y=0$ boundary. Our standard choice of $W$ in the bulk is that, it first penetrates into the ``deep enough'' $y=1+1/2$ cube layer, then runs in the $x$-direction and then in the $t$-direction, and finally comes out of the boundary.
\\[.05cm]
On the right we illustrate the intermediate steps in the derivation of \eqref{boundary_action_W}. Integrating out $a_{\pl^t}$ in the bulk, we find the yellow spatial links have $b_{\lk^x}=(2\pi w/n)\, \sgn(t_2-t_1) +(d\chi)_{\lk^x}$, as opposed to just $(d\chi)_{\lk^{x,y}}$ on other spatial links. Next, integrating out $b_{\lk^t}$ in the bulk, we find the red spatial plaquettes have $a_{\pl^x}= -(\pi w k /n^2) \, \sgn(t_2-t_1) +(d^\star\phi)_{\pl^x}$, as opposed to just $(d^\star \phi)_{\pl^{x,y}}$ on other spatially oriented plaquettes. Substituting these solutions into the remaining action, after some cancellations, the only terms in which the charge $w$ appears are its couplings to the $\phi$ at the two cubes indicated by red edges. This explains the insertions in \eqref{boundary_action_W} (after redefining $\phi_\cb$ as $\phi_{\vt=\cb-\hat{x}/2-\hat{y}/2-\hat{z}/2}$).
\\[.05cm]
Those yellow links and red plaquettes touching the boundary also make $w$ dependent contributions to \eqref{iep_prescription_a_b}, leading to the last term of \eqref{boundary_action_W}. This term is negligible as we will justify at the end of Appendix \ref{app_scaling}.
\\[.05cm]
Looking at the red plaquettes, we can explain why the $y=1+1/2$ layer is considered ``deep enough'' for the bulk placement of $W$, but not $y=1/2$. As long as the bulk placement is at $y>1/2$, there will be two columns of red plaquettes touching the boundary; however, if the placement is at $y=1/2$, there will be only the right column of red plaquettes. The missing of half of the red plaquettes will affect the boundary momenta operators \eqref{momenta_a_b} defined in Section \ref{ssect_modular}.}
\label{f5-1_corr}
\end{figure}

Our primary goal is to compute the boundary anyon correlation function and reproduce the anomalous scaling. For simplicity, let's consider the $W$ anyons only and assume $k\neq 0$.
\footnote{For $k=0$, the gapless boundary condition may be protected for another reason: the symmetry of exchanging the $a$ and $b$ fields, which is not on-site \cite{Gaiotto:2015zta}. More explicitly, our gapless boundary condition corresponds to $a_t=0, b_t=0$ in the continuum, which is indeed symmetric between $a$ and $b$.}
From the continuum theory we expect $|\langle e^{iw\phi(t,x)} \: e^{-iw\phi(0,0)} \rangle| \sim |x|^{-|k|w^2/n^2}$ regardless of their time separation. The physical process we have in mind is a pair of charge $\pm w$ anyons being created in the bulk and brought to two different plaquettes on the boundary. Therefore, we should insert a Wilson line $W$ starting and ending on two different boundary plaquettes. (Recall its charge $w$ does not have to be integer in doubled $\R$ CS for now, but we may still think of integer $w$ to make easy connection to doubled $U(1)$ CS later.) We do not care much about how the Wilson line extends in the bulk, because changing its placement in the bulk only changes the expectation value by a complex phase as we know from Section \ref{ssect_RRCS}. However, to interpret $W$ as the physical process described above, we \emph{do} require $W$ to penetrate beneath the boundary ``deep enough''. In our construction, $W$ should \emph{at least} penetrate to the second cube layer (whose $y$ coordinate is centered at $1+1/2$) before running in the $x, t$-directions; see Figure \ref{f5-1_corr} for an illustration. The reason will become clearer in Section \ref{ssect_modular}; roughly speaking, we have viewed the $y=0$ and $y=1/2$ layers as the boundary system, so we want the bulk placement of $W$ to be deeper than that.

For definiteness, let's therefore assume $W$ runs in the bulk on the $y=1+1/2$ cube layer, first running in the $x$-direction and then in the $t$-direction; see Figure \ref{f5-1_corr}. We carry out the same steps that lead to \eqref{boundary_action}, but with the presence of $W$; intermediate steps are explained by Figure \ref{f5-1_corr}. The resulting exponent $iS_{bd} - (\epsilon , \epsilon' \,\mathrm{ terms})$ is
\begin{align}
& \sum_{\mathrm{boundary \: vert.} \: \vt} \left( i\left(\frac{k}{4\pi} \chi_\vt + \frac{n}{2\pi}\phi_\vt \right) \left( -\chi_{\vt+\hat{x}+\hat{t}} + \chi_{\vt+\hat{t}} + \chi_{\vt+\hat{x}} - \chi_{\vt}  \right) \right. \nonumber \\[.2cm]
& \hspace{5cm} \left. - \epsilon \left( \phi_\vt - \phi_{\vt-\hat{x}} \right)^2 - \epsilon' \left( \chi_\vt - \chi_{\vt-\hat{x}} \right)^2 \phantom{\frac{1}{1}} \!\!\! \right) \nonumber \\[.2cm]
& + \ \: i w \left( \phi_{(t_1, x_2)} - \phi_{(t_1, x_1)} \right) \ + \ \mathcal{O}(\epsilon\, |t_2-t_1| \phi, \: \epsilon' \, |t_2-t_1| \chi)
\label{boundary_action_W}
\end{align}
where $(t_1, x_1)$ and $(t_2, x_2)$ are the coordinates of (the lower-left corners of) the starting and ending plaquettes, respectively. Note that in \eqref{boundary_action_W}, both $\phi$ insertions turn out to have the same time coordinate $t_1$, regardless of $t_2$, so they are only separated in space. The expectation value is thus independent of the time separation of the two boundary anyons, as is expected from the continuum theory. The last term, whose origin is also explained in Figure \ref{f5-1_corr}, is negligible as long as $\epsilon, \epsilon' \ll |t_2-t_1|$. We will address this at the end of Appendix \ref{app_scaling}.

The partition function weighted by \eqref{boundary_action_W} can be computed in the momentum space (which is why we choose cubic lattice). We assume the $x, t$-directions are infinite; the result will not change substantially if we make them periodic, as long as the periods are much larger than the separation. We Fourier transform
\begin{align}
\phi_{(t, x)} = \int_{-\pi}^\pi \frac{dp}{2\pi} \int_{-\pi}^\pi \frac{d\omega}{2\pi} \ e^{-i\omega t + ipx} \ \phi_{(\omega, p)}, \ \ \ \ \ \ \ \phi_{(\omega, p)} = \sum_{x\in \Z} \sum_{t\in \Z} \ e^{i\omega t - ipx} \ \phi_{(t, x)}
\end{align}
and likewise for $\chi$. The Gaussian exponent becomes
\begin{align}
& \int_{-\pi}^\pi \frac{dp}{2\pi} \int_{-\pi}^\pi \frac{d\omega}{2\pi} \nonumber \\[.2cm]
& \left(
-\frac{1}{2} \ \left[ \begin{array}{c} \chi \\ \phi \end{array} \right]^T_{(-\omega, -p)} \ M(\omega, p) \ \left[ \begin{array}{c} \chi \\ \phi \end{array} \right]_{(\omega, p)} + \ \left[ \begin{array}{c} 0 \\ iw\, e^{-i\omega t_1}\left(e^{ipx_2} - e^{ipx_1}\right) \end{array} \right]^T \: \left[ \begin{array}{c} \chi \\ \phi \end{array} \right]_{(\omega, p)} \right)
\label{boundary_momentum_space}
\end{align}
where $M(\omega, p)$ is given by
\begin{align}
\left[ \begin{array}{cc} 
i\frac{k}{4\pi} \left( \left( e^{-i\omega} - 1 \right) \left( e^{ip} - 1 \right) + c.c. \right) + 2\epsilon' \left( e^{ip} - 1\right) \left(e^{-ip}-1\right) & i\frac{n}{2\pi}\left( e^{i\omega} - 1 \right) \left( e^{-ip} - 1 \right) \\[.2cm]
i\frac{n}{2\pi}\left( e^{-i\omega} - 1 \right) \left( e^{ip} - 1 \right) &  2\epsilon \left( e^{ip} - 1\right) \left(e^{-ip}-1\right) \end{array} \right].
\label{boundary_momentum_space_matrix}
\end{align}
We perform the Gaussian integral to obtain the expectation value. The computation is presented in Appendix \ref{app_scaling}. The resulting expectation value is
\begin{align}
\frac{Z[W]}{Z[0]} = e^{I(x_2-x_1)}, \ \ \ \ I(x) = -\frac{|k|\, w^2}{n^2} \left( \sum_{m=1}^{|x|-1} \frac{1-(-1)^m}{m} + \frac{1-(-1)^{|x|}}{2|x|} \right) - \frac{i\pi k w^2}{2n^2}
\label{anomalous_scaling_result}
\end{align}
(and $I(0)=0$ is understood). For $|x| \gg 1$, the real part of $I(x)$ approaches
\begin{align}
\Re\, I(x) \ \longrightarrow \ -\frac{|k|\, w^2}{n^2} \left(\ln |x| + \gamma_{\phantom{}_{\mathrm{EM}}} + \ln 2 + \mathcal{O}(1/|x|) \right)
\end{align}
where $\gamma_{\phantom{}_{\mathrm{EM}}}$ is the Euler-Mascheroni constant. Therefore, the expectation value approaches
\begin{align}
\frac{Z[W]}{Z[0]} \ \longrightarrow \ \ e^{-i\pi k w^2\, /2n^2} \ \left(2e^{\gamma_{\phantom{}_{\mathrm{EM}}}} \ |x_2-x_1| \right)^{-|k|\, w^2 /n^2}
\end{align}
with an anomalous scaling dimension $-|k|\, w^2 /n^2$ as desired. As we will show later, the expression stays the same when we reduce the doubled $\R$ CS to doubled $U(1)$ CS, and by then, $n, k, w$ must be integers.

It is also worth noting that the complex phase in \eqref{anomalous_scaling_result} is $1/2 \mod \Z$ in units of the anyon statistics $\pi k\, w^2 /n^2$. Hence the phase cannot be removed no matter how we change the bulk placement of $W$, which only changes the phase in units of the anyon statistics. This result is related to the Kac-Moody algebra \cite{Elitzur:1989nr,Wen:1995qn}, the commutation relations of the boundary fields, even though our quantization procedure does not seem to have directly employed it. In the continuum theory, if we compute the Gaussian path integral carefully, the precise dependence on the spacetime separation should be
\begin{align}
& \left\langle \: :\! e^{iw\phi(t,x)} \!\! : \ : \! e^{-iw\phi(0,0)}\!\! : \: \right\rangle \ = \ \exp\left(w^2 \left\langle \phi(t,x) \, \phi(0,0) \right\rangle\right) \nonumber \\[.2cm]
=& \ (\mbox{real const.}) \ \left( x+i\epsilon\, \sgn(k) \, t \right)^{-|k|\, w^2/2n^2}\left(-x-i\epsilon\, \sgn(k) \, t \right)^{-|k|\, w^2/2n^2}
\label{continuum_boundary_details}
\end{align}
where the branch cut of the fractional power is placed along the negative real axis. The correlation is invariant under the simultaneous flips of $x, t$ and $w$, as it should. The chosen branch cut is so that $\Im \langle \phi(t,x)\, \phi(0,0)\rangle = (\pi k/2n^2)\, \sgn(xt)$, consistent with the Kac-Moody commutation relation $[\phi(x), \phi(0)]=(i\pi k/n^2)\, \sgn(x)$. As a result of this consistency with Kac-Moody algebra, the complex phase of \eqref{continuum_boundary_details} is $e^{\pm i\pi k w^2/2n^2}$ for $\sgn(xt)=\pm 1$, which, in either case, is half in size of the anyon statistics $\pi k\, w^2 /n^2$. We can make the phase of the lattice result \eqref{anomalous_scaling_result} the same as the continuum result if we place $W$ in the bulk differently for $\sgn(xt)=\pm 1$; the crucial element we already have, however, is the half unit part. The detailed placement prescription is obviously not unique, but one may check that such extra geometrical prescription just describes the familiar idea that the bulk anyon statistics gives rise to the boundary Kac-Moody algebra.

We have computed the boundary correlation of the $W$ observable in the lattice doubled $\R$ theory, and reproduced the anomalous scaling and complex phase as expected. The same can be done for the $V$ observable through similar calculations, which we will not repeat.

Now we discuss the reduction of the doubled $\R$ CS to doubled $U(1)$ CS with gapless boundary condition. One may verify that the boundary condition \eqref{boundary_condition} is sufficient to ensure the gauge invariances \eqref{s_gauge_transf}\eqref{l_gauge_transf}\eqref{b_gauge_transf}\eqref{a_gauge_transf} in the bulk to hold on the boundary. It would then be natural to use the $1$-form $\Z$ gauge invariances to reduce the $a$ and $b$ fields to $U(1)$ variables, as we did when there is no boundary. One would obtain the theory \eqref{theory_U1U1_f} with the boundary condition \eqref{boundary_condition}, and the $i\epsilon$ prescription \eqref{iep_prescription_a_b} with $d^\star a \rightarrow d^\star a - 2\pi l$. However, this theory would be consistent only if the electric charge $p=0$. When $p=0$, the background $U(1)$ gauge invariance is respected on each lattice vertex including the boundary ones; in particular, it is not violated by the $q\int l\cdot A$ term because $l$ is conserved even on the boundary vertices. 

When $p\neq 0$, we need to be more careful with the boundary condition. While the $l$ Dirac string is conserved on each lattice vertex and forms closed loops within the system, the $s$ Dirac string, being conserved on each lattice cube, might end on the $y=0$ boundary plaquettes. This causes problem when $s$ is electrically charged. First, mathematically, the $p\int s\cp A$ term in \eqref{theory_U1U1_f} violates the background $U(1)$ gauge invariance when $s$ has ends on the boundary. Second, conceptually, suppose the $x, t$-directions are periodic, and in the $y$-direction there is another boundary at $y=Y\gg 1$. Consider a $W$ Wilson line insertion of charge $w=n$ that runs from a plaquette on $y=0$ to $y=Y$. Since such $W$ can be absorbed by the dynamical $s$ Dirac string, the expectation value is non-vanishing. However, this process corresponds to taking the overlap between the ground state and a state with one boson / fermion (for even / odd $k$ respectively) brought from one edge to the other. The boson / fermion carries electric charge $p$. If we assume the ground state to conserve electric charge on the boundary, the overlap should vanish. Therefore, these two problems show the conservation of the electric charge on the boundary (assuming $p\neq 0$) is incompatible with the naive gapless boundary condition.

One may certainly use the two possible gapped boundary conditions we introduced before. But in either case, for generic values of $p, q$, the electric field is screened on the boundary (for instance, if we use the $b=0$ boundary condition on $y=0$, then the expectation value would be non-vanishing only if $q\, dA=0$ on the boundary plaquettes), and thus corresponds to a superconducting boundary \cite{Levin:2013gaa}. How to write down a boundary condition that is gapless and respects the electromagnetic $U(1)$ for generic values of $p, q$? The key is to constrain the $s$ loops to stay within the system. We find the following boundary condition can be used: In addition to \eqref{boundary_condition}, we impose
\begin{align}
& s_{\pl^y} = 0 \ \ \mbox{for any $y$-facing plaquette $\pl^y$ on the $y=0$ boundary,} \nonumber \\[.2cm]
& b_{\lk^x} \in \R \ \ \mbox{for any $x$-direction link $\lk^x$ on the $y=0$ boundary.}
\label{boundary_condition_U1}
\end{align}
In the bulk $y>0$ we just reduce the $b$ field to $U(1)$ as usual. Moreover, the $a$ field is reduced to $U(1)$ everywhere as usual. The $i\epsilon$ prescription \eqref{iep_prescription_a_b} is still understood with $d^\star a \rightarrow d^\star a - 2\pi l$.

The computation of expectation values involving boundary insertions is straightforward. Exploiting the $1$-form $\Z$ gauge invariances, we may replace $d^\star a - 2\pi l$ with $d^\star a' - 2\pi l_{\gamma'}$ and $db-2\pi s$ with $db'-2\pi s_\gamma$ for real valued $a', b'$ (and on the $\lk^x$ links on the boundary, $b'=b$), as we did above \eqref{PF_U1U1_alt}. The topological representatives $s_\gamma$ and $l_{\gamma'}$ can be neglected: They cannot thread through the open $y$-direction since they must be conserved within the system; while if they thread through the $x, t$-directions (assuming periodic), we may place them at $y\rightarrow \infty$ anyways. Thus, the theory becomes locally the same as the doubled $\R$ CS. The computation of boundary expectation values is therefore identical to that of a doubled $\R$ CS, except the real valued $W, V$ are now replaced with $W+q\, dA/2\pi$ and $V+p\, dA/2\pi$ in which $W, V$ are integer valued. Associated electromagnetic phenomena are manifest, such as that an electric field parallel to the boundary accumulates electric charges from the bulk onto the boundary, due to the Hall conductivity.

In other geometries we may see other topological properties of the doubled $U(1)$ CS. For instance consider a spacetime with periodic $t$-direction and finite $x, y$-directions, so that it forms a solid torus, with a ``circular'' spatial boundary. Then we may consider the bulk having a $W$ loop of charge $w$ threading around the periodic $t$-direction. As a result, one finds the boundary solution of $b$ must have a $2\pi w \in 2\pi\Z$ holonomy around the ``circular'' spatial boundary, manifesting the relation between bulk charge insertion and boundary mode excitation \cite{Elitzur:1989nr, Wen:1995qn}.

The boundary conditions \eqref{boundary_condition} along with \eqref{boundary_condition_U1} are also applicable to the $U(1)\times \R$ CS we introduced (and $a$ is real valued everywhere). Restricting to the $U(1)$ sector observables, i.e. $p=0$ and $V=0$, the expectation values are indeed identical to those in a single $U(1)$ CS theory.

\subsection{Modular Invariance and Counter-Propagating Momenta}
\label{ssect_modular}

The realizability of the doubled CS theory as an exactly soluble lattice model is based on its non-chiral nature. The gapless boundary theory \eqref{boundary_action} can be presented as an intrinsically $2D$ lattice model, with two chiral equations of motion of opposite chiralities. In the below, we will first demonstrate the modular-$T$ invariance when the $x, t$-directions are periodic, a property of non-chiral theories \cite{Verlinde:1988sn}. Then, we will derive the expression of the boundary momentum operator as the generator of the modular transformation. We show it has two counter-propagating modes $P=P_+ + P_-$, which, under the gapless boundary condition, take the universal zero-point expectation values
\begin{align}
\langle P_\pm \rangle = \mp\frac{\pi\, \sgn(k)}{12\, C},
\label{chiral_central_charges}
\end{align}
which correspond to chiral central charge of $\pm\sgn(k)$ respectively; here $C$ is the circumference of the periodic $x$-direction. If we swap the $x$- and $t$-directions, the momentum becomes the energy current, and the circumference $C$ becomes the inverse temperature $\beta=1/T$. The physical phenomenon is then known as the thermal Hall conductivity \cite{Kane:1997fda, Kitaev:2006lla}
\begin{align}
\beta \langle j^E_\pm \rangle = \mp\frac{\pi\, \sgn(k)}{12\, \beta} \ \ \ \ \ \ \ \mbox{or} \ \ \ \ \ \ \partial_T \langle j^E_\pm \rangle = \mp \frac{\pi\, \sgn(k) \, T}{6}
\end{align}
where $j^E_\pm$ are the modes of energy current density. The counter-propagating gapless modes are protected not to mix and gap out if we impose the electromagnetic $U(1)$ invariance on the boundary.

Consider the $x$- and $t$-directions being periodic, with periods $C$ and $\beta$ respectively. Now we consider twisting this torus. We identify the points
\begin{align}
(t, x)\sim (t, x+C) \sim (t+\beta, x+\delta).
\label{twisting}
\end{align}
When $\delta=0$, this is the original untwisted torus. $\delta=1$ is the generator of the twisting \cite{Tu:2012nj} which we will use the derive the boundary momentum operator in the below. For now we consider $\delta=C$, which corresponds to a modular-$T$ transformation. For simplicity we let the lattice be so that $\beta=C$. The modular-$T$ transformation with $\delta=C$ is \emph{not} identical to $\delta=0$. Instead, it is equivalent to $\delta=0$ but with a change of the notion of $\cp$, so that the $\hat{x}/2+\hat{y}/2+\hat{t}/2$ in Figure \ref{f2-1_bdb} is replaced by $-\hat{x}/2+\hat{y}/2+\hat{t}/2$. Alternatively, $\delta=C$ is equivalent to $\delta=0$, keeping the notion of $\cp$ unchanged, but changing the boundary condition \eqref{boundary_condition} to
\begin{align}
b_{\lk^t}= -b_{{\lk'}^x}, \ \ \ \ \mbox{where} \ \ \lk^t - {\lk'}^x =  \hat{x}/2+\hat{t}/2
\label{boundary_condition_alt}
\end{align}
and then redefining $x-t$ as $x$. In either interpretation, the boundary theory \eqref{boundary_action} becomes
\begin{align}
S'_{bd} = \sum_{\mathrm{boundary \: vert.} \: \vt} \left(\frac{k}{4\pi}\chi_\vt+\frac{n}{2\pi}\phi_\vt\right)\left(-\chi_{\vt+\hat{t}}+\chi_{\vt+\hat{t}-\hat{x}}+\chi_{\vt}-\chi_{\vt-\hat{x}}\right).
\end{align}
The expectation of modular-$T$ transformation is given by the ratio between the partition function using $S'_{bd}$ versus the original one using $S_{bd}$. In the momentum space, this is to compute 
\begin{align}
\left\langle \ \mbox{modular-}T \ \right\rangle = \prod_{0<p<\pi} \: \prod_{\omega \neq 0} \: \frac{\det M(\omega, p)}{\det M(\omega, -p)}
\end{align}
where $M(\omega, p)$ is the matrix \eqref{boundary_momentum_space_matrix}, and $p, \omega$ are multiples of $2\pi/C$ (we set $\beta=C$ for convenience). It is easy to see
\begin{align}
\frac{\det M(\omega, p)}{\det M(-\omega+p, -p)} = 1 + \epsilon (\mbox{bounded quantity}).
\end{align}
Therefore, by shuffling the $\omega$ product, the expectation is $1$ as $\epsilon \rightarrow 0^+$, i.e. the theory is modular-$T$ invariant as it must be.

Now we perform a small twisting $\delta=1$ \cite{Tu:2012nj} to extract the boundary momentum operator via a lattice analogue of the usual Noether procedure, and demonstrate its two counter-propagating modes have the universal zero-point expectation values. When $\delta=0$ in \eqref{twisting}, the momentum $p$ is a multiple of $2\pi/C$, the frequency $\omega$ is a multiple of $2\pi/\beta$, so that the Fourier transformation by $e^{-i\omega t+ipx}$ is unambiguous. After twisting by $\delta=1$ in \eqref{twisting}, while $p$ is still a multiple of $2\pi/C$, the frequency is so that $\omega\beta-p\delta \in 2\pi\Z$. Let's redefine $\omega$ so that it takes the original values in multiples of $2\pi/\beta$, but then $e^{-i\omega t+ipx}$ is shifted to $e^{-i(\omega + p/\beta) t+ipx}$ in the Fourier transformation. When $|p|\ll\pi$, the shift by $p/\beta$ is small compared to $2\pi/\beta$, the step size of $\omega$. Linearizing the shift of the Guassian weight \eqref{boundary_momentum_space}\eqref{boundary_momentum_space_matrix} in $p/\beta$ for small $p$, we find
\begin{align}
&\frac{1}{C}\sum_{p \in (2\pi/ C) \Z_C} \: \frac{1}{\beta}\sum_{\omega\in(2\pi/\beta)\Z_\beta} \nonumber \\[.2cm]
&\left(-\frac{1}{2}\right) \ \left[ \begin{array}{c} \chi \\ \phi \end{array} \right]^T_{(-\omega, -p)} \ 
\left[ \begin{array}{cc} 
i\frac{k}{4\pi} \left( \left(-i\frac{p}{\beta} \right) \left( e^{ip} - 1 \right) + c.c. \right)  & i\frac{n}{2\pi}\left( i\frac{p}{\beta} \right) \left( e^{-ip} - 1 \right) \\[.2cm]
i\frac{n}{2\pi}\left( -i\frac{p}{\beta} \right) \left( e^{ip} - 1 \right) & 0 \end{array} \right]
\ \left[ \begin{array}{c} \chi \\ \phi \end{array} \right]_{(\omega, p)}.
\end{align}
Following the Noether procedure in the continuum, we may identify this shift as an insertion of the operator $e^{-i P(t) (\delta/\beta)}$ at some arbitrary time $t$ on the \emph{untwisted} torus, where $P$ is the generator of spatial translation on the boundary, i.e. the boundary momentum operator. This expression of $P$ is only valid for small $p$. We may UV complete the definition of $P$ by promoting $-ip$ to $\left(e^{-ip}-1\right) f(p)$ where the regularization function $f(p)$ is a slowing varying, even function, that approaches $1$ for small $|p|\ll \pi$, but vanishes fast enough for large $|p|\sim \pi$ (we will comment more on this point). As we will see soon, the details of $f(p)$ does not matter.

In summary, through the lattice analogue of the Noether procedure, we find the boundary momentum operator
\begin{align}
P(t) \ = \ \frac{1}{C}\sum_{p \in (2\pi/ C) \Z_C}
\frac{f(p) \ (e^{-ip}-1)(e^{ip}-1)}{4\pi} \ \left[ \begin{array}{c} \chi \\ \phi \end{array} \right]^T_{(t, -p)} \ 
\left[ \begin{array}{cc} 
k  & n \\[.2cm]
n & 0 \end{array} \right]
\ \left[ \begin{array}{c} \chi \\ \phi \end{array} \right]_{(t, p)}.
\end{align}
The small $p$ expansion, $p^2 f(p)/4\pi$, agrees with the boundary momentum operator in the continuum, with $f(p)$ providing a smooth UV cutoff just as in the continuum. For definiteness we may let $f(p)$ be a Gaussian regularization
\begin{align}
f(p) \: \sim \: \exp(-(R\, p)^2/2)
\end{align}
for some smearing range $1\ll R\ll C$. One may diagonalize the matrix and write $P=P_+ + P_-$ where
\begin{align}
P_+(t) &= \frac{1}{C}\sum_{p \in (2\pi/ C) \Z_C} \frac{f(p) \ (e^{-ip}-1)(e^{ip}-1)}{4\pi} \ \frac{1}{k} \ \left(k\chi + n\phi\right)_{(t, -p)} \left(k\chi + n\phi\right)_{(t, p)}, \nonumber \\[.2cm]
P_-(t) &= \frac{1}{C}\sum_{p \in (2\pi/ C) \Z_C} \frac{f(p) \ (e^{-ip}-1)(e^{ip}-1)}{4\pi} \ \frac{-1}{k} \ n\phi_{(t, -p)} \: n\phi_{(t, p)},
\label{momenta}
\end{align}
which are, respectively, positive and negative (negative and positive) definite when $\sgn(k)>0$ ($\sgn(k)<0$). These are the two counter-propagating momenta operators on the spatial boundary.

We may Fourier transform their expressions back in the coordinate space and use the original $a, b$ variables instead of the $\phi, \chi$ variables. We find
\begin{align}
P_+(t) \ &= \ \sum_{x=1}^C \sum_{r=-\lfloor C/2 \rfloor}^{\lfloor C/2 \rfloor} \frac{1}{4\pi k} \ \left(kb_{(t, x-1/2)} + nd^\star a_{(t+1/2, x)}\right) \: f(r) \: \left(kb_{(t, x+r-1/2)} + nd^\star a_{(t+1/2, x+r)}\right)  \nonumber\\[.2cm]
P_-(t) \ &= \ \sum_{x=1}^C \sum_{r=-\lfloor C/2 \rfloor}^{\lfloor C/2 \rfloor} \frac{-1}{4\pi k} \ \left(nd^\star a\right)_{(t+1/2, x)} \: f(r) \: \left(nd^\star a\right)_{(t+1/2, x+r)} 
\label{momenta_a_b}
\end{align}
where $y=0$ is understood in the coordinates. The coordinate space smearing function
\begin{align}
f(r) \ \equiv \ \frac{1}{C}\sum_{p\in(2\pi/C)\Z_C} f(p) \: e^{ipr} \sim \frac{\exp(-r^2/2R^2)}{\sqrt{2\pi} \: R}
\end{align}
ensures the definition of the physical momentum operator is insensitive to the lattice details, as it should.
\footnote{If we had chosen, say, $f(r)=\delta_{r=0}$ (i.e. $f(p)=1$), then the expectations $\langle P_\pm \rangle$ will turn out to be sensitive about whether $C$ is even or odd, which is unphysical. (The last term of \eqref{P_expectation} changes the $-1/12$ term to $-1/6$ for even $C$ and to $+1/12$ for odd $C$.)}
The expressions \eqref{momenta_a_b} in terms of $a$ and $b$ are more general than \eqref{momenta} in terms of $\phi$ and $\chi$. In particular, \eqref{momenta_a_b} is applicable when there are $W, V$ insertions, in which case the relations between $a$ and $\phi$, and $b$ and $\chi$, are modified (Figure \ref{f5-1_corr}).

One natural question to ask is in what sense can $P_+$ and $P_-$ be individually called a chiral momentum mode, as opposed to the non-chiral total momentum $P$? We may answer this question by applying one of them, say $P_-$, on certain observables and show $P_-$ spatially translates the observables as the full momentum operator $P$ would do. The observables which see $P_-$ as an effective momentum operator are the ones coupled to the $a$ gauge field only, i.e. the $W$ loops. To demonstrate this point concretely, again consider Figure \ref{f5-1_corr}, whose expectation value is given by \eqref{anomalous_scaling_result}. Now, let's turn on a spatial translation by $v$ at each time slice, $\exp\left(-i\sum_t v P_-(t)\right)$. We expect the anomalous scaling \eqref{anomalous_scaling_result} becomes 
\begin{align}
\left|\frac{\left\langle \exp(-i\int a\cdot W) \ \exp\left(-i\sum_t v P_-(t)\right) \right\rangle}{\left\langle \exp\left(-i\sum_t v P_-(t)\right) \right\rangle}\right| \ \sim \ |(x_2-x_1) - v\, (t_2-t_1)|^{-|k|\, w^2/n^2}
\label{transl_scaling}
\end{align}
at large spacetime separations, because a length $v$ spatial translation at each time slice accumulates to a total translation by a distance of $v \, (t_2-t_1)$. In Appendix \ref{app_P_eff} we confirm this is indeed the case, given the bulk placement of $W$ is ``deep enough'' into $y\geq 1+1/2$ as we demanded. Therefore $P_-$ is the effective momentum operator for the sector of observables coupled to $a$ only. In turn, $P_+$ acts almost trivially on $W$ at large distances. Similarly, one can show that at large distances, the observables coupled to the combination $kb+na$ are translated by $P_+$, while $P_-$ acts almost trivially on them.

Finally, we compute the zero-point expectation values of $P_-$. We let the $x$-direction circumference $C\gg 1$ be finite; on the other hand, for simplicity we may take the $t$-direction to be infinite. The computation is similar to that in the continuum. We present the details in Appendix \ref{app_zpm}. The result is
\begin{align}
\left\langle P_- \right\rangle \ &= \ C \int_{-\pi}^\pi \frac{dp}{2\pi} \ f(p) \left( \frac{\sin p}{2} \ \theta(\sin p) - \frac{i}{4} \left(1-\cos p\right) \right) \nonumber \\[.3cm]
& \ \ \ \ - \ \sgn(k)\, \frac{\pi}{C} \left( -\frac{1}{12} \: + \: \mathcal{O}(R/C, \, \mbox{real}) \right) \ + \ \mathcal{O}(f(p=\pi)/C, \, \mbox{real}).
\label{P_expectation}
\end{align}
The first line is extensive in $C$, but we can exactly remove it via a shift of $P_-$ by a local counter term $\sum_{x=1}^C \bar\rho$.
\footnote{The real and imaginary parts of $\bar\rho$ are order $1/R^2$ and $1/R^3$ respectively.}
The second line are the finite size effects, since they vanish as $C\rightarrow \infty$. It is important that there is no imaginary contribution from the finite size corrections, so by an appropriate choice of $\bar\rho$, the expectation value is strictly real as it should for a real operator.
\footnote{As we can see from the derivation in Appendix \ref{app_zpm}, the extensive imaginary part (to be cancelled off by $\bar\rho$) is a reminiscent of the imaginary part of \eqref{anomalous_scaling_result}, which is important for reproducing the Kac-Moody algebra, as commented below \eqref{continuum_boundary_details}.}
Taking the physical limits $1\ll R\ll C$, we thus reproduce the zero-point momentum $\langle P_- \rangle$ in \eqref{chiral_central_charges} with chiral central charge $-\sgn(k)$. Via a similar calculation, one can show $\langle P \rangle$ is exactly linear in $C$ and can be completely removed by a local counter term (the reason is mentioned at the end of Appendix \ref{app_zpm}). Subsequently, the finite size effect and the chiral central charge of $P_+$ must be opposite to those of $P_-$. Thus, we have confirmed that the momentum operator can be unambiguously separated into two counter-propagating modes $P_\pm$, each having the universal zero-point expectation value \eqref{chiral_central_charges}.

The discussions above apply to the doubled $\R$, doubled $U(1)$ and $U(1)\times \R$ CS we constructed on the lattice, using the gapless boundary condition \eqref{boundary_condition}, along with \eqref{boundary_condition_U1} in the latter two cases. For doubled $U(1)$ CS, the $d^\star a$ in \eqref{momenta_a_b} should be understood as $d^\star a - 2\pi l$, but there is no essential change in the computations of expectation values. Notably, the universal zero-point expectation \eqref{chiral_central_charges} of $P_-$ is individually measurable if the doubled $U(1)$ CS is coupled to the electromagnetic background with $q\neq 0, p=0$, since the $d^\star a - 2\pi l$ that would appear in the definition \eqref{momenta_a_b} of $P_-$ is $(-2\pi/q) (\delta S/ \delta A)$, proportional to the electric current operator on the boundary.
\footnote{But the values of $q, p$ and $k$ must be known in order to make the connection between $P_-$ and the electric current, therefore this connection is not considered universal.}
For $U(1)\times \R$ CS, $P_-$ is the boundary momentum operator associated with the $U(1)$ sector, and again its behaviors match with the boundary momentum operator of a single $U(1)$ CS at level $-k$.

\section{Towards Microscopic Hamiltonian}
\label{sect_Hamiltonian}

The ultimate goal of our study is to systematically construct \emph{microscopic} lattice models for these topological phases coupled to electromagnetic field that are soluble -- maybe not exactly soluble but at least in controllable approximation when the electromagnetic field is weak at the lattice scale. Although we have presented a systematic spacetime lattice construction, there remain two gaps to be filled between the present construction and the ultimate goal.
\begin{enumerate}
\item
Our construction is a topological Lagrangian description on the $3D$ spacetime lattice. A physical microscopic model should be a Hamiltonian description on the $2D$ spatial lattice that is only topological in its low energy sector. Usually there is a systematic procedure to generate the latter from the former \cite{Kirillov:2011mk, Bhardwaj:2016clt, Cong:2017ffh}. However, with the electromagnetic field there is the extra issue below.

\item
More importantly, our constructed theory on the lattice \emph{must} be viewed as an effective description in the deep IR limit, but not a microscopic one. While these two interpretations usually lead to no essential difference in the formulation of the theory (e.g. \cite{Levin:2004mi, Chen:2011pg}), in the presence of the electromagnetic field, the consistency conditions imposed by these two interpretations on the theory are \emph{different}, as discussed at the beginning of Section \ref{sssect_EM_Hall}.
\end{enumerate}
The technical manifestation of the ``effective versus microscopic'' issue, as described in Section \ref{sssect_EM_Hall}, is that
\begin{itemize}
\item[--]
either the $1$-form $\Z$ gauge symmetries of the dynamical $a, b$ fields stay exact, but the $U(1)$ compactness of the electromagnetic field $A$ is only preserved up to another $1$-form $\Z$ transformation involving the Wilson loop insertions,
\item[--]
or the $U(1)$ compactness of $A$ is exact, but the dynamical $1$-form $\Z$ gauge symmetries are not.
\end{itemize}
In the previous sections we have adopted the former condition, which is the right condition for the theory to be a deep IR effective theory. In a microscopic lattice model, the latter condition should be imposed.

Although our construction is not a microscopic Hamiltonian, it may guide us towards the systematic design of such a Hamiltonian. Notably, such Hamiltonian \emph{must} not be a local commuting projector Hamiltonian if the Hall conductivity coefficient is non-zero, due to the non-trivial constraint proven in \cite{kapustin2018local}. However, the Hamiltonian may still be soluble by designing a controlled separation of energy scales.

We need a procedure to restore the microscopic $U(1)$ compactness of $A$ and relax the dynamical $1$-form $\Z$ invariances from fundamental to emergent. A key idea appeared in the ``charging Hamiltonian'' in \cite{Levin:2011hq}, although this Hamiltonian corresponds to $k=0$, $q=0$, $p=2$ in our notations, which has vanishing Hall conductivity (see \eqref{theory_U1U1_shifted_A}). In fact, started with our effective Lagrangian construction with $k=0, q=0$ and carrying out the procedure to be described, we will automatically arrive at (a slight generalization of) the charging Hamiltonian in \cite{Levin:2011hq}.  It will therefore be instructive for us to explain this natural mapping in details first. After that we will consider more general values of $k, q$ which may host non-vanishing Hall conductivity.

To connect from Lagrangian to Hamiltonian, it is convenient to consider a spacetime being built out of cubic lattice or, more generally, prisms. In the case of prism lattice, the triangular side of the prism triangulates the spatial manifold, and the height discretizes the time; $\int \cdot$ is defined naturally, and $\int \cp$ can also be defined given the vertex ordering on the spatial manifold (we skip the details here). The spacetime lattice action \eqref{theory_U1U1_no_A} at $k=0$, with \eqref{closedness_Lagrange_multiplier} explicitly included, is $S = n\int a\cdot \left(db/2\pi - s \right) + \int l \cdot \left( d\theta - nb \right) + \int \varpi \cdot ds$. We will include electromagnetic coupling later, but since we will set $q=0$ anyways, we may integrate out $l$ first, so that we have the variable $\bar{b}_\lk \equiv (-d\theta+nb)_\lk/2\pi \in \{0, 1, \cdots, n-1\}$ as in \eqref{theory_U1Zn} (this requires an orientation for each link to be specified). The remaining spacetime lattice action can be written as $S=\sum_{t\in \Z} L(t)$, with the Lagrangian $L(t)$ at each integer time $t$ given by
\begin{align}
L(t) \ = \ & \sum_{\mathrm{spatial \: links\: } \mathbf{l}} a_{\mathbf{l}, (t+1/2)} \left( \bar{b}_{\mathbf{l}, (t+1)} - \bar{b}_{\mathbf{l}, t} \right) + \sum_{\mathrm{spatial \: plaq.\: } \mathbf{p}} \varpi_{\mathbf{p}, (t+1/2)} \left( s_{\mathbf{p}, (t+1)} - s_{\mathbf{p}, t} \right) \nonumber \\[.2cm]
& + \sum_{\mathrm{spatial \: vert. \: } \mathbf{v}} \bar{b}_{\mathbf{v}, (t+1/2)} (\mathbf{d}^\star a)_{\mathbf{v}, (t+1/2)} + \sum_{\mathrm{spatial \: links \: } \mathbf{l}} s_{\mathbf{l}, (t+1/2)} \left( (\mathbf{d}^\star \varpi)_{\mathbf{l}, (t+1/2)} - n a_{\mathbf{l}, (t+1/2)} \right) \nonumber \\[.2cm]
& + \sum_{\mathrm{spatial \: plaq.\: } \mathbf{p}} a_{\mathbf{p}, t} \left( (\mathbf{d}\bar{b})_{\mathbf{p}, t} - n s_{\mathbf{p}, t} \right)
\label{charging_Lagrangian}
\end{align}
In the notations, $\bar{b}_{\mathbf{l}, t}$ denotes $\bar{b}$ on the spacetime link at time $t$ that projects to $\mathbf{l}$ on the space, while $\bar{b}_{\mathbf{v}, (t+1/2)}$ denotes $\bar{b}$ on the spacetime link between time $t$ and $t+1$ that projects to $\mathbf{v}$ on the space, etc. Moreover, $\mathbf{d}$ is $d$ restricted to the spatial directions.

The first line of $L(t)$ occurs between two time slices. As usual, we can recognize the raising and lowering commutation relations
\begin{align}
\left[ \bar{b}_{\mathbf{l}}, e^{-i a_{\mathbf{l}}} \right] = e^{-i a_{\mathbf{l}}}, \ \ \ \left[ \bar{b}_{\mathbf{l}}, e^{i a_{\mathbf{l}}} \right] = -e^{i a_{\mathbf{l}}}, \ \ \ \ \ \left[ s_{\mathbf{p}}, e^{-i \varpi_{\mathbf{p}}} \right] = e^{-i \varpi_{\mathbf{p}}}, \ \ \ \left[ s_{\mathbf{p}}, e^{i \varpi_{\mathbf{p}}} \right] = - e^{i \varpi_{\mathbf{p}}}.
\label{charging_commutations}
\end{align}
Here we are using the basis in the $\bar{b}_{\mathbf{l}}, s_{\mathbf{p}}$ variables. The third line of $L(t)$ occurs at an integer time slice and corresponds to a potential energy for the $\bar{b}_{\mathbf{l}}, s_{\mathbf{p}}$ variables. It is written as a Lagrange multiplier constraint; to map it to a potential energy, we must relax the constraint to let it be only energetically favored:
\begin{align}
H_{\mathcal{U}} \ = \ \frac{\mathcal{U}}{2} \sum_{\mathbf{p}} \left( (\mathbf{d}\bar{b})_{\mathbf{p}} - n s_{\mathbf{p}} \right)^2.
\label{potential_energy}
\end{align}
The second line of $L(t)$ occurs at a half-integer time slice and corresponds to a kinetic energy. Again its two terms are Lagrange multiplier constraints, that together imposes the Gauss's constraints associated with the $1$-form $\Z$ gauge invariance and the $0$-form $\Z_n$ gauge invariance of the $\bar{b}$ variable. Again we may relax these gauge invariance constraints to be only energetically favored, hence the notion of ``emergent gauge field'' \cite{Kitaev:1997wr}. There is a small technical issue here. If the $\bar{b}_{\mathbf{l}}$ variable were valued in $\Z$, then the kinetic energy can be
\begin{align}
H'_{\mathcal{K}} = \mathcal{K} \sum_{\mathbf{v}} \left(1-\cos(\mathbf{d}^\star a)_{\mathbf{v}} \right) + \mathcal{K}' \sum_{\mathbf{l}} \left( 1 - \cos(\mathbf{d}^\star \varpi - na)_{\mathbf{l}} \right)
\label{kinetic_energy_Z}
\end{align}
which energetically imposes the two Gauss's constraints separately. However, we actually have $\bar{b}_{\mathbf{l}}$ valued in $\{0, 1, \cdots, n-1\}$ (fixed an orientation of the link $\mathbf{l}$), which means we have already fixed the $1$-form $\Z$ gauge in our local Hilbert space.
\footnote{Doing so makes the theory gapped. If $\bar{b}$ is $\Z$ valued, the  conjugate $a$ variable will turn out to be gapless.}
We must project the time evolution $e^{-i H'_{\mathcal{K}} T}$ onto this gauge fixed Hilbert space. This results in the kinetic energy:
\begin{align}
& H_{\mathcal{K}} \: \mbox{ is given by } \: \mathcal{K} \sum_{\mathbf{v}} \left(1-\cos(\mathbf{d}^\star a)_{\mathbf{v}} \right) \: \mbox{ but with the replacements in $\cos(\mathbf{d}^\star a)_{\mathbf{v}}$:} \nonumber \\[.2cm] 
& e^{ia_{\mathbf{l}}} \ \longrightarrow \ e^{ia_{\mathbf{l}}} + e^{ia_{\mathbf{l}}} e^{i(\mathbf{d}^\star \varpi - na)_{\mathbf{l}}}, \ \ \ \ e^{-ia_{\mathbf{l}}} \ \longrightarrow \ e^{-ia_{\mathbf{l}}} + e^{-ia_{\mathbf{l}}} e^{-i(\mathbf{d}^\star \varpi - na)_{\mathbf{l}}}.
\label{kinetic_energy}
\end{align}
\footnote{The coefficients in front of $e^{ia_{\mathbf{l}}}$ and $e^{ia_{\mathbf{l}}} e^{i(\mathbf{d}^\star \varpi - na)_{\mathbf{l}}}$ can be different, but we just let them be the same for convenience.}
This projection is easily understood. Consider $e^{ia_{\mathbf{l}}} + e^{ia_{\mathbf{l}}} e^{i(\mathbf{d}^\star \varpi - na)_{\mathbf{l}}}$. For a basis state in which $\bar{b}_{\mathbf{l}} = 1, 2, \cdots, n-1$, the $e^{ia_{\mathbf{l}}}$ term decreases $\bar{b}_{\mathbf{l}}$ by $1$, while the $e^{ia_{\mathbf{l}}} e^{i(\mathbf{d}^\star \varpi - na)_{\mathbf{l}}}$ term annihilates the state; for a basis state in which $\bar{b}_{\mathbf{l}}=0$, the $e^{ia_{\mathbf{l}}}$ term annihilates the state, while the $e^{ia_{\mathbf{l}}} e^{i(\mathbf{d}^\star \varpi - na)_{\mathbf{l}}}$ term changes $\bar{b}_{\mathbf{l}}=0$ to $n-1$ and at the same time hops an $s$ particle from the plaquette on the right of $\mathbf{l}$ to the plaquette on the left of $\mathbf{l}$. Likewise for $e^{-ia_{\mathbf{l}}}+e^{-ia_{\mathbf{l}}} e^{-i(\mathbf{d}^\star \varpi - na)_{\mathbf{l}}}$.

Piecing up the above, our spacetime lattice Lagrangian \eqref{charging_Lagrangian} has been naturally realized by the commutation relations \eqref{charging_commutations} and the spatial lattice Hamiltonian $H_{\mathcal{U}}+H_{\mathcal{K}}$. This Hamiltonian is a local commuting projector Hamiltonian that can be defined on any discretization of the space,
\footnote{Although we primarily assumed the spatial plaquettes to be squares or triangles, we did not make use of this assumption, since we have not use the cup product.}
as long as an orientation on each link is specified. If the discretization is a bipartite lattice, one may perform some redefinition of variables, and take the dual spatial lattice, after which our $H_{\mathcal{U}}+H_{\mathcal{K}}$ becomes the ``charging Hamiltonian'' in \cite{Levin:2011hq}; but the bipartite lattice is not necessary for us to define $H_{\mathcal{U}}+H_{\mathcal{K}}$. One can show the Hamiltonian $H_{\mathcal{U}}+H_{\mathcal{K}}$ is gapped by $\mathcal{U}$ and $2\mathcal{K}(1-\cos(2\pi/n))$, corresponding to pair creations of $W$ and $V$ anyon excitations respectively \cite{Levin:2011hq}.

Now, following the idea of \cite{Levin:2011hq}, we couple the Hamiltonian to the electromagnetic background on the spatial lattice, with $q=0$, $p\neq 0$.
\footnote{In \cite{Levin:2011hq}, $p=2$, but this value is not crucial. Also, when $k=0$, we can switch $p$ and $q$, as long as we switch $a$ and $b$ from the beginning.}
In \eqref{theory_U1U1}, $A$ lives on the links, while for now it is more convenient to consider $A^\star$ that lives on the plaquettes (links of the dual lattice); we may set $A^\star = \mathrm{M}\capp A$ or alternatively (in general inequivalently) $A= A^\star \capp \mathrm{M}$. In the spacetime effective theory, $A^\star$ couples to the dynamical fields as $p \int \left( s-d\bar{b}/n\right) \cdot A^\star$. This coupling respects the dynamical $1$-form $\Z$ gauge invariance, at the cost that $A^\star_\pl$ is not identified with $A^\star_\pl + 2\pi$, due to the fractional coefficient of the $d\bar{b}$ term. As we move towards a microscopic model, $A^\star_\pl$ must be identified with $A^\star_\pl + 2\pi$. We do so by just dropping the $d\bar{b}$ term, so $A^\star$ couples as $p\int s\cdot A^\star$, with $p s$ being the integer valued local electric current in spacetime.

The $1$-form $\Z$ gauge invariance between $s$ and $\bar{b}$ becomes non-exact. However, remarkably, the $W$ anyon still has fractional electric charge \cite{Levin:2011hq}, despite the absence of the explicit $(p/n) d\bar{b}$ coupling. Let's see this in the Hamiltonian formalism. We first set $A^\star=0$ and look at the expectation value of the electric charge density at some spatial plaquette $\mathbf{p}$:
\begin{align}
\left\langle p\, s_{\mathbf{p}} \right\rangle = \frac{p}{n} \left\langle \left(ns_{\mathbf{p}} - (\mathbf{d}\bar{b})_{\mathbf{p}} \right) \right\rangle + \frac{p}{n} \left\langle (\mathbf{d}\bar{b})_{\mathbf{p}} \right\rangle.
\end{align}
The operator $\left(ns_{\mathbf{p}} - (\mathbf{d}\bar{b})_{\mathbf{p}} \right)$ is diagonalized in the eigenstates of the (local commuting projector) Hamiltonian $H_\mathcal{U} + H_\mathcal{K}$. It is $0$ in the ground state, and equals to $-w$ if a $W$ anyon excitation of charge $w$ is present at $\mathbf{p}$. On the other hand, each $\langle \bar{b}_{\mathbf{l}} \rangle$ that appears in $\langle (\mathbf{d}\bar{b})_{\mathbf{p}} \rangle$ is \emph{independent} of the particular energy eigenstate, because the Gauss's constraint requires any energy eigenstate to be an equal weight superposition of basis states with $\bar{b}_{\mathbf{l}} = 0, 1, \cdots n-1$,
\footnote{Even if a $V$ anyon is present in the vicinity of $\mathbf{l}$, only the relative phases between the weights are changed, but the magnitudes of their weights are still the same.}
which leads to $\langle \bar{b}_{\mathrm{l}} \rangle = (0+1+\cdots (n-1))/n = (n-1)/2$ regardless of the particular energy eigenstate \cite{Levin:2011hq}. We can therefore remove the ground state expectation value $(p/n) \langle (\mathbf{d}\bar{b})_{\mathbf{p}} \rangle_0$ from the electric charge density operator, and obtain the normal ordered expectation
\begin{align}
\left\langle p \, :\! s_{\mathbf{p}} \!: \right\rangle \equiv  \left\langle p \left(s_{\mathbf{p}} - \langle s_{\mathbf{p}} \rangle_0\right) \right\rangle = \left\langle p \left(s_{\mathbf{p}} - \frac{1}{n} \langle (\mathbf{d}\bar{b})_{\mathbf{p}} \rangle_0\right) \right\rangle = \frac{p}{n} \left\langle \left(ns_{\mathbf{p}} - (\mathbf{d}\bar{b})_{\mathbf{p}} \right) \right\rangle = -\frac{p}{n} \, w_{\mathbf{p}}.
\label{eff_fractional_charge}
\end{align}
We thus recover the fractional electric charge of the $W$ anyon (compare with \eqref{theory_U1U1_shifted_A}) as a reminiscent of the $1$-form $\Z$ gauge invariance in $H_\mathcal{U} + H_\mathcal{K}$, even though the electric charge density operator $p\, s_{\mathbf{p}}$ itself is integer valued and not $1$-form $\Z$ invariant.

Now consider $A^\star \neq 0$. It is convenient to fix the gauge so that $A^\star$ only has spatial components. Clearly the Lagrangian coupling $p\int s\cdot A^\star$ appears in the Hamiltonian as $e^{\pm i (\mathbf{d}^\star \varpi)_{\mathbf{l}}} \rightarrow e^{\pm i(\mathbf{d}^\star \varpi + p A^\star)_{\mathbf{l}}}$ in the definition \eqref{kinetic_energy} of $H_{\mathcal{K}}$ (if the spatial lattice is a bipartite lattice, after redefining some variables and taking the dual spatial lattice, this coupling is equivalent to that in \cite{Levin:2011hq}.) It is easy to see that when $|\left(\mathbf{d}^\star A^\star\right)_{\mathbf{v}} | < 2\pi/ 2p$ and $|\partial_t A^\star_{\mathbf{l}}| \ll \mathcal{U}, \mathcal{K}$, the ground state and low energy states are unchanged from $A^\star=0$  \cite{Levin:2011hq}. This is sufficient to concretely claim that the Hamiltonian describes the desired topological phase in the presence of the electromagnetic background, which is normally assumed to be weak at the lattice scale. One may also consider, e.g. when $|\left(\mathbf{d}^\star A^\star\right)_{\mathbf{v}}|$ increases beyond $2\pi/ 2p$ and approaches $2\pi/ p$. A level crossing occurs
\footnote{A small non-universal term can be added to open up an energy gap at the level crossing \cite{Levin:2011hq}.}
so that the excited state with a $V$ anyon at $\mathbf{v}$ becomes the ground state \cite{Levin:2011hq} -- this is obviously the microscopic realization of the $1$-form $\Z$ ``indistinguishability'' \eqref{EMU1_compactness} in our effective spacetime lattice description.
\footnote{These properties are studied in \cite{Levin:2011hq} via a ``gauge'' transformation that violates the $U(1)$ compactness of the microscopic $A$. This trick restores the dynamical $1$-form $\Z$ gauge invariance between $\bar{b}$ and $s$, and essentially converts the microscopic interpretation to our effective theory interpretation.}

Having made this concrete connection between our effective spacetime lattice theory and the microscopic ``charging Hamiltonian'' in \cite{Levin:2011hq}, we shall now discuss more general values of $k$ and $q$. First, let's set $k\neq 0$ but still $q=0$, so there is no Hall conductivity according to \eqref{theory_U1U1_shifted_A}. Since $q=0$, we can still integrate out $l$ first and use the variable $\bar{b}$ (as in \eqref{theory_U1Zn}), i.e. the emergent gauge field is still $\Z_n$ with the $\Z$ central extension by $s$. The only change is we need to implement the Dijkgraaf-Witten term of coefficient $k$ from the spacetime Lagrangian onto the spatial Hamiltonian. For the even $k$ bosonic phases, this has been done in \cite{hu2013twisted, mesaros2013classification, Lin:2014aca} without the electromagnetic background and the $s$ variable; since we have the central extension variable $s$ now, the rather technical spatial implementation of the Dijkgraaf-Witten term would actually be simplified (similar to how the central extension helps us understand the computation in Section \ref{sssect_WL_by_central_Z}). For the odd $k$ fermionic phases, besides the Dijkgraaf-Witten term, the fermion sign factor introduced in Section \ref{sect_fermionic} must be also implemented on the spatial lattice using the systematic methods of \cite{Gu:2013gma, Tarantino:2016qfy, Bhardwaj:2016clt}. Importantly, since $A^\star$ is coupled to the integer valued $s$ particle only but not to $\bar{b}$, we can use the more stringent notion of lattice spin-c structure mentioned at the end of Section \ref{sect_fermionic}, that the lattice spin structure data $\Sigma$ -- also coupled to $s$ only -- can be entirely absorbed into a redefinition of $A^\star$.
\footnote{As a result of the redefinition, when the vertex ordering on the spatial lattice changes, $A^\star$ must change accordingly by $\pi$ fluxes, playing its role as a spin-c connection.}
We will present the relatively technical implementation of these ideas in future works.

Finally, we shall turn on $q\neq 0$ which gives rise to non-trivial Hall conductivity according to \eqref{theory_U1U1_shifted_A}. By the conclusion of \cite{kapustin2018local}, any Hamiltonian realization must have some spatial correlations and hence not a local commuting projector one. But a Hamiltonian realization may still be soluble given a controlled separation of energy scales. Although concretely constructing such a Hamiltonian is beyond the scope of the present paper, in the below we discuss the ingredients and make proposals towards such a Hamiltonian realization, motivated by our spacetime effective Lagrangian. To see the key issues, it suffices to focus on $k=0$ but $q\neq 0, p\neq 0$.

The main changes that $q\neq 0$ brings in is we cannot integrate out $l$ before \eqref{charging_Lagrangian}, because the $l$ particle is also charged under the electromagnetic field as $q\int l\cdot A$. Recall we may set $A= A^\star \capp \mathrm{M}$ or $A^\star = \mathrm{M}\capp A$. In the present discussion is convenient to think of a cubic spacetime lattice, whose associated spatial lattice is a square lattice; on the square lattice, $A^\star_{\mathbf{l}} = A_{\mathbf{l}-\hat{x}/2-\hat{y}/2}$.
\footnote{Since the cap product is a one-to-one mapping between links on the square lattice, small $\mathbf{d}A$ is equivalent to small $\mathbf{d}^\star A^\star$. On more general triangulation, extra care needs to be taken to ensure the smallness of both of them (essentially using the indistinguishability \eqref{EMU1_compactness}).}
Since the $l$ field is not integrated out, the $b$ field takes continuous rather than discrete values. For now let's start with $b$ taking $\R$ value as in Section \ref{sssect_HS_PF_DW}, and then its canonical conjugate $a$ is also $\R$ valued -- in fact this is the crucial issue that should be taken care of, as will be revealed soon. Just as we dropped the coupling between $db$ and $A^\star$, we also drop the coupling between $d^\star a$ and $A$, so that $A$ is $U(1)$ compact on the microscopic lattice. The commutation relations are
\begin{align}
& \left[ b_{\mathbf{l}}, a_{\mathbf{l}} \right] = \frac{i 2\pi}{n}, \ \ \ \ \ \left[ s_{\mathbf{p}}, e^{-i \varpi_{\mathbf{p}}} \right] = e^{-i \varpi_{\mathbf{p}}}, \ \ \ \left[ s_{\mathbf{p}}, e^{i \varpi_{\mathbf{p}}} \right] = - e^{i \varpi_{\mathbf{p}}}, \nonumber \\[.2cm]
& \left[ l_{\mathbf{v}}, e^{-i \theta_{\mathbf{v}}} \right] = e^{-i \theta_{\mathbf{v}}}, \ \ \ \left[ l_{\mathbf{v}}, e^{i \theta_{\mathbf{v}}} \right] = - e^{i \theta_{\mathbf{v}}}.
\end{align}
As before, the lattice Hamiltonian should energetically impose the $0$- and $1$-form Gauss's constraints for the $a$ and $b$ fields, so we would consider the Hamiltonian
\begin{align}
& \frac{\mathcal{U}}{2} \sum_{\mathbf{p}} \left( \mathbf{d}b - 2\pi s \right)_{\mathbf{p}}^2 + \mathcal{U}'  \sum_{\mathbf{l}} \left( 1 - \cos(\mathbf{d} \theta - nb + qA)_{\mathbf{l}} \right) \nonumber \\[.2cm]
+ & \frac{\mathcal{\mathcal{K}}}{2} \sum_{\mathbf{v}} \left( (\mathbf{d}^\star a)_{\mathbf{v}} - 2\pi l_{\mathbf{p}} \right)^2 + \mathcal{K}'  \sum_{\mathbf{l}} \left( 1 - \cos(\mathbf{d}^\star \varpi - na + pA^\star)_{\mathbf{l}} \right).
\label{gapless_Hamiltonian}
\end{align}
The four terms commute with each other so this Hamiltonian is exactly soluble. If we assume $\mathcal{U}', \mathcal{K}' \gg \mathcal{U}, \mathcal{K}$ and compute the ground state expectation value of the electric charge density operator $q l_{\mathbf{v}} + p s_{\mathbf{v}+\hat{x}/2-\hat{y}/2}$, we can see its dependence on the magnetic field $\mathbf{d} A$ indeed corresponds to the desired Hall conductivity coefficient $2pq/n$.
\footnote{The $\mathcal{U}', \mathcal{K}'$ terms set $\mathbf{d}b - (q/n)\mathbf{d}A$ and $\mathbf{d}^\star a - (p/n) \mathbf{d}^\star A^\star$ to be close to multiples of $2\pi/n$. As long as $\mathbf{d}A<2\pi/2q$ and $\mathbf{d}^\star A^\star<2\pi/2p$, the $\mathcal{U}, \mathcal{K}$ terms selects the ground state so that $\mathbf{d}b$ is closest to $2\pi s$ and $\mathbf{d}^\star a$ is closest to $2\pi l$. Then the $\mathbf{d}A$ dependence of $s$ and $l$ can be extracted as in \eqref{eff_fractional_charge}.}
The problem, however, is this Hamiltonian is gapless, because the $a$ and $b$ are continuous now and the commuting terms do not open an energy gap. This would be the crucial issue to be resolved.

In the discussion below \eqref{Higgsing_potential} we could add a small term quadratic in $a$ that smears out $b$ to open up a Higgsing gap for $b$. The low energy sector of the Higgs mechanism is controllable there because the $0$- and $1$-form gauge invariances of $b$ are still \emph{fundamental}, even though those of $a$ become emergent after the treatment \eqref{Higgsing_potential}. In the Hamiltonian language, this means to use the following Hamiltonian and the \emph{fundamental} Gauss's constraints on physical states:
\begin{align}
H \ = \ \frac{K}{2} \sum_{\mathbf{l}} a_{\mathbf{l}}^2 + \frac{\mathcal{U}}{2} \sum_{\mathbf{p}} \left( \mathbf{d}b - 2\pi s \right)_{\mathbf{p}}^2 + \mathcal{U}'  \sum_{\mathbf{l}} \left( 1 - \cos(\mathbf{d} \theta - nb + qA)_{\mathbf{l}} \right), \nonumber \\[.2cm]
e^{\pm i(\mathbf{d}^\star \varpi - na + pA^\star)_{\mathbf{l}}} \ | \mbox{phys} \rangle \ = \ 0, \ \ \ \ \ \left( n\mathbf{d}^\star a - 2\pi n l - p\mathbf{d}^\star A^\star\right)_{\mathbf{v}} \ | \mbox{phys} \rangle \ = \ 0.
\end{align}
These fundamental Gauss's constraints are self-compatible and commute with the Hamiltonian. The Hamiltonian can be controllably solved in perturbation theory by setting the energy gaps $\sqrt{K\mathcal{U}'} \gg \mathcal{U} \gg 1$ and the smearing $\sqrt{K/\mathcal{U}'} \ll 1$. The Hall conductivity coefficient in the long distance limit is reliably $2pq/n$ -- it cannot have small deviations from this value because the system is a gapped topological phase. Since the $\mathcal{U}$ term does not commute with the $K$ terms on neighboring links, a small spatial correlation is generated. This scenario is already in agreement with \cite{kapustin2018local}, although we still have a \emph{fundamental} dynamical gauge field $b$ which is beyond the physical assumptions of \cite{kapustin2018local}.

In a physical microscopic Hamiltonian, the gauge invariances of $b$ cannot be fundamental. Thus, the small changes in the $a$ configuration still makes the system gapless. In order to resolve this, it would be natural to consider adding small terms that are quadratic in $a$ and $b$ respectively. This means the terms \eqref{gapless_Hamiltonian} are occurring on top of a wide and shallow harmonic oscillator on each link, which regularizes the large values of $a$ and $b$ (and hence they give each other a small smearing around each potential minima). More sophisticated analysis is needed to understand whether this simple proposal is controllably soluble in some limit of parameters and yield the desired topological order and Hall conductivity. This proposal is very closely related to the Hamiltonian proposed in \cite{geraedts2013exact},
\footnote{If we limit to the $c=1$ and $d\neq 0$ case in the model of \cite{geraedts2013exact}, it corresponds to properly quantized doubled CS with $k=0, p=q=1$ and $n=d$ in our notations. The interpretation of $c\neq 1$ is more obscure.}
whose physical properties has been examined with the aid of Monte Carlo numerics.

Another way to regularize the large values of $a$ and $b$ would be to let $e^{i Nb}$ be a $\Z_{NN'n}$ variable and hence so is $e^{iN' a}$, for some $N, N' \gg 1$. While the $a, b$ degrees of freedom are generally gapped, small $\cos$ terms for $Nb$ and $N' a$ on each link are still needed to gap the level crossing when $A$ changes across odd multiples of $2\pi/2N$ and $2\pi/2N'$. The $\mathcal{K}$ and $\mathcal{U}$ terms must be rewritten in terms of $\cos$ of $N\mathbf{d}b$ and $N'\mathbf{d}^\star a$. While this will lead to extra (approximate) degeneracies $s_{\mathbf{p}}\sim s_{\mathbf{p}} + N$, $l_{\mathbf{v}}\sim l_{\mathbf{p}} + N'$, we may add strong potentials to energetically restrict $s$ and $l$ to $\{0, 1, \cdots, N-1\}$ and $\{0, 1, \cdots, N'-1\}$ respectively. When we change the energy eigenstate or change $A$, the expectation value of $s$ (or $l$) may sometimes jump by $\pm(N-1)$ (or $\pm(N'-1)$) across neighboring plaquettes (or vertices), but we expect that would not happen between $A=0$ and small $A$ at the lattice scale. 
It would again take more sophisticated analysis to check whether this proposal works as desired.

In summary, we can naturally construct local commuting projector Hamiltonians from our effective spacetime lattice Lagrangians when $q=0$, i.e. when the Hall conductivity vanishes. In particular, when $k=0$, we recover the ``charging Hamiltonian'' in \cite{Levin:2011hq}. The details of the $k\neq 0$ cases will be presented in future works, in which the odd $k$ cases are particularly interesting for manifesting the role of a microscopic spin-c structure. On the other hand, when $q\neq 0$, i.e. with Hall conductivity, any Hamiltonian realization must  not be local commuting projector according to \cite{kapustin2018local}. If we allow the $b$ field to be a \emph{fundamental} gauge field, we can construct a gapped lattice Hamiltonian realization that is soluble in controlled approximation by a separation of energy scales; it has non-zero spatial correlations, manifesting the conclusion of  \cite{kapustin2018local}. More physically, $b$ must be an \emph{emergent} gauge field. We proposed some simple possibilities of gapped Hamiltonian realizations; we will examine these proposals with more sophisticated analysis in future works.

\section{Conclusion}

In this paper we systematically constructed spacetime lattice descriptions for the twisted doubled abelian topological orders in $3D$ in an electromagnetic background. Our method is based on the gauging of $1$-form $\Z$ symmetries, similar to the idea of the Villain model. The resulting spacetime lattice description must be viewed as a coarse-grained effective description of gapped systems, and the Hall conductivity is reproduced. If the topological order is fermionic, the spin-c nature of the electromagnetic field is also manifest. Explicit connection between this lattice construction and the doubled $U(1)$ Chern-Simons path integral in the continuum is made. We also reproduced the anomalous scaling and the zero-point momenta modes under a gapless boundary condition on the lattice. Finally, for the cases with vanishing Hall conductivity, our coarse-grained spacetime Lagrangians can be naturally mapped to exactly soluble microscopic Hamiltonians, while for the cases with non-vanishing Hall conductivity, we made proposals for Hamiltonian realizations that might be controllably soluble in certain limits, as will be scrutinized in future works.

We would like to note that the form \eqref{doubledU1CS_cont} we started with does not encompass all boundary gappable abelian topological orders. There is another category of fascinating phases that detect configurations of Borromean rings and host anyons with non-abelian braiding \cite{deWildPropitius:1995cf,ferrari2015topological, He:2016xpi, Putrov:2016qdo}. In a future work we will extend our method to these theories.

Our general theme is to study the lattice realizations of those topological phases which do not admit the usual exactly soluble lattice construction. In this paper the phases we considered become ungappable on the boundary because of the enrichment by the global $U(1)$ symmetry. However, more generally, there are also phases whose intrinsic topological order itself is ungappable on the boundary; some simplest examples include the single, chiral abelian Chern-Simons theories of arbitrary levels. At least for these abelian examples, we believe lattice constructions that are controllably (though not exactly) soluble are possible, for instance by the introduction of Maxwell terms \cite{DeMarco:2019pqv} as mentioned in Section \ref{ssect_RRCS}. We will study these abelian examples in details in forthcoming works. On the other hand, the lattice implementation of boundary ungappable non-abelian topological orders is much more obscure, and the problem is related to some old puzzles in lattice gauge theory; it would be of great interest if progress can be made in this direction in the future.


\vspace{.8cm} 

\noindent \emph{Acknowledgement.} The author thanks Weiyan Chen, Clay C\'{o}rdova, Anton Kapustin, Xiao-Liang Qi, Ryan Thorngren,  Kantaro Ohmori, Shu-Heng Shao, Juven Wang, Edward Witten, and especially Lukasz Fidkowski and Michael Levin for valuable discussions, and thanks Juven Wang for hospitality at the Institute of Advanced Studies. The author is supported by the Gordon and Betty Moore Foundation's EPiQS Initiative through Grant GBMF4302.

\vspace{.8cm}

\appendix

\section{Summary of Homology and Cohomology}
\label{app_alg_topo}

In this appendix we summarize the basic aspects of the homology and cohomology of $3D$ oriented manifolds. The universal coefficient theorem states that the short exact sequence holds for each $n=1,2,3$ for any abelian coefficient group $\mathcal{G}$:
\begin{align}
0 \ \rightarrow \ \mathrm{Ext}(H_{n-1}(M; \Z), \mathcal{G}) \ \rightarrow \ H^n(M; \mathcal{G}) \ \rightarrow \ \mathrm{Hom}(H_n(M; \Z), \mathcal{G})  \ \rightarrow \ 0.
\end{align}
(The $\mathrm{Ext}$ functor can be efficiently computed using the basic properties: $\mathrm{Ext}(\Z, \mathcal{G})=0$, $\mathrm{Ext}(\Z_n, \mathcal{G})=\mathcal{G}/n\mathcal{G}$ and $\mathrm{Ext}(H\oplus H', \mathcal{G})=\mathrm{Ext}(H, \mathcal{G})\oplus\mathrm{Ext}(H', \mathcal{G})$.) Moreover, as the manifold $\M$ is oriented, the Poincar\'{e} duality holds for each $n=0,1,2,3$ for any abelian coefficient group $\mathcal{G}$:
\begin{align}
H_{3-n}(M; \mathcal{G})\cong H^n(M; \mathcal{G}).
\end{align}
We will primarily consider homology and cohomology with $\mathcal{G}=\Z$, since the information of any other $\mathcal{G}$ can be deduced from there via the theorems above. We suppose the manifold $\M$ has $B_0$ connected components. Then $H_0(M; \Z)\cong H^0(M; \Z)\cong H_3(M; \Z)\cong H^3(M; \Z)\cong \Z^{B_0}$ trivially. We can solve the general form of $H_1, H^1, H_2, H^2$ using the two theorems above. We can use the general ansatz for finitely generated abelian groups:
\begin{align}
H_1(M; \Z) \ \cong \ H^2(M; \Z) \ \cong \ \mathcal{F} \oplus \mathcal{T}, \ \ \ \ \ \mathcal{F}=\Z^{B_1}, \ \ \ \mathcal{T} = \bigoplus_j \Z_{\mathrm{p}_j}.
\end{align}
Here $\mathcal{F}$ is a free abelian group, with $B_n$, the number of free generators of $H_n$, known as the $n$th Betty number; here $B_1=B_2$ due to Poincar\'{e} duality. On the other hand, $\mathcal{T}$ is a finite abelian group, known as the ``torsion'', which in general can be written as a direct sum of a few cyclic groups as we did.
\footnote{This follows from the fundamental theorem of finite abelian groups (which has nothing in particular to do with algebraic topology). Note that the same finite abelian group $\mathcal{T}$ may be presented using different sets of $\{p_i\}$, and the group stays the same. The fundamental theorem further states that one (standard) choice is such that each $\mathrm{p}_j$ is a power of a prime number, i.e. $\mathrm{p}_j=\mathfrak{p}_j^{m_j}$, $\mathfrak{p}_j$ is prime.}
We may identify $\mathcal{F} = \mathrm{Hom}(H_2(M; \Z), \Z)$ and $\mathcal{T} = \mathrm{Ext}(H_1(M; \Z), \Z)$. To be consistent with the universal coefficient theorem and the Poincar\'{e} duality, we find
\begin{align}
H_2(M; \Z) \ \cong \ H^1(M; \Z) \ = \ \mathrm{Hom}(H_1(M; \Z), \Z) \ \cong \ \Z^{B_1},
\end{align}
and $\mathcal{F}$ and $\mathcal{T}$ are not subjected to any further constraint.

Due to the natural map from $H^2(M; \Z)$ onto $\mathcal{F}=\mathrm{Hom}(H_2(M; \Z), \Z)$, for any $\gamma\in H^2(M; \Z)$ we can naturally identify its ``free part'', denoted as $[\gamma] \in \mathcal{F}$. On the other hand, there is no natural map from $H^2(M; \Z)$ back to $\mathcal{T}$, so there is no natural identification of the ``torsion part'' of an element $\gamma\in H^2(M; \Z)$, unless $\gamma$ has trivial free part so that itself is a torsion element. Let's introduce some notations to describe this. For a $2$-cocycle $s\in Z^2(M; \Z)$, we denote its class in $H^2(M; \Z)$ as $\gamma=[s]$, and denote its class in $\mathcal{F}$ as $\eta=[\gamma]=[[s]]$. When and only when $\gamma$ has vanishing free part, i.e. $[\gamma]=0$, it can be naturally identified as a torsion element $\gamma=\tau\in\mathcal{T}$, using the natural map from $\mathcal{T}$ into $H^2(M; \Z)$.

\begin{figure}
\centering
\includegraphics[width=.2\textwidth]{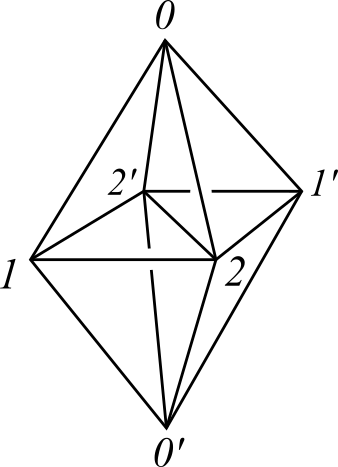}
\caption{The manifold $\R\mathrm{P}^3$ is obtained by a solid ball with antipodal points on the $S^2$ surface identified. We can discretize the solid ball as two pyramids in the above, and to obtain a discretized $\R\mathrm{P}^3$, we identify the surface points $0\sim 0'$, $1\sim 1'$, $2\sim 2'$, the surface links $(12)\sim (1'2')$, $(2'1)\sim (21')$ etc., and the surface triangles $(012)\sim(0'1'2')$ etc. (note the orientations).
\\[.05cm]
A closed loop $l$ consisting of $(12)$ and $(21')$, which is identified with the closed loop $l'$ consisting of $(1'2')$ and $(2'1)$, is non-contractible, and hence a non-trivial element of $H_1(\R\mathrm{P}^3; \Z)=\Z_2$. Twice of this loop $l$ is contractible due to the said identification $l\sim l'$, as $l$ and $l'$ together bound the square formed by the triangles $(122')$ and $(1'2'2)$ inside the solid ball.
\\[.05cm]
A non-trivial element of $H^2(\R\mathrm{P}^3; \Z)=\Z_2$ can be, say, a closed integer field $X$ that is $1$ on the triangles $(122')$, $(022')$ and $(01'2')$ (which is identified with $(0'12)$). Twice of the field $X$ is exact, because $2X=dY$ for an integer field $Y$ that is $1$ on the links $(22')$, $(2'1)$ (which is identified with $(21')$), $(2'0)$ (which is identified with $(20')$) and $(01')$ (which is identified with $(0'1)$).
\\[.05cm]
This $Y$ (actually the $\mod 2$ reduction of it), in turn, is a non-trivial element of $H^1(\R\mathrm{P}^3; \Z_2) = \Z_2$ which classifies the spin structure data, since it maps the non-contractible loop $l$ (or any other non-contractible loop) to $1\in \Z_2$.}
\label{fA_RP3}
\end{figure}

A familiar example with non-trivial torsion is $\R\mathrm{P}^3$, the manifold of $SO(3)$, for which $H_1(\R\mathrm{P}^3; \Z) \cong H^2(\R\mathrm{P}^3; \Z) = \mathcal{T} = \Z_2$. See Figure \ref{fA_RP3}. More general examples that have trivial $\mathcal{F}$ but non-trivial $\mathcal{T}$ include the Lens spaces, which can be discretized in a similar way.

Next we consider $\mathcal{G}=\R$. The $\mathrm{Ext}$ functors are trivial since $\R$ is divisible (i.e. $\R/n\R$ is trivial), so $H_1, H^1, H_2, H^2$ with $\R$ coefficient are all given by $\R^{B_1}$. Turning back to the universal coefficient theorem with $\mathcal{G}=\Z$, the surjective map from $H^2(M; \Z)$ to $\mathcal{F}$ can also be viewed as the natural map from $H^2(M; \Z)$ into $H^2(M; \R)=\R^{B_1}$, with $\mathcal{F}$ the image of the map. 

The spin structure is classified by $H^1(M; \Z_2)$, therefore we are also interested in $\mathcal{G}=\Z_2$. The universal coefficient theorem gives
\begin{align}
H^1(M; \Z_2) \ = \ \mathrm{Hom}(H_1(M; \Z), \Z_2) \ \cong \ \left(\Z_2\right)^{B_1} \oplus \bigoplus_{j \ (\mathrm{p}_j \: \mathrm{even})} \Z_2.
\label{spin_structure_info}
\end{align}
We can naturally identify the ``torsion part'' of an $H^1(M; \Z_2)$ element, by acting it (as a $\mathrm{Hom}(H^2(M; \Z), \Z_2)$ element, since $H^2\cong H_1$) on torsion elements $\gamma=\tau\in\mathcal{T}$. On the other hand, there is no natural identification of its ``free part'' unless the torsion part is trivial. Let's emphasize that when $\mathcal{T}\neq 0$ and has some even $\mathrm{p}_j$'s, there can exist $Y\in C^1(M; \Z)$ such that $dY \in 2C^2(M; \Z)$ but $Y \notin 2 C^1(M; \Z)$; see Figure \ref{fA_RP3} for example. This is why $H^1(M; \Z_2)$ contains the torsion part in addition to the $\mod 2$ reduction of $H^1(M; \Z)$ which would have been $(\Z_2)^{B_1}$ only.

\section{Anomalous Scaling on Boundary}
\label{app_scaling}

In this appendix we compute the Gaussian integral weighted by \eqref{boundary_momentum_space} to show \eqref{anomalous_scaling_result}. We shall also see that in the $i\epsilon$ prescription \eqref{iep_prescription_phi_chi} the discretized difference does not have to be taken at the lattice scale. For now we drop the last term of \eqref{boundary_action_W}; we will confirm its contribution to $I$ is of order $(\epsilon, \epsilon')\, |t_2-t_1|$ at the end of this appendix.

Performing the Gaussian integration \eqref{boundary_momentum_space} leads to $Z[W]/Z[0] = e^{I(x)}$ where $x\equiv x_2-x_1$ and
\begin{align}
I(x) &= -\frac{w^2}{2} \int_{-\pi}^\pi\frac{dp}{2\pi} \int_{-\pi}^\pi \frac{d\omega}{2\pi} \ \ \frac{\left(e^{ipx}-1\right) \left(e^{-ipx}-1\right)}{\left( e^{ip}-1 \right) \left( e^{-ip}-1 \right)} \nonumber \\[.25cm]
& \hspace{4.4cm} \frac{i\frac{k}{4\pi} \left( \left(e^{-i\omega}-1\right)\left(e^{ip}-1\right) + \left(e^{i\omega}-1\right) \left(e^{-ip}-1\right) \right)}{\left(\frac{n}{2\pi}\right)^2\left(e^{-i\omega}-1\right) \left(e^{i\omega}-1\right) + 2\epsilon\, (\mbox{numerator})}.
\label{I_omega_p}
\end{align}
The denominators have zero modes at $\omega=0$ and $p=0$, but the integrand is always well-behaved. For $p=0$, the observables are decoupled and the integrand vanishes; this is related to the fact that the $p=0$ mode corresponds to the global symmetry \cite{Elitzur:1989nr}. For $\omega=0$, the divergence of the integrand is regularized by $1/\epsilon$. We have safely dropped the $\epsilon'$ in \eqref{boundary_momentum_space_matrix}; if $V$ were present as well, the $\epsilon$ term alone is not sufficient to regularize the $\omega=0$ mode, then the $\epsilon'$ term cannot be neglected -- we will encounter a similar issue in Section \ref{ssect_modular}. For now we consider $W$ only for simplicity.

We perform the $\omega$ integral by denoting $\zeta \equiv e^{i\omega}$, $d\omega/2\pi = d\zeta/ 2\pi i \zeta$ and integrating $\zeta$ around the unit circle. The above reads
\begin{align}
I(x) = -\frac{w^2}{2}\int_{-\pi}^\pi \frac{dp}{2\pi} \oint \frac{d\zeta}{2\pi i \zeta} \ \ \frac{\left(e^{ipx}-1\right) \left(e^{-ipx}-1\right)}{\left( e^{ip}-1 \right) \left( e^{-ip}-1 \right)} \ \frac{i\frac{\pi k}{n^2} \left(1-e^{-ip}\right) \left(e^{ip}+\zeta\right)}{(\zeta-1) + 2\epsilon \, (\mbox{numerator})}.
\label{I_zeta_p}
\end{align}
There are two poles of $\zeta$ located at
\begin{align}
\zeta =0, \ \ \ \ \ \zeta -1 = 2\epsilon\: \frac{2\pi k}{n^2} \sin p \ + \ \mathcal{O}(\epsilon^2).
\end{align}
The second pole, corresponding to the $\omega\simeq 0$ mode, is picked up if $k \sin p < 0$. Therefore
\begin{align}
I(x) &= -\frac{w^2}{2}\int_{-\pi}^\pi \frac{dp}{2\pi} \ \frac{\left(e^{ipx}-1\right) \left(e^{-ipx}-1\right)}{\left( e^{ip}-1 \right) \left( e^{-ip}-1 \right)} \ \frac{\pi k}{n^2} \left( -2\sin p \ \theta(-k \sin p) - i(e^{ip}-1) \right) \nonumber\\[.2cm]
&= -\frac{\pi |k|\, w^2}{n^2}\int_0^\pi \frac{dp}{2\pi} \ \frac{1-\cos(px)}{1-\cos p} \: \sin p \ - \ \frac{i\pi k w^2}{2n^2}.
\label{I_p}
\end{align}
Note that the pole at $0$ gives rise to the imaginary part, while the pole near $1$ gives rise to the real part which will show the anomalous scaling behavior. Clearly, the role of $\epsilon$ is only to determine whether the pole near $1$ is inside or outside of the unit circle. Therefore we may consider replace $\epsilon$ with $\epsilon h(p)$ where $h(p)$ is any smooth, positive, even function. Physically this means we are allowed to smear out the lattice second derivative in \eqref{iep_prescription_phi_chi}.

We can analogously perform the $p$ integral. But before we do so, we can readily estimate the $x$ dependence at large $x$. The large $x$ behavior should be dictated by the small $p$ expansion of the integrand of $\Re\, I(x)$, which reduces to the corresponding expression in the continuum theory, and hence the scaling must be the same. Let's confirm this intuition explicitly. Let's shift $x\rightarrow x+1$, under which
\begin{align}
I(x+1)-I(x) = -\frac{\pi |k|\, w^2}{n^2}\int_0^\pi \frac{dp}{2\pi} \ \left(\cos(px) + \frac{\sin p}{1-\cos p}\: \sin(px) \right) \sin p.
\label{I_difference}
\end{align}
Here $\sin(px)$ and $\cos(px)$ are rapidly oscillating at large $x$, and the integrand is finite for all values of $p$, so the integral vanishes as $x\rightarrow \infty$. We now extract the leading $1/x$ behavior. This can be extracted by computing $x \left(I(x+1)-I(x)\right)$ with $x\rightarrow \infty$. The multiplication of $x$ can be expressed as
\begin{align}
x \left(I(x+1)-I(x)\right) = -\frac{\pi |k|\, w^2}{n^2}\int_0^\pi \frac{dp}{2\pi} \ \left(\partial_p \sin(px) - \frac{\sin p}{1-\cos p}\: \partial_p\cos(px) \right) \sin p.
\end{align}
Now we integrate $p$ by parts, so there is a boundary term and another integral. The integral involving $\sin(px)$ and $\cos(px)$ with finite integrand again vanishes as $x\rightarrow \infty$, but the boundary term remains finite:
\begin{align}
x\left(I(x+1)-I(x)\right) \ \longrightarrow \ -\frac{|k|\, w^2}{2n^2} \left. \left( \sin p \, \sin(px) - \frac{(\sin p)^2}{1-\cos p} \, \cos(px) \right) \right|_{p=0}^{p=\pi} \ = \ -\frac{|k|\, w^2}{n^2}.
\end{align}
This means $\Re \: I(x) \sim -\frac{|k|\, w^2}{n^2} \ln |x|$ at large $x$, and therefore $|e^{I(x)}| \sim |x|^{-\frac{|k|\, w^2}{n^2}}$ with the expected anomalous scaling dimension.

For completeness, let's evaluate the real part of \eqref{I_p} explicitly. We use the Fourier series
\begin{align}
\theta(\sin p) = \frac{1}{2} + \sum_{n\neq 0} \frac{-i}{\pi (2n-1)} \ e^{ip(2n-1)} 
\end{align} 
and the polynomial division $(e^{ip|x|}-1)/(e^{ip}-1) = \sum_{m=0}^{|x|-1} e^{ipm}$. We find
\begin{align}
\Re\: I(x) &= -\frac{\pi |k|\, w^2}{2n^2}\int_{-\pi}^\pi \frac{dp}{2\pi} \left( 2\sum_{m=1}^{|x|-1} \left( e^{ipm} - e^{-ipm} \right) + e^{ip|x|}-e^{-ip|x|} \right) \sum_{n\neq 0} \frac{-1}{\pi (2n-1)} \: e^{ip(2n-1)} \nonumber \\[.25cm]
&= -\frac{|k|\, w^2}{n^2} \left( \sum_{m=1}^{|x|-1} \frac{1-(-1)^m}{m} + \frac{1-(-1)^{|x|}}{2|x|} \right)
\label{long_step}
\end{align}
where in the second equality we picked up terms of zeroth power in $e^{ip}$.

Finally, we shall confirm that the last term of \eqref{boundary_action_W} yields order $\epsilon\, |t_2-t_1|$ contributions which are negligible. Including this omission (arising from the $w$ dependent contributions to the $\epsilon$ prescription, as explained in Figure \ref{f5-1_corr}) leads to extra terms in the numerator of the first factor of \eqref{I_zeta_p}, with $\zeta$ dependence
\begin{align}
\epsilon\, \zeta^j, \ \ \ \epsilon\, \zeta^{-j'}, \ \ \ \epsilon^2\, \zeta^{j-j'}
\end{align}
for $j, j'= 1, \cdots, |z_2-z_1|$. They change the behavior of the $\zeta=0$ pole. However, for both the $\zeta=0$ pole and the $\zeta\simeq 1$ pole, the contribution to $I$ due to each $\epsilon\, \zeta^{\pm j}$ term is of the form $\epsilon\, (1+\mathcal{O}(\epsilon))^{\pm j}$. Therefore, the overall contribution to $I$ is order $\mathcal{O}(\epsilon\, |z_2-z_1|)$. Likewise if we restore $\epsilon'$.

\section{Effective Spatial Translation on Boundary}
\label{app_P_eff}

In this appendix we confirm the scaling behavior \eqref{transl_scaling}. We take $\beta, C\rightarrow \infty$ for simplicity because the scaling is the leading behavior insensitive to whether $\beta, C$ are finite. 

The result without the spatial translation is given in the previous appendix. With the $P_-$ insertion, the solution of $a$ explained by Figure \ref{f5-1_corr} must be inserted into the expression \eqref{momenta_a_b} of $P_-$. This has two effects. First, the $\phi\phi$ matrix element in \eqref{boundary_momentum_space_matrix} is modified:
\begin{align}
\left[ \begin{array}{cc} 
i\frac{k}{4\pi} \left( \left( e^{-i\omega} - 1 \right) \left( e^{ip} - 1 \right) + c.c. \right) & i\frac{n}{2\pi}\left( e^{i\omega} - 1 \right) \left( e^{-ip} - 1 \right) \\[.2cm]
i\frac{n}{2\pi} \left( e^{-i\omega} - 1 \right) \left( e^{ip} - 1 \right) &  \left(-i\frac{n^2}{2\pi k} vf(p) + 2\epsilon\right) \left( e^{ip} - 1\right) \left(e^{-ip}-1\right) \end{array} \right].
\end{align}
\footnote{As mentioned in the previous appendix, when considering the $W$ insertion only, we may ignore $\epsilon'$. Note that $v$ in the expression is the ``velocity'' introduced in \eqref{transl_scaling}, \emph{not} the charge of a $V$ Wilson loop.}
Second, in the linear term in \eqref{boundary_momentum_space}, there is an additional contribution
\footnote{The origin of this term is due to the substitution of the solution on the red plaquettes in Figure \ref{f5-1_corr} into the expression \eqref{momenta_a_b} of $P_-$ (there is also a constant term that we dropped). The correct realization of this term is why we demanded the bulk placement of $W$ to be at least into $y=1+1/2$: as explained via Figure \ref{f5-1_corr}, this ensures there are two columns of red plaquettes touching the $y=0$ boundary; while if the bulk placement is at $y=1/2$, the left column is missed.}
\begin{align}
& iw\, e^{-i\omega t_1} (e^{ipx_2}-e^{ipx_1}) \phi_{(\omega, p)} + iv f(p) \frac{n^2}{4\pi k} \sum_{t'=t_1}^{t_2-1} \frac{-2\pi k w}{n^2} \left(e^{ip-i\omega}-e^{-i\omega}+1-e^{-ip}\right) e^{ipx_1-i\omega t'} \phi_{(\omega, p)} \nonumber \\[.2cm]
&= \ iw \left( \frac{e^{ipx}-1}{e^{ip}-1} - vf(p) \, \frac{e^{-i\omega}+e^{-ip}}{2} \, \frac{e^{-i\omega t}-1}{e^{-i\omega}-1} \right) \left(e^{ip}-1\right) e^{ipx_1-i\omega t_1}\: \phi_{(\omega, p)}.
\end{align}
Here we assumed $t\equiv t_2-t_1>0, \ x\equiv x_2-x_1>0$ for definiteness, and we used the polynomial division $(e^{-i\omega t}-1)/(e^{-i\omega}-1) = \sum_{j=0}^{t-1} e^{-i\omega j}$. The two terms in the parenthesis are very similar. More particularly, at small $p$, we expect there to be an energy mode $\omega\simeq -vp$, and subsequently, $vf(p) (e^{ip}+e^{-ip})/2 \simeq v$ to leading order. This intuition suggests the result \eqref{transl_scaling}. In the below we verify this intuition via explicit calculations.

Performing the Gaussian integral, the left-hand-side of \eqref{transl_scaling} is $|e^{I(t, x)}|$ where
\begin{align}
I(t, x) &= -\frac{i\pi k w^2}{2n^2} \int_{-\pi}^\pi\frac{dp}{2\pi} \int_{-\pi}^\pi \frac{d\omega}{2\pi} \ \left| \frac{e^{ipx}-1}{e^{ip}-1} - vf(p) \, \frac{e^{-i\omega}+e^{-ip}}{2} \, \frac{e^{-i\omega t}-1}{e^{-i\omega}-1} \right|^2 \nonumber \\[.25cm]
& \hspace{3cm} \frac{\left( \left(e^{-i\omega}-1\right)\left(e^{ip}-1\right) + \left(e^{i\omega}-1\right) \left(e^{-ip}-1\right) \right)}{\left(e^{-i\omega}-1\right) \left(e^{i\omega}-1\right) +\left( vf(p) / 2 + i\epsilon \, \sgn(k)\, \right) (\mbox{numerator})}
\label{Ical_omega_p}
\end{align}
where we have rescaled $\epsilon$ by $n^2/2\pi |k|$ for convenience; clearly $I(t, x)$ reduces to the previous $I(x)$ when $v=0$. We are only interested in the real part $\Re\, I(t, x)$ responsible for the scaling behavior. We note $I(t, x)$ would be purely imaginary were it not for the $i\epsilon$ term. Similar to the previous appendix, we can extract the effect of the $i\epsilon$ by looking at whether the pole of $\zeta\equiv e^{i\omega}$ in the last fraction of \eqref{Ical_omega_p} is inside the unit circle (this is equivalent to performing the principle function decomposition $\Im (x\pm i\epsilon)^{-1} = \mp\pi\delta(x)$, but more explicit). The last fraction can be decomposed as
\begin{align}
\left( \frac{(1-e^{-ip})}{1+ vf(p) (1-e^{-ip})/2} \ + \ \frac{\frac{\left(1-e^{-ip}\right) \left( e^{ip} + 1\right)}{\left(1+ (vf(p)/2) (1-e^{-ip}) \right)^2}}{\zeta - \frac{1+(vf(p)/2 - i\epsilon \, \sgn(k))(1-e^{ip})}{1+ (vf(p)/2 -i\epsilon \, \sgn(k))(1-e^{-ip})}} \right).
\label{big_factor}
\end{align}
Note that $\epsilon$ is only important in the pole position of $\zeta$, because this pole has magnitude $1+\mathcal{O}(\epsilon)$, almost on the unit circle on which $\zeta$ resides. More particularly, this pole is picked up inside the unit circle if and only if $k\sin p<0$ (similar to that in \eqref{I_zeta_p}). Let's denote this pole as $\zeta = \exp(-iE(p)) + \mathcal{O}(\epsilon)$, corresponding to the $\omega= -E(p)$ mode. One may check the properties that $-E(p)<0$ for $k\sin p<0$ (so we pick up the negative energy modes as expected, because the positive energy modes are viewed as excitations), and that $E(p) \sim v|p|$ at small $p$. The real part $\Re\, I(t, x)$ is the contribution from picking up this pole (note that there is a factor of $\zeta^{-1}$ from the Jacobian of $d\omega$):
\begin{align}
\Re\, I(t, x) &= -\frac{\pi |k| w^2}{n^2} \int_0^\pi\frac{dp}{2\pi} \ \left| \frac{e^{ipx}-1}{e^{ip}-1} - vf(p) \, \frac{e^{iE(p)}+e^{-ip}}{2} \, \frac{e^{iE(p) t}-1}{e^{iE(p)}-1} \right|^2 \nonumber \\[.25cm]
& \hspace{3.6cm} \frac{\sin p}{\left|1+ vf(p) (1-e^{-ip})/2 \right|^2}
\end{align}
which is a generalized version of the real part of \eqref{I_p}.
\footnote{Note the denominator $\left(1+ vf(p) (1-e^{-ip})/2 \right)$ does not vanish given our requirements on $f(p)$.}

Now consider $x, t\gg 1$. From the steps following \eqref{I_difference} we learn that the fast oscillating integral, and hence the scaling behavior, is controlled by the integrand at small $p$. The expansion of the integrand at small $p$ is $(x-vt)^2\, p$, in comparison to $x^2\, p$ when $v=0$. Therefore \eqref{transl_scaling} must hold at large spacetime separations.

\section{Zero-Point Momenta on Boundary}
\label{app_zpm}

In this appendix we compute the expectation value of $P_-$ given by \eqref{momenta}, and obtain \eqref{P_expectation}. In the end we comment on the expectation value of the total momentum $P$.

With finite $C$ and infinite $\beta$, the Fourier transformation becomes
\begin{align}
\phi_{(t, x)} = \frac{1}{C} \sum_{p \in (2\pi/C) \Z_C} \int_{-\pi}^\pi \frac{d\omega}{2\pi} \ e^{-i\omega t + ipx} \ \phi_{(\omega, p)}, \ \ \ \ \ \phi_{(\omega, p)} = \sum_{x\in \Z_C} \sum_{z\in \Z} \ e^{i\omega t -ipx} \ \phi_{(t, x)}.
\label{discrete_Fourier}
\end{align}
According to \eqref{momenta}, $P_-$ has $\phi_{(\omega, p)} \phi_{(\omega', p')}$ matrix element
\begin{align}
-\frac{n^2}{4\pi k} \ C\, \delta_{p+p'} \ \left( e^{ip}-1 \right) \left( e^{-ip}-1 \right) f(p) \ e^{-i(\omega+\omega')t},
\end{align}
while the $\phi_{(\omega', p')} \phi_{(\omega, p)}$ component of the inverse of \eqref{boundary_momentum_space_matrix} is
\footnote{Again we can safely drop $\epsilon'$ since $P_-$ has $\phi\phi$ entry only.}
\begin{align}
C\, \delta_{p'+p} \: 2\pi \, \delta(\omega'+\omega) \: \frac{1}{\left( e^{ip}-1 \right) \left( e^{-ip}-1 \right)} \frac{i\frac{k}{4\pi} \left( \left(e^{-i\omega}-1\right)\left(e^{ip}-1\right) + \left(e^{i\omega}-1\right) \left(e^{-ip}-1\right) \right)}{\left(\frac{n}{2\pi}\right)^2 \left(e^{-i\omega}-1\right) \left(e^{i\omega}-1\right) + 2\epsilon\, (\mbox{numerator})}.
\end{align}
Tracing them with the measure specified in \eqref{discrete_Fourier} leads to the expectation value 
\begin{align}
-\frac{i}{4} \sum_{p \in (2\pi/C) \Z_C} \int_{-\pi}^\pi\frac{d\omega}{2\pi} \ \ f(p) \ \frac{ \left( \left(e^{-i\omega}-1\right)\left(e^{ip}-1\right) + \left(e^{i\omega}-1\right) \left(e^{-ip}-1\right) \right)}{ \left(e^{-i\omega}-1\right) \left(e^{i\omega}-1\right) + i\epsilon \, \sgn(k) \, (\mbox{numerator})}
\end{align}
where we have rescaled $\epsilon$ by $n^2/2\pi |k|$ for convenience. We perform the $\omega$ integral as we did for \eqref{I_omega_p}, and find
\begin{align}
& \ \frac{1}{4} \sum_{p \in (2\pi/C) \Z_C} f(p) \left( 2\sin p \ \theta(-k \sin p) + i(e^{ip}-1) \right) \nonumber \\[.2cm]
=& \ -\sgn(k) \sum_{p \in (2\pi/C) \Z_C} \theta(\sin p) \: f(p) \: \frac{\sin p}{2} - \sum_{p \in (2\pi/C) \Z_C} f(p) \: \frac{i}{4} \left(1-\cos p\right).
\label{expectation_finite_size}
\end{align}
We can estimate the $C\rightarrow \infty$ behavior using the standard Euler-Maclaurin approximation. Let's consider the real part first, which is supported on $0<p<\pi$. Given the assumed properties of $f(p)$, the function to be summed over is approximately $p/2$ for $p\gtrsim 0$ and smoothly vanishes for $p\lesssim \pi$;
\footnote{Near $p\lesssim \pi$, $f(p)$ is suppressed as $\exp(-(\pi R)^2/2)$ which is much smaller than any scale of interest. It is actually consistent to just set $f(p)$ to $0$ for order $1$ values of $p$.}
in particular, a kink is seen across $p=0$, but not across $p=\pi$. We may approximate the summation by an integration when $C\gg 1$, but there is a universal deviation due to the kink near $p=0$. More particularly, for $p\gtrsim 0$, the summation is approximately $\pi/C$ times $\left(1+2+3+ \cdots\right)$, whose said deviation is well-known to be $-1/12$:
\begin{align}
\sum_{p \in (2\pi/C) \Z_C} \theta(\sin p) \: f(p) \: \frac{\sin p}{2} \  = \ C\int_0^\pi \frac{dp}{2\pi} \ f(p) \ \frac{\sin p}{2} \ + \ \frac{\pi}{C} \left( -\frac{1}{12} \ + \ \mathcal{O}(R/C) \right).
\end{align}
The integration approximation term is of order $1/R^2$. Now let's consider the imaginary part. The function to be summed over is smooth over the entire $2\pi$ domain without a kink, so the summation is well approximated by integration: 
\begin{align}
\sum_{p \in (2\pi/C) \Z_C} f(p)\: \frac{i}{4} \left(1-\cos p\right) \ = \ C\int_{-\pi}^{\pi} \frac{dp}{2\pi} \ f(p)\: \frac{i}{4} \left(1-\cos p\right) \ + \ \frac{i\pi}{2C} \ \mathcal{O}(R/C) .
\end{align}
The integration approximation term is of order $1/R^3$.

We can study the summation \eqref{expectation_finite_size} more closely and show the finite size correction to its imaginary part vanishes exactly. Recall that $f(p) = \sum_{r=-\lfloor C/2\rfloor}^{\lfloor C/2\rfloor} f(r) \, e^{-ipr}$. In the imaginary part of \eqref{expectation_finite_size}, $f(p) \left(1-\cos p\right)$ is a finite Laurent series in terms of $e^{ip}$ with power bounded by $\pm (\lfloor C/2\rfloor +1)$, so both the summation over $p \in (2\pi/C) \Z_C$ and the integration over $p$ picks out the zeroth power term in the Laurent series only:
\begin{align}
\sum_{p \in (2\pi/C) \Z_C} f(p) \: \frac{i}{4} \left(1-\cos p\right) \ = \ C\int_{-\pi}^{\pi} \frac{dp}{2\pi} \ f(p) \: \frac{i}{4} \left(1-\cos p\right) \ = \ \frac{iC}{4} \left(f(r=0)-f(r=1)\right)
\end{align}
which is exactly linear in $C$. On the other hand, in the real part,
\begin{align}
f(p) \: \sin p \ \left(\theta(\sin p) - 1/2\right) = \sum_{r=-\lfloor C/2\rfloor}^{\lfloor C/2\rfloor} f(r) \, e^{-ipr} \ \frac{-i\left(e^{ip}-e^{-ip}\right)}{2} \ \sum_{n\neq 0} \frac{-i}{\pi (2n-1)} \ e^{ip(2n-1)}
\end{align}
(we subtract the $1/2$ from the step function since its contribution is odd in $p$ and vanishes anyways). The integration over $p$ only picks out the terms of zeroth power in $e^{ip}$, i.e. those with $2n-1 \pm 1 = r$, whilst the summation over $p \in (2\pi/C) \Z_C$ picks out infinitely many terms with $2n-1 \pm 1 = r \mod C$, giving rise to the finite size corrections.

Finally we comment on the total $\langle P \rangle$. One may check that in picking the poles, there is \emph{no} appearance of $\theta(k \sin p)$ as in \eqref{expectation_finite_size}. Hence there is no finite size effect, and the summation over $p$ is exactly the same as the integral over $p$ which is linear in $C$, for the same reasons as above. This is just the expected property that the theory is non-chiral.

\bibliography{AbTopo}{}
\bibliographystyle{utphys}

\end{document}